\documentclass[authoryear,preprint, 12pt]{elsarticle}
\usepackage[utf8]{inputenc}
\usepackage[utf8]{inputenc}
\usepackage[english]{babel}
\usepackage{blindtext}
\usepackage[left=2.5cm, right=2.5cm, top=2.5cm, bottom=2.5cm]{geometry} %% sets margins
\usepackage{amsmath,amsfonts}
\usepackage{amssymb}
\usepackage{algorithm}
\usepackage{algcompatible}
\usepackage{algpseudocode}
\usepackage{natbib}
\usepackage{setspace}
\usepackage{graphicx}
\usepackage{caption}
\usepackage{subcaption}
\usepackage{hyperref}
\hypersetup{
    colorlinks,
    allcolors=black
}
\usepackage{mathrsfs}
\usepackage{amsthm}
\usepackage{tikz}
\usepackage[section]{placeins}
\usepackage{float}
\usepackage{mathtools}
\usepackage{bm}
\newcommand{\RR}{\mathbb{R}}

\newcommand{\shortx}{{\mathbf{y}}}
\newcommand{\longx}{{\mathbf{x}}}
\newcommand{\shortX}{{\mathbf{Y}}}
\newcommand{\longX}{{\mathbf{X}}}
\newcommand{\longE}{{\mathcal{E}_\longX}}

\newcommand{\wrt}[1]{\text{d}#1}

\newcommand{\Prb}[1]{\Pr\left(#1\right)}
\newcommand{\Prbb}[1]{\mathbb{P}\left(#1\right)}
\DeclareMathOperator*{\argmax}{argmax}

\theoremstyle{definition}

\journal{Ocean Engineering}

\begin{document}

\begin{frontmatter}

%% Title, authors and addresses
\title{{Sequential Design for the Efficient Estimation of Offshore
Structure Failure Probability}}

\author{Matthew Speers}
\author{Jonathan Tawn}
\author{Philip Jonathan}
\address{School of Mathematical Sciences, Lancaster University LA1 4YF, United Kingdom.}	

\begin{abstract}
Estimation of the failure probability of offshore structures exposed to extreme ocean environments is critical to their safe design and operation. The conditional density of the environment (CDE) quantifies regions of the space of long term environment responsible for extreme structural response. Moreover, the probability of structural failure is obtained by simply integrating the CDE over the environment space. In this work, two methodologies for estimation of the CDE and failure probability are considered. The first (IS-PT) combines parallel tempering MCMC ({Markov chain - Monte Carlo}, for CDE estimation) with important sampling (for eventual estimation of failure probability). The second (AGE) combines adaptive Gaussian emulation with Bayesian quadrature. {Both approaches provide large reductions in the number of function evaluations of the complex structural response to fluid loading, relative to naive brute-force calculations}. We evaluate IS-PT and two variants of the AGE procedure in application to a simple synthetic structure with multimodal CDE, and a monopile structure exhibiting non-linear resonant response. IS-PT provides reliable results for both applications. The AGE procedures require balancing exploration and exploitation of the environment space, using a typically-unknown weight parameter, $\lambda$. {When $\lambda$ is known, perhaps from prior engineering knowledge, AGE provides an order of magnitude reduction in computational cost relative to IS-PT. However, with $\lambda$ unknown, IS-PT is more reliable.}

\end{abstract}

\begin{keyword}
%% keywords here, in the form: keyword \sep keyword
Structural design, extreme, full probabilistic analysis, contour, importance sampling, bridge sampling, Gaussian process, active learning, significant wave height, wave steepness, monopiles.
%% PACS codes here, in the form: \PACS code \sep code
\PACS 0000 \sep 1111
%% MSC codes here, in the form: \MSC code \sep code
%% or \MSC[2008] code \sep code (2000 is the default)
\MSC 0000 \sep 1111
\end{keyword}

\end{frontmatter}

% % \setlength{\parskip}{0pt plus 1fil}
% \newpage
% \tableofcontents
% \newpage
% % \setlength{\parskip}{1em}

\section{Introduction} \label{section:introduction}

\subsection{Background}

An offshore structure (such as an oil platform or wind turbine) is subject to environmental loading, e.g., from winds, waves and currents. The ocean engineer seeks to evaluate the risk posed to structural integrity by the environment, enabling the structure to be designed and maintained to the required level of reliability. Often, this involves computationally demanding fluid loading and structural response calculations. Therefore, the design of computationally efficient approaches for assessment of structural risk is a topic of considerable importance.   

Take an environmental variable $\longX$ (such as significant wave height $H_S$) characterising the long term metocean environment on space $\mathcal{E}_\longX$. The short term environment variable $\shortX$ (such as individual wave height $H$) defined on space $\mathcal{E}_\shortX$, depends on $\longX$. This $\shortX$ is stochastic given $\longX$, in the sense that many values of $\shortX$ may be summarised by a single value $\longx$ of $\longX$, in terms of the distribution for $\shortX|\{\longX=\longx\}$. Given complete knowledge of the short term conditions $\shortX$, along with a physical model for the response $\mathbf{R} \in {\mathcal{E}_{\mathbf{R}}}$ induced on the structure by $\shortX$, it is possible to characterise the \textit{multivariate} structural response induced on the structure fully. Typically, ${\mathcal{E}_{\mathbf{R}}}=\RR^d$ for some dimension $d>0$. 

In our setting, we assume the existence of a deterministic function $g_{\mathbf{R}}(\shortx) : \mathcal{E}_\shortX \mapsto {\mathcal{E}_{\mathbf{R}}}$ for the structural response $\mathbf{R}=\mathbf{r}$ induced by the environment $\shortX =\shortx$. Practitioners do not typically have knowledge of the full short term environment $\shortX$, but instead have information on the long term summary variable $\longX$. Since $\shortX$ is not a deterministic function of $\longX$, practitioners estimate the density function $f_{\shortX|\longX}(\shortx|\longx):  \mathcal{E}_\shortX \times \longE \mapsto \RR^+$ of the short term environment $\shortX|\{\longX=\longx\}$. In practice, evaluation of $g_{\mathbf{R}}$ and $f_{\shortX|\longX}$ can be computationally expensive, involving complex load calculations and the simulation of 3-dimensional wave and wind fields.

Given the functions $g_{\mathbf{R}}$ and $f_{\shortX|\longX}$, we can evaluate the density $f_{\mathbf{R}|\longX}(\mathbf{r}|\longx):{\mathcal{E}_{\mathbf{R}}}\times\longE \mapsto \RR^+$ as
\begin{equation} \label{equation:resp_density}
    f_{\mathbf{R}|\longX}(\mathbf{r}|\longx) = \int_{\mathcal{E}_\shortX} g_\mathbf{R}(\shortx) f_{\shortX|\longX}(\shortx|\longx) \wrt \shortx,
\end{equation}
for $\mathbf{R}|\{\longX=\longx\}$, the multivariate response conditioned on the long term environment. Again, evaluating $f_{\mathbf{R}|\longX}$ can be prohibitively expensive, due the potential complexities of $f_{\shortX|\longX}(\longx)$ and $g_\mathbf{R}(\shortx)$. 
A natural approach to quantify the risk to a structure is then to estimate the probability of failure $p$ due to response $\mathbf{R}$ and environment $\longX$. For $\mathbf{R} = (R_1, \ldots, R_d)\in {\mathcal{E}_{\mathbf{R}}}$, this can be written
\begin{equation}
    p = \Prbb{\bigcup_{i=1}^d (R_i >r_\text{Cr}^{(i)}) } = 1- \Prbb{\bigcap_{i=1}^d (R_i <r_\text{Cr}^{(i)}) },
\end{equation}
the probability that at least one response component $R_i$, $i=1,\ldots, d$, exceeds its critical level $r_\text{Cr}^{(i)} \in \RR$. This can be written using \eqref{equation:resp_density} as
\begin{equation}\label{equation:multi_d_matt_target}
    p = \int_\longE \left\{ \int_{\mathcal{E}_{\mathbf{R}}}  \Bigg[ 1 -   \left( \prod^d_{i=1} I(R_i<r_\text{Cr}^{(i)}|\{\longX=\longx\})\right)\Bigg] f_{\mathbf{R}|\longX}(\mathbf{r}|\longx) \wrt \mathbf{r} \right\} f_\longX(\longx) \wrt \longx,
\end{equation}
{where $I$ is the indicator function which takes the value unity when its argument is true, and value zero otherwise,} and the integral evaluated numerically by sampling repeatedly from models for $\mathbf{R}|\{\longX=\longx\}$ and $\longX$. Throughout, we assume the density $f_\longX$ of the long term environment $\longX$ is either known or estimable, possibly using extreme value techniques (e.g., as in Section~\ref{section:application}). Evaluation of \eqref{equation:multi_d_matt_target} is thus solely made difficult by the computational expense required to obtain draws of $\mathbf{R}|\{\longX=\longx\}$.

We aim to minimise the uncertainty in estimating \eqref{equation:multi_d_matt_target} given a budget of a set number of realisations of $\mathbf{R}|\{\longX=\longx\}$. We use the available budget efficiently by making informed choices about the values of $\longX$ at which to sample from $\mathbf{R}|\{\longX=\longx\}$. Typically, methods for the efficient evaluation of \eqref{equation:multi_d_matt_target} target values of $\longX$ contributing most to the integral. In the simplest terms, this is achieved by targetting regions where the value of the integrand
\begin{align}\label{equation:cde_mult}
    \tilde{f}_{\longX}(\longx ; \mathbf{r}_\text{Cr}) &= \left\{ \int_{\mathcal{E}_{\mathbf{R}}}  \bigg[ 1 -   \left( \prod^d_{i=1} I(R_i<r_\text{Cr}^{(i)}|\{\longX=\longx\})\right)\bigg] f_{\mathbf{R}|\longX}(\mathbf{r}|\longx) \wrt \mathbf{r} \right\} \times f_\longX(\longx) \\
    &= \Prbb{\text{`failure'}|\left\{\longX=\longx\right\}} \times f_\longX(\longx), 
\end{align}
in \eqref{equation:multi_d_matt_target} is large, where $\mathbf{r}_\text{Cr} = \left(r_\text{Cr}^{(1)}, \ldots, r_\text{Cr}^{(d)}\right)$ is the vector of critical values of responses. That is, it is beneficial to target values of the long-term environmental variables that are both likely to occur (large $f_\longX(\longx)$) and to induce structural failure (large $ \Prbb{\text{`failure'}|\left\{\longX=\longx\right\}}$). We subsequently refer to $ \tilde{f}_{\longX}(\longx ; \mathbf{r}_\text{Cr})$ defined in \eqref{equation:cde_mult} as the conditional density of the environment (CDE), as it is the unnormalised long-term environment density conditional on the occurrence of structural failure; we use $\tilde f$ (rather than $f$) to indicate an unnormalised density.

\cite{peherstorfer2016multifidelity}, \cite{Yang2018} and \cite{Wang2021} show that minimising the uncertainty in \eqref{equation:multi_d_matt_target} can be achieved for an arbitrary multi-dimensional response. We restrict ourselves to $d=1$, with $\mathbf{R}=R$ and $\mathbf{r}_\text{Cr} = r_\text{Cr}$, for brevity and ease of presentation. In this case, equation \eqref{equation:cde_mult} reduces to
\begin{equation}\label{equation:cde}
    \tilde{f}_{\longX}(\longx ; r_\text{Cr}) = \Prbb{R>r_\text{Cr}|\{\longX=\longx\}} \times f_\longX(\longx).
\end{equation}
Existing methodologies reducing uncertainty in \eqref{equation:multi_d_matt_target} by targetting \eqref{equation:cde} include: \textit{sampling} methods such as importance sampling (see e.g., \citealt{castellon2022investigations}) and bridge sampling \citep{0f232807-5558-3c25-9446-537f6b18f086}; \textit{adaptive Gaussian emulation} (e.g., \citealt{gramstad2020sequential} and \citealt{lystad2023full}); and approaches combining sampling and adaptive emulation (e.g., \citealt{castellon2023full} and \citealt{Xiao2020}). 
Good sampling methods reduce the variance of a target integral for a given sampling budget, whilst emulation provides a cheaper approximate route to otherwise expensive complex function evaluation. Relevant recent reviews are given by \cite{Moustapha2022}, \cite{wang2022recent}, \cite{tabandeh2022review} and \cite{Marrel2024}.

In simple cases, we might expect that the CDE $\tilde{f}_{\longX}$ is approximately elliptically-contoured (e.g., \citealt{speers2024estimating}), and therefore well-approximated by a unimodal Gaussian-like density in $\mathcal{E}_\longX$. However, in reality there are good reasons to expect this not to be the case in general, due to e.g., the presence of multiple failure modes or resonant responses. In the current work, we are particularly interested in investigating methodologies to estimate such complex CDE structures well. 

We choose to investigate the efficient estimation of \eqref{equation:cde} in the context of designing monopile structures. We choose this structural type for two reasons: firstly, because it provides a useful template structure for generic studies of fluid loading; and secondly, it is of itself a relevant structural type for e.g., offshore wind applications. This thinking motivates the synthetic study of Section \ref{section:synthetic_study}, and the wind turbine application of Section \ref{section:application}.

\subsection{Objectives and outline}

The objective of the current work is to explore methodologies based on efficient sampling or adaptive Gaussian emulation, to estimate the conditional density of the environment (CDE) and thereby failure probabilities for synthetic and real-world monopile structures. In Section~\ref{section:methodology}, we first describe an approach, termed IS-PT, coupling importance sampling with parallel tempering Markov chain-Monte Carlo (MCMC) \citep{earl2005parallel} for estimation of multi-modal CDEs, a scenario which has received little attention in the offshore reliability literature. Secondly, building on \cite{gramstad2020sequential} and \cite{cohn1993neural}, we consider two variants of an alternative approach, termed AGE, based on adaptive Gaussian emulation, adopting an acquisition function promoting sampling which balances exploration and exploitation of regions of $\longE$ contributing to the CDE. In Section~\ref{section:synthetic_study}, the approaches from Section~\ref{section:methodology} are applied for a synthetic monopile structure exhibiting different resonant responses, to evaluate their respective performance. We find that all approaches provide good estimation of failure probability, but that AGE approaches require fewer expensive function evaluations provided that the required balance between exploitation and exploration of $\longE$ is assumed known. If this balance is unknown, IS-PT provides a more reliable procedure. In Section~\ref{section:application}, we demonstrate good performance of all approaches in a more realistic setting, estimating the structural failure probability for oscillating monopiles, with harmonic response modelled using the T-FNV (Transformed - Faltinsen, Newman and Vinje, \citealt{Faltinsen_Newman_Vinje_1995}) model of \cite{taylor2024transformed}. Our findings here regarding the relative computational complexities of IS-PT and AGE approaches are similar to those for the synthetic case. Discussion and conclusions are provided in Section~\ref{section:discussion}. Online supplementary material (SM) provides supporting description of methodology and results.

% {\color{red}
% \begin{itemize}
%     \item there's also a lot of literature on general reliability analysis using emulation and/or importance sampling \citep{bichon2008efficient, Echard2011, Echard2013, Yun2018, Zhang2019, Ameryan2022, Zuniga2021, Dubourg2013, kurtz2013cross, lelievre2018ak, Liu2019,  Gaspar2017, bugallo2017adaptive, Xiong2021, Fauriat2014, cheng2020structural, Guo2021, renganathan2023camera, lu2023agp, cheng2023rare, cole2023entropy, mohamad2018sequential}
%     \item Early mention of the idea of using MCMC sample to give kernel smoothed proposal (https://www.sciencedirect.com/science/article/pii/S0167473099000144)
% \end{itemize}}
\section{Methodology} \label{section:methodology}

\subsection{Overview of methodologies} \label{section:method_overview}

We begin by discussing two methods for the efficient evaluation of integral \eqref{equation:multi_d_matt_target}. In Section~\ref{section:monte_carlo}, we introduce an importance sampling scheme coupled with an adaptive parallel tempering MCMC algorithm, designed for scenarios where the CDE $\tilde{f}_{\longX}$ is multimodal. In Section~\ref{section:emulation}, we describe an emulator replacing expensive draws of the structural response $R|\{\longX=\longx\}$ with predictions from a Gaussian process, and provide methods for the adaptive design of the emulator training set. 

We emphasise that these methods are introduced as \textit{alternative} options for the efficient estimation of \eqref{equation:multi_d_matt_target}, both seeking to minimise the target error within some set budget of expensive function evaluations. These approaches will then be compared in \eqref{section:synthetic_study} to see which performs better.

\subsection{MCMC-informed importance sampling}  \label{section:monte_carlo}

In offshore reliability, importance sampling methods select values $\longx_1^*, \ldots, \longx_{n_\text{IS}}^*$, $n_\text{IS}>0$, of $\longX$ at which to evaluate $R|\{\longX=\longx\}$, to make efficient use of limited computational resources (see e.g., \citealt{castellon2022investigations}). These include traditional importance sampling techniques, or extensions such as bridge sampling (e.g., \citealt{0f232807-5558-3c25-9446-537f6b18f086}). In this article, we focus our attention on the former, since our initial investigations of bridge sampling showed no improvement in performance, despite increased computational cost.

Evaluation points are drawn from a proposal distribution $g_\text{Pr}$, chosen so that values $\longx$ with higher density $g_\text{Pr}(\longx)$ are more informative to the target quantity. Our target quantity is the marginal structural failure probability \eqref{equation:multi_d_matt_target}, which may be written
\begin{equation}\label{equation:IS_true_int}
    p = \int_{\mathcal{E}_\longX} \Prbb{R>r_\text{Cr}|\{\longX=\longx\}}  \frac{f_\longX(\longx)}{g_\text{Pr}(\longx)} g_\text{Pr}(\longx) \wrt{\longx},
\end{equation}
approximated by the importance sampling estimate
\begin{equation} \label{equation:paper_p1IS}
      \hat{p}_\text{IS} = {\frac{1}{n_\text{IS}}}\sum^{n_\text{IS}}_{i=1}\Prbb{R>r_\text{Cr}|\{\longX=\longx^*_i\}} \frac{f_\mathbf{X}(\mathbf{x}^*_i)}{g_\text{Pr}(\mathbf{x}_i^*)},
\end{equation}
for $\longx_1^*, \ldots, \longx_{n_\text{IS}}^*\sim g_\text{Pr}$. The variance of $\hat{p}_\text{IS}$ is dependent on the proposal density $g_\text{Pr}$, with the optimal choice of proposal minimising the variance in the estimate for the fixed budget $n_\text{IS}$. Here, the choice of $g_\text{Pr}$ minimising this variance is given by the CDE (\ref{equation:cde}, \citealt{rubinstein2016simulation}), so methods typically attempt to find proposal densities approximately equal to the CDE, either by using MCMC (e.g., \citealt{Xiao2020}) or surrogate modelling of the response function (e.g., \citealt{lystad2023full}). 

We choose to develop methodology to estimate the CDE for use as proposal density $g_\text{Pr}$ utilising an MCMC scheme with the CDE $\tilde{f}_{\longX}(\longx ; r_\text{Cr}) $ as the posterior target distribution from Bayes' rule $\tilde{\pi}(\longx|\theta) = {\pi}(\theta|\longx) \times \pi(\longx)$, where ${\pi}(\theta|\longx)$ is an empirical estimate of the probability $\Prbb{R>r_\text{Cr}|\{\longX=\longx\}} $ obtained by repeated sampling of $R|\{\longX=\longx\}$, and $\pi(\longx) =f_{\longX} (\longx)$. Using this approach, we obtain a sample from $\tilde{f}_{\longX}(\longx ; r_\text{Cr})$, and adopt a Gaussian smoothed version of $\tilde{\pi}(\longx|\theta)$ as the proposal density $g_\text{Pr}$, see supplementary material SM3.1.

In simple scenarios, with lower dimensional environment space $\mathcal{E}_\longX$ and unimodal, approximately elliptically-contoured $\tilde{f}_{\longX}(\longx ; r_\text{Cr})$, MCMC samples can be obtained using traditional algorithms such as Metropolis-Hastings (see e.g., \citealt{chib1995understanding}). In practice, however, the posterior $\tilde{f}_{\longX}(\longx ; r_\text{Cr}) $ may be more complex, e.g., exhibiting multi-modality or obvious departures from an elliptically-contoured density. Here we adopt parallel tempering MCMC as a more robust approach to estimate CDEs of arbitrary complexities.
 
Parallel tempering MCMC allows jumps between disjoint positive-density regions of the target distribution $\tilde{\pi}$ by combining some $n_\text{Tm}>1$ MCMC chains, each targetting scaled forms of $\tilde{\pi}$. These chains are evaluated at different `temperatures' $T_1, \ldots, T_{n_\text{Tm}}>0$, with the $j$th chain, $ j = 1,\ldots, n_\text{Tm}$, sampling from $\tilde{\pi}^{1/T_{j}}$; chains with a higher temperature target a `flatter' form of the target posterior density $\tilde{\pi}$, allowing movement between otherwise disjoint regions of positive density. Individual chains are sampled using a Metropolis-Hastings scheme with proposal density $\longx'|\longx \sim N(\longx, \sigma^2_\text{MH})$, $\sigma_\text{MH}>0$, and acceptance probability $\alpha_\text{MH}$. Swaps between chains $i$ and $j$, ($i, j = 1,\ldots, n_\text{Tm}$ , $i\neq j$), are periodically proposed with acceptance probability $\alpha_\text{Sw}(i, j)$, allowing chains of lower temperature to move between disjoint high-density regions in $\mathcal{E}_\longX$. \cite{10.1093/gji/ggt342} shows that, for a parallel tempering algorithm to satisfy detailed balance, individual-chain moves from $\longx$ to $\longx'$ should be accepted with probability $\alpha_\text{MH} = \min \{1, \tilde{\pi}(\longx'|\theta)/\tilde{\pi}(\longx|\theta)\}$, and swaps between the $i$th chain at state $\longx_i$ and the $j$th chain at state $\longx_j$ should be accepted with probability
\begin{equation}\label{equation:swap_prob}
    \alpha_\text{Sw}(i, j) = \min\left\{1, \left(\frac{\tilde{\pi}(\longx_j|\theta)}{\tilde{\pi}(\longx_i|\theta)}\right)^{1/T_i}\left(\frac{\tilde{\pi}(\longx_i|\theta)}{\tilde{\pi}(\longx_j|\theta)}\right)^{1/T_j}\right\}.
\end{equation}
Typically swaps are proposed only between adjacent chains, at predetermined set intervals.  We use the approach of \cite{Vousden2015}, adaptively selecting the temperature ladder $T_1,\ldots,T_{n_\text{Tm}}$, as well as the step size standard deviation $\sigma_\text{MH}$ for each individual chain. This method is implemented in the Python \texttt{pyPESTO} module \citep{10.1093/bioinformatics/btad711}, employed for all MCMC sampling in this work.

\begin{figure}
    \centering
\includegraphics[width=0.55\linewidth]{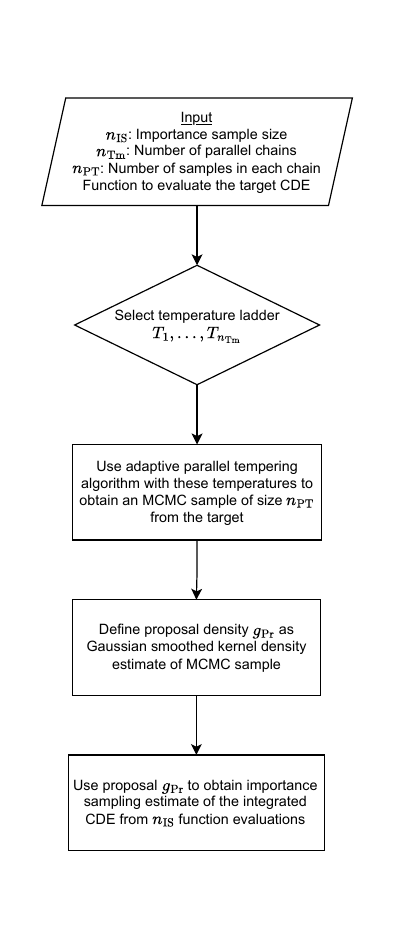}
    \caption{Schematic for the importance sampling-parallel tempering (IS-PT) sequential design algorithm of Section~\ref{section:monte_carlo}.}
    \label{fig:ispt-flowchart}
\end{figure}

A schematic of the resulting sequential design algorithm, henceforth referred to as importance sampling-parallel tempering (IS-PT), is given in Figure~\ref{fig:ispt-flowchart}.

% The choice of temperature scale is important, see {\color{red} [References in \cite{10.1093/gji/ggt342} point to the importance of temperature ladder choice.]}.

\subsection{Adaptive Gaussian emulation}\label{section:emulation}
\subsubsection{Gaussian emulation}

Importance sampling reduces the number of evaluations of the expensive response function needed for the calculation of failure probability \eqref{equation:multi_d_matt_target}. It does, however, still require some number of expensive evaluations, with this number being dependent on the convergence rate of the MCMC required for proposal distribution estimation. An alternative approach further reducing the need for computationally expensive evaluations is to replace draws from the true response function with estimates provided by a surrogate model, such as a Gaussian process emulator.

Various approaches to Gaussian process (GP) emulation have been reported in the offshore literature. \cite{gramstad2020sequential}, \cite{castellon2023full} and \cite{lystad2023full} assume a parametric form for the distribution of the structural response $R|\{\longX=\longx\}$, and so model realisations as draws from this parametric distribution, with unknown parameters modelled as a GP. In our case, we choose to target the (logarithm of the) CDE \eqref{equation:cde}. To do so, we make repeated draws from $R|\{\longX=\longx\}$ at $n_\text{Tr1}$ training points $\longx_1, \ldots, \longx_{n_\text{Tr1}}$, obtaining estimates of the conditional failure probability $\Prbb{R>r_\text{Cr}|\{\longX=\longx\}}$ at each of these values for $\longX$. These values form the training set $\mathcal{D} \subset \mathcal{E}_\longX$, the selection of which is discussed in Section~\ref{section:actv_learn}. After training the GP emulator on $\mathcal{D}$, we can then emulate the CDE at un-observed values $\longx \notin \mathcal{D}$.

We define the GP emulator for the log-CDE as
\begin{equation} \label{equation:gp_log}
    w(\longx) = \log \{\Prbb{R>r_\text{Cr}|\{\longX=\longx\}}f_\longX(\longx)\} \sim \mathcal{GP}(\mu_\text{GP}(\longx), k(\longx, \longx')), \quad w: \mathcal{E}_\longX \mapsto \RR,
\end{equation}
for mean and covariance functions
\begin{align} \label{equation:mean_cov}
\mu_\text{GP}(\longx)&=\mathbb{E}[w(\longx)], \quad \mu_\text{GP}: \mathcal{E}_\longX \rightarrow \RR, \\
     k\left(\longx, \longx'\right)&=\mathbb{E}\left[\left(w(\longx)-\mu(\longx)\right)\left(w(\longx')-\mu\left(\longx'\right)\right)\right], \quad k(\longx, \longx'): \mathcal{E}_\longX \times \mathcal{E}_\longX \rightarrow \mathbb{R},
\end{align}
where the log transform is used to ensure positivity of the predicted CDE. Compared to direct emulation of the failure probability (see SM1.2) this approach is advantageous when the density $f_\longX$ is itself not modelled continuously; see for instance the gridded density estimated in Section \ref{section:TFNV_wave}. Under this model, estimation of \eqref{equation:multi_d_matt_target} is an application of Bayesian quadrature using a Gaussian process prior (e.g., \citealt{hennig2022probabilistic}).

For the kernel function $k$, we use the Mat\'ern kernel (e.g., \citealt{genton2001classes})
\begin{equation}\label{eq:matern}
k_\text{Mt}(\longx, \longx') = \sigma_\text{Mt}^2 \frac{2^{1-\nu}}{\Gamma(\nu)} \left( \sqrt{2\nu} \frac{\|\longx- \longx'\|}{\ell} \right)^\nu K_\nu \left( \sqrt{2\nu} \frac{\|\longx - \longx'\|}{\ell} \right),
\end{equation}
with variance and length scale parameters $\sigma_\text{Mt}^2, \ell > 0$, and smoothness parameter $\nu>0$, where $||\cdot||$ is the Euclidean norm, $\Gamma:\RR\mapsto\RR$ is the gamma function, and $K_\nu:\RR^+ \mapsto \RR^+$ is the modified Bessel function of the second kind (e.g., \citealt{abramowitz1965handbook}); this kernel is chosen for its improved ability to capture sudden changes in the target function relative the squared exponential kernel. We combine the Mat\'ern kernel (weighted by a multiplicative constant $C_\text{Kr} > 0$) with additive white noise kernel $k_\text{WN}(\longx, \longx') = \kappa$ when $\longx = \longx'$, and $=0$ otherwise, for constant white noise variance $\kappa>0$, yielding the full kernel function
\begin{equation}
k(\longx, \longx') = C_\text{Kr} k_\text{Mt}(\longx, \longx') + k_\text{WN}(\longx, \longx').
\end{equation}
The addition of white noise allows the model to perform well when evaluations of the true target are made with some uncertainty, as is typically the case in application, see Section~\ref{section:TFNV_wave}. Kernel parameters $\sigma_\text{Mt}^2, \ell, C_\text{Kr}$ and $\kappa$ are jointly estimated via maximum likelihood at each posterior update using the L-BFGS-B algorithm (see \citealt{JMLR:v12:pedregosa11a}). The remaining parameter $\nu$ must be fixed; a brief sensitivity analysis suggests $\nu=2.5$ as a sensible choice. 

We assume a flat prior $\mu_\text{GP}(\longx) = 0 $ for all $\longx \in \longE$. Given covariance function $k$ as defined above, and training data $\mathbf{w} = (w(\longx_1), \ldots, w(\longx_{n_\text{Tr1}}))$, the posterior predictive mean $\mu_\text{GP}^*$ and covariance function $k^*$ for regression \eqref{equation:gp_log} can be found as,
\begin{align}\label{equation:post_update}
    \mu_\text{GP}^* (\longx) &= \mu_\text{GP}(\longx) + k(\mathcal{D},\longx)^T (k(\mathcal{D}, \mathcal{D}) + \alpha_\text{Ng}I_{n_\text{Tr1}})^{-1}(\mathbf{w} - \mu_\text{GP}(\mathcal{D})), \\
     k^*(\longx, \longx') &= k(\longx, \longx') - k(\mathcal{D}, \longx)^T (k(\mathcal{D}, \mathcal{D}) + \alpha_\text{Ng}I_{n_\text{Tr1}})^{-1}k(\mathcal{D}, \longx'),
\end{align}
 where $\alpha_\text{Ng}$ is an assumed observational nugget variance. In practice, we take $\alpha_\text{Ng}=10^{-5}$. Given this trained GP emulator, the target marginal failure probability estimate $\hat{p}_\text{GP}$ can be calculated using
\begin{align}\label{equation:gp_target}
       \hat{p}_\text{GP} = \mathbb{E}_{W,\longX}(\{\exp(w(\longx))\})
       &= \int_{\mathcal{E}_\longX} \left\{ \int_{\RR}  \exp(w) \phi(w ; \mu^*_\text{GP}(\mathbf{\longx}), k^*(\longx, \longx)) \wrt{w} \right\}  \wrt\longx \\
       &= \int_{\mathcal{E_\longX}} \exp\left(\mu^*_\text{GP}(\longx) + \frac{k^*(\longx, \longx')}{2}\right) \wrt{\longx},
\end{align}
using the expression for the expectation of log-normal random variable $\exp(w)$, for $W|\{\longX=\longx\} \sim N(\mu^*_\text{GP}(\mathbf{\longx}), k^*(\longx, \longx))$ with parameters obtained from \eqref{equation:gp_log}. For display purposes in the figures of Section~\ref{section:synthetic_study}, we evaluate the performance of our GP regression \eqref{equation:gp_log} in terms of the absolute difference $\Delta_\text{GP}$ between the true failure probability \eqref{equation:multi_d_matt_target} and this estimate
\begin{equation}\label{equation:gp_error}
    \Delta_\text{GP} = |p-\hat{p}_\text{GP}|.
\end{equation}

\subsubsection{Active learning}\label{section:actv_learn}

% \noindent \textit{Bias-variance weighted acquisition function}

The surrogate model \eqref{equation:gp_log} must be trained to provide reliable estimates of the CDE. Often, this training is carried out iteratively, with iteration $n$ training the surrogate against true evaluations of $w(\longx)$ for all $\longx$ in a training set $\mathcal{D}_n$, chosen inductively: at iteration $n+1$, we update training set $\mathcal{D}_n$ to $\mathcal{D}_{n+1}=\{\mathcal{D}_n, \longx_{n+1}\}$, where $\longx_{n+1} = \argmax_{\longx \in \mathcal{E}_\longX} U_n(\longx)$, for acquisition function $U_n$ taking the form 
\begin{equation}\label{eq:acq_general}
    U_n(\longx;\lambda) = \lambda \Sigma_n(\longx) + (1-\lambda) \text{M}_n(\longx),
\end{equation}
for $\lambda \in [0,1]$. Specification of the initial training set $\mathcal{D}_0$ is discussed in Section~\ref{section:synthetic_study}. Here $\Sigma_n: \mathcal{E}_\longX \mapsto \RR$ is an exploration term encouraging sampling at points far from existing members of $\mathcal{D}_n$, and $\text{M}_n: \mathcal{E}_\longX \mapsto \RR$ is an exploitation term encouraging sampling close to high values of the target function; see \cite{pollatsek1970theory} for an early discussion of this utility form. In our setting, $\text{M}_n: \mathcal{E}_\longX \mapsto \RR$ is large at values $\longx$ with high contributions to the integral \eqref{equation:gp_target}, motivating our first acquisition function 
\begin{equation} \label{equation:paper_acq_function}
    U^{(1)}_n(\longx;\lambda) = \lambda\log k_n^*(\longx, \longx) + (1-\lambda)\log\hat{f}^{(n)}_{\longX}(\longx ; r_\text{Cr}),
\end{equation}
where $k_n^*$ is the posterior GP kernel function obtained via \eqref{equation:post_update} with training set $\mathcal{D}_n$, and $\hat{f}^{(n)}_{\longX}(\longx ; r_\text{Cr})$ is the CDE estimate at iteration $n$. \cite{gramstad2020sequential}, \cite{lystad2023full} and \cite{wang2024comparison} provide examples of iterative schemes using Gaussian process emulation with acquisition functions similar to \eqref{equation:paper_acq_function}, for their respective forms of GP emulator \eqref{equation:gp_log}. Following \eqref{equation:multi_d_matt_target} and \eqref{equation:cde_mult}, we estimate the CDE $\hat{f}^{(n)}_{\longX}(\longx ; r_\text{Cr})$ as the integrand in \eqref{equation:gp_target}, namely
\begin{equation} \label{equation:cde_est}
    \hat{f}^{(n)}_{\longX}(\longx ; r_\text{Cr})= { \exp\left(\mu^*_n(\longx) + \frac{k_n^*(\longx, \longx')}{2}\right) },
\end{equation}
where $\mu_n^*$ is the posterior GP mean function obtained via posterior update \eqref{equation:post_update} with training set $\mathcal{D}_n$.

% , since
% \begin{equation}
%     \log \left(\frac{\int_{\RR}  g_\text{Lg}(w) \phi(w ; \mu_n(\mathbf{\longx}), k_n(\longx, \longx)) \wrt{w}}{\int_{\mathcal{E}_\longX} \left\{ \int_{\RR}  g_\text{Lg}(w) \phi(w ; \mu_n(\mathbf{\longx}), k_n(\longx, \longx)) \wrt{w} \right\}  f_{\longX} (\longx)\wrt\longx}\right) =  \log\{\int_{\RR}  g_\text{Lg}(w) \phi(w ; \mu_n(\mathbf{\longx}), k_n(\longx, \longx)) \wrt{w}\} - C,
% \end{equation}
% for a constant $C<0$,

% \noindent \textit{Active learning Cohn-based acquisition function}

A similar approach is the active learning Cohn (ALC) scheme of \cite{cohn1993neural}, aiming to reduce the overall variance of the GP surrogate on $\mathcal{E}_\longX$. They find the deduced reduction in variance across the \textit{entire space} $\longE$, given the addition of a new query point $\longx$ to the training set $\mathcal{D}_n$ at training iteration $n$. This is approximated over a reference set $\mathcal{P}=\{\mathbf{x}_j\}_{j=1}^{n_\text{Rf}}$ on $\mathcal{E}_\longX$ as 
\begin{equation} \label{equation:alc}
    \text{ALC}(\mathbf{x}) = \frac{1}{n_\text{Rf}} \sum^{n_\text{Rf}}_{j=1} k_n^*(\longx_j, \longx_j)  - \tilde{k}_{n+1}^*(\longx_j, \longx_j; \longx) , \quad \mathbf{x} \in \mathcal{P},
\end{equation}
for positive semi-definite $\text{ALC}(\longx)$, where $\tilde{k}_{n+1}^*(\longx_i, \longx_i; \longx)$ is the variance of the GP \eqref{equation:gp_log} at iteration $n+1$, given that query point $\longx$ is chosen as the next training point, thereby making $\mathcal{D}_{n+1} = \{\mathcal{D}_n, \longx\}$. The summand in \eqref{equation:alc} can be written
\begin{equation}\label{eq:alc_summand}
   k_n^*(\longx_j, \longx_j)  - \tilde{k}_{n+1}^*(\longx_j, \longx_j; \longx)  = \frac{(\mathbf{k}^*_{n, j}{\mathbf{C}^*_n}^{-1} \mathbf{m}_n^* - k_{n}^*(\mathbf{x}, \mathbf{x}_j))^2}{(k_n^*(\mathbf{x}, \mathbf{x}) - {\mathbf{m}_{n}^*}^T{\mathbf{C}_n^*}^{-1}\mathbf{m}^*_{n})},
\end{equation} 
see \citealt{861310}, where $\mathbf{C}_n^* = k_n^*(\mathcal{D}_n, \mathcal{D}_n)$ is the covariance matrix over current design points, $\mathbf{k}^*_{n, j}= k_n^*(\mathcal{D}_n, \mathbf{x}_j) $ is the vector of covariances between the training data and reference point $\mathbf{x}_j$ and $\mathbf{m}^*_{n}=k_n^*(\mathcal{D}_n, \mathbf{x}) $ is the covariance vector between the training data and the query point $\mathbf{x}$. 

\cite{861310} recommend selecting the best next query point $\longx$ by maximising a \textit{weighted} sum of \eqref{eq:alc_summand} over the reference grid $\mathcal{P}$. Instead, we employ an acquisition function of the form \eqref{eq:acq_general} utilising the ALC criterion. We find that the acquisition function 
\begin{equation} \label{eq:u2}
    U^{(2)}_n(\longx;\lambda) = \lambda \log \text{ALC}(\longx) + (1-\lambda) \log \hat{f}^{(n)}_{\longX}(\longx ; r_\text{Cr}),
\end{equation}
performs well for careful choice of $\lambda$. This is similar to the acquisition function \eqref{equation:paper_acq_function}, except that in \eqref{eq:u2} the exploration term $\Sigma_n(\longx)= \log \text{ALC}(\longx)$ considers the effect of including a query point $\longx$ in reducing error on the whole candidate space $\longE$, rather than just at the query point itself.

\begin{figure}
    \centering
    \includegraphics[width=0.7\linewidth]{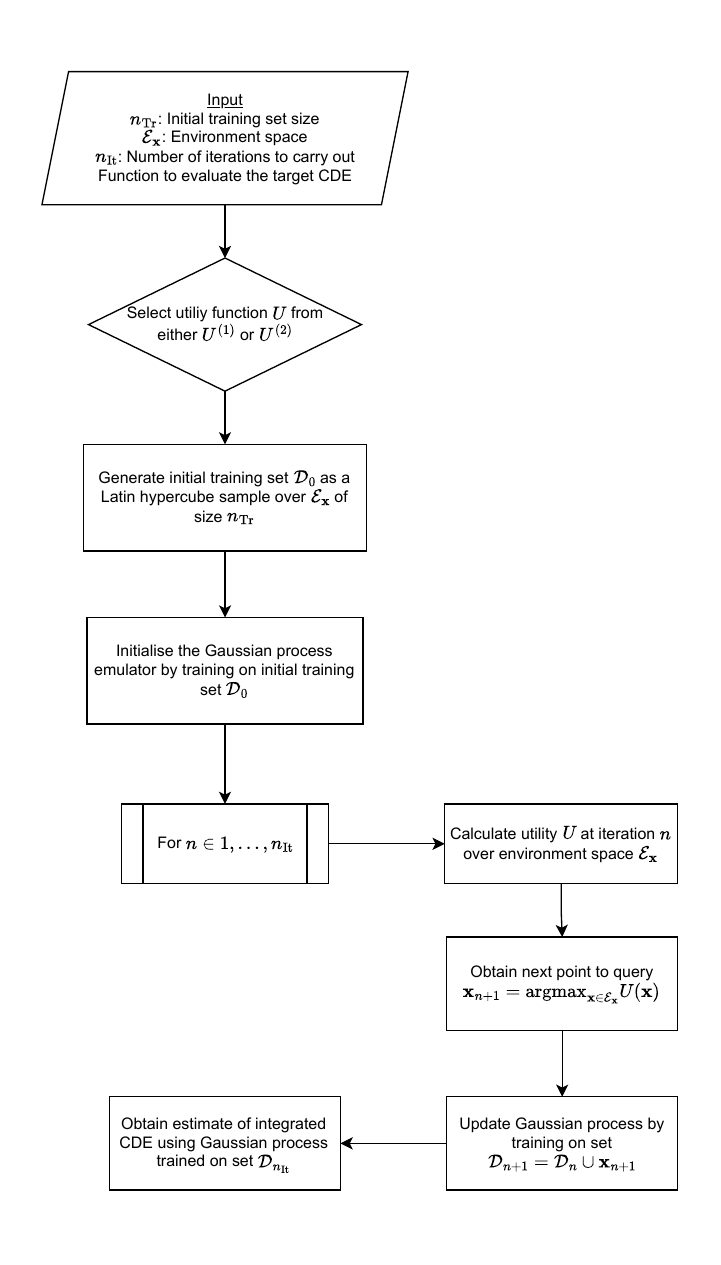}
    \caption{Schematic for the adaptive Gaussian emulation (AGE) sequential design algorithm of Section~\ref{section:emulation}.}
    \label{fig:age_flowchart}
\end{figure}

A schematic of the adaptive Gaussian emulation (AGE) procedure for sequential design is given in Figure~\ref{fig:age_flowchart}. 

% In practice, we only evaluate \eqref{eq:alc_summand} over values $\longx$ where $f_\longX(\longx) > \epsilon$, for $\epsilon >10^{-7}$, thus avoiding regions of $\mathcal{E}_\longX$ which are very unlikely to occur. This allows the use of a reference set $\mathcal{P}$ which is denser over the region of feasible values of the environment $\longX$.

\section{Synthetic simulation study} \label{section:synthetic_study}

\subsection{Synthetic scenario design}

We now compare the methods introduced in Section~\ref{section:methodology} under a synthetic test scenario, designed to be simple enough to yield a valuable comparison, {whilst being sufficiently methodologically challenging, and representative of a real-world structure}. We construct a synthetic response with an artificially bimodal CDE, intended to represent the extreme case of non-convex failure regions discussed in Section~\ref{section:introduction}. We follow the approaches of the likes of \cite{gramstad2020sequential} and \cite{castellon2023full}, who model structural responses (approximately) described by some parametric distribution function. The structural response $R$ given long term environment $\longX$ is modelled as a Weibull random variable, with distribution function
\begin{equation}\label{equation:weibull_cdf}
    F_{R|\longX}(r|\longx) = 1- \exp\left\{-\left(\frac{r}{\eta(\mathbf{x})}\right)^k\right\}, \quad r>0, 
\end{equation}
for fixed shape parameter $k = 2$ and scale parameter $\eta: \longE \mapsto \RR$ dependent on the long term environment. Adoption of this conditional Weibull form allows straightforward sampling from $R|\{\longX=\longx\}$, as well as exact evaluation of the conditional failure probability $\Prbb{R>r_\text{Cr} |\{\longX=\longx\}}$ . The environment $\longX$ is assumed bivariate $\longX = (X_1, X_2)$, with density function $f_\longX(\longx) = f_{(X_1, X_2)}(x_1, x_2) = f_{X_2|X_1}(x_2|x_1) f_{X_1}(x_1)$, where
\begin{align}\label{equation:synth_density}
    {f_{X_1}(x_1) = \frac{x_1}{\sigma_R^2} \exp(-\frac{x_1^2}{2 \sigma_R^2},) \quad x_1 \geq 0}
\end{align}
{is the Rayleigh density with scale parameter $\sigma_R>0$ set to $\sigma_R=12$ in this case}, and 
\begin{align}
    f_{X_2|X_1}(x_2|x_1) = \frac{1}{x_2 \sigma_\text{LN}(x_1) \sqrt{2\pi}} \exp\left\{ -\frac{1}{2}\left(\frac{\log(x_2) - \mu_\text{LN}(x_1)}{\sigma_\text{LN}(x_1)}\right)^2\right\}
\end{align}
is the log-normal density with  
\begin{align}\label{equation:lognorm_pars}
    \mu_\text{LN}(x_1) &= 0.933 + 0.578  x_1^{0.395}, \\
    \sigma_\text{LN}(x_1) &= 0.055 + 0.336 + \exp(-0.585x_1),
\end{align}
as in \cite{MATHISEN199093} {for wave period given significant wave height. The reader might therefore choose to consider this environmental specification as an example of an extreme environment of significant wave height (Rayleigh) and conditional significant wave period (log-normal), although this interpretation is not necessary.} 

\begin{figure}[t]
    \centering
    \includegraphics[width=1\linewidth]{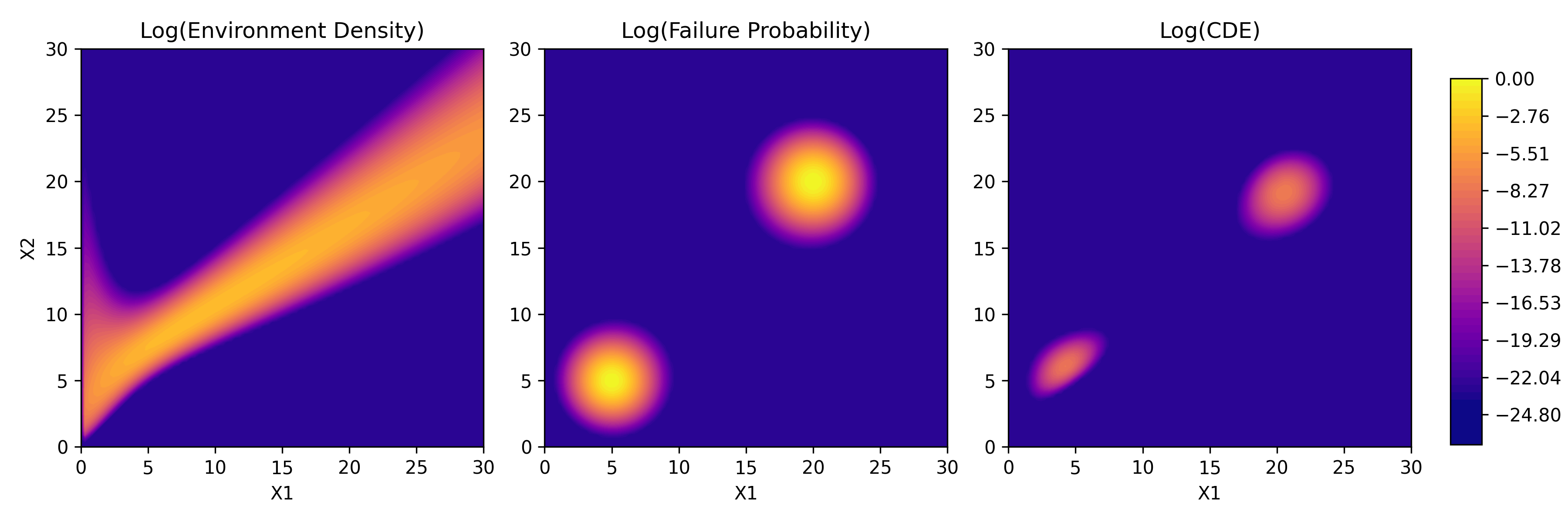}
     \caption{Panels summarising the synthetic response case study used in this section. From left to right the panels show: bivariate environment log-density \eqref{equation:synth_density}; structural log-failure probability as Weibull exceedance probability of $r_\text{Cr}=175$ \eqref{equation:synth_fail}; and log-CDE \eqref{equation:cde} obtained by multiplying failure probability by environment density.}
    \label{fig:bimodal_synth}
\end{figure}

The function $\eta$ is constructed to provide a structural response with the desired multimodal behaviour. To achieve this, we define a scenario with scale $\eta(\longx)$ increasing around values $\longx_{\text{Pk1}}$ and $\longx_{\text{Pk2}}$, modelling the scale parameter using the multimodal function 
\begin{equation} \label{equation:synth_lambda}
    \eta(\mathbf{x}) =C\left\{ A\max(||\longx - \longx_{\text{Pk1}}||, \nu) \ + B\max(||\longx - \longx_{\text{Pk2}}||, \nu) \right\},
\end{equation}
for scaling parameters $A, B, C>0$, and peak radius $\nu>0$ surrounding `resonant' $\longX$ values $\longx_{\text{Pk1}}=(5,5)$ and $\longx_{\text{Pk2}}=(20, 20)$. We set constants, $A=1.3$, $B=1.5$, $C=100$, $\nu=0.5$, and critical response $r_\text{Cr}=175$, chosen in order to yield a true `synthetic' failure probability 
\begin{equation} \label{equation:synth_fail}
    p_\text{Sn} = \Prbb{R>r_\text{Cr}} = \int_\longE \exp \left\{ -\left(
    \frac{r_\text{Cr}}{\eta(\longx)}\right)^2 \right\} f_\longX(\longx) \wrt{\longx} = 1.3 \times 10^{-3}.
\end{equation}
This yields a failure probability in order of magnitude comparable to the failure probabilities discussed in Section~\ref{section:TFNV_wave}. Figure~\ref{fig:bimodal_synth} shows the environment density, failure probability and CDE for this synthetic scenario, over the bivariate environment space $\longE = [0, 30]^2$.

In practice, estimates of conditional failure probability $\Prbb{R>r_\text{Cr} |\{\longX=\longx\}}$ are found empirically using realisations of $R|\{\longX=\longx\}$. We introduce further uncertainty in this synthetic case in the form of the conditional distribution for $R|\{\longX=\longx\}$ by making $\eta$ stochastic, with
\begin{equation} \label{equation:noisy_scale}
    \eta_\delta(\longx) = \eta(\longx) \cdot (1 + \epsilon_\delta), \quad \text{for} \quad \epsilon_\delta \sim N(0, \delta^2),
\end{equation}
where we set $\delta=0.05$. This applies an additive white noise to the scale of our observations with variance proportional to the value of the scale function, meaning that larger values of the scale function will correspond to `more uncertain' observations. In the absence of uncertainty in $\eta_\delta$ (i.e., with $\delta=0$), the expected distribution of $R|\{\longX=\longx\}$ is relatively easily identified from a smaller number of realisations of fluid loading simulation. However, for uncertain $\eta_\delta$, the number of realisations required to be confident about the expected distribution of $R|\{\longX=\longx\}$ increases. That is, particularly with $\delta>0$, we expect to need to sample from the same regions of $\longE$ multiple times to build confidence in our estimate of CDE. 

{Note that the IS-PT and AGE procedures are expected to give good performance for any reasonable combination of values for parameters $\sigma_R$, $\mu_{LN}$, $\sigma_{LN}$, $A$, $B$, $C$, $\nu$, $x_{\text{Pk1}}$ and $x_{\text{Pk2}}$ in this simulation study, and therefore that the actual values of parameters used here are of little direct relevance. The critical feature of the current simulation study is that the general characteristics of the distribution of environmental variables and those of the structural response to the environment, reflect the general characteristics of actual environments and structural responses. It is for this reason that we refer to the environmental variables in this section as $X_1$ and $X_2$ rather than e.g., $H_s$ and $T_p$.} 

\subsection{Results of synthetic study} \label{section:synth_study}

\subsubsection{Overview}

Here we apply the methods introduced in Section~\ref{section:methodology} to the synthetic scenario discussed above. We first present the results of the importance sampling-parallel tempering (IS-PT) approach of Section~\ref{section:monte_carlo} in Section~\ref{section:is_results}, followed by those of the adaptive Gaussian emulation (AGE) procedure of Section~\ref{section:emulation} in Section~\ref{section:age_results}. We adjust the number of expensive function evaluations $n_\text{Ev}$ used for each of IS-PT and AGE methods so they yield the same order of magnitude of root mean squared error 
\begin{equation}\label{eq:rmse}
    \text{RMSE}(\hat{p}) = \sqrt{\sum_{r=1}^{n_\text{Rp}}\frac{(\hat{p}_{r} - p_\text{Sn})^2}{n_\text{Rp}}},
\end{equation}
over some number $n_\text{Rp}$ of replicate analyses, where $\hat{p}_r$ is the estimate provided by either IS-PT or AGE at replicate $r$. We also evaluate the bias

\begin{equation}\label{eq:bias}
    \text{Bias}(\hat{p}) = \sum_{r=1}^{n_\text{Rp}}\frac{(\hat{p}_{r} - p_\text{Sn})}{n_\text{Rp}},
\end{equation}
for each of the methods.

\subsubsection{Importance sampling coupled with parallel tempering MCMC (IS-PT)}\label{section:is_results}

We apply the IS-PT framework of Section~\ref{section:monte_carlo} under the synthetic scenario in three stages: first (a) parallel tempering MCMC sampling with the CDE \eqref{equation:cde} as target posterior density; followed by (b) kernel smoothing of the resulting sample to obtain proposal density $p_\text{Pr}$; and finally (c) evaluation of importance sampling estimate \eqref{equation:paper_p1IS} using $n_\text{IS}=100$ draws from this proposal. Steps (a)-(c) are repeated $n_\text{Rp}=100$ times, to estimate $\hat{p}_\text{IS}$.

\begin{figure}[t]
    \centering
    \includegraphics[width=\linewidth]{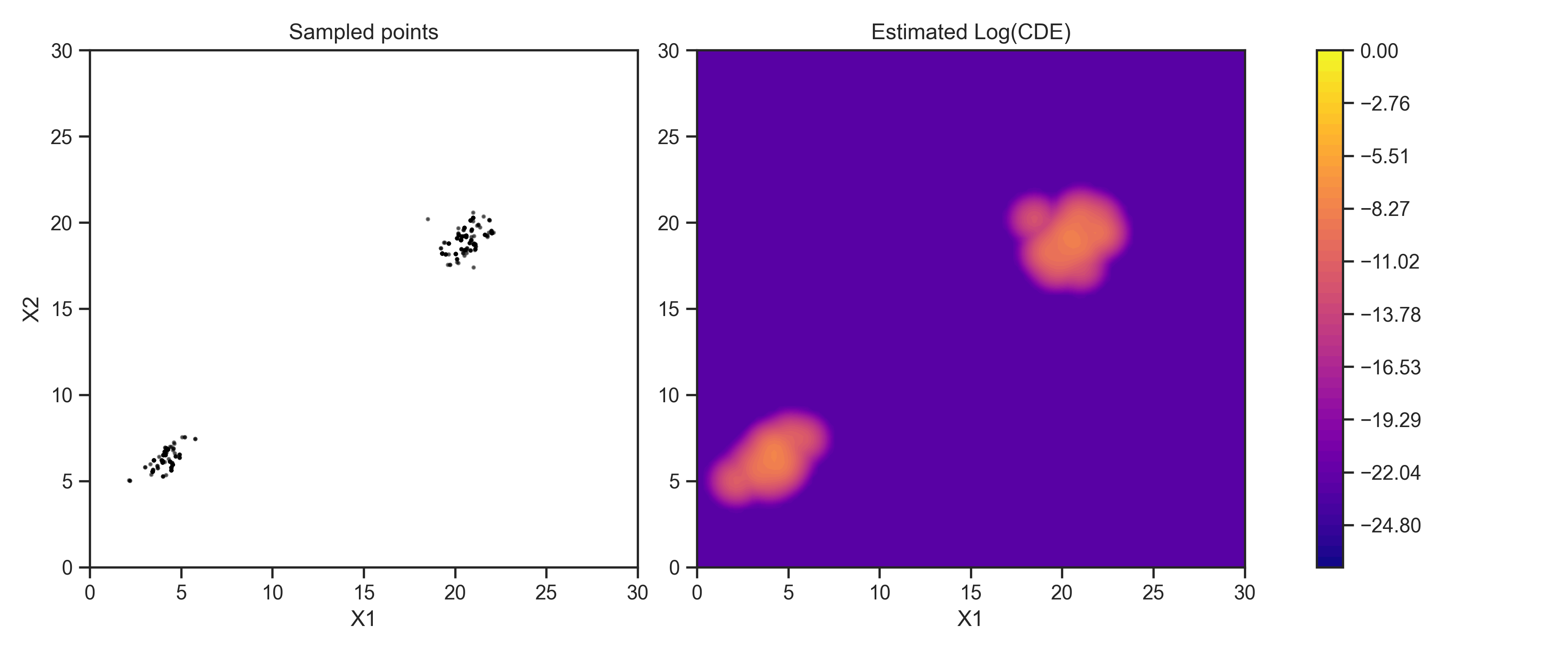}
    \caption{Example sample from CDE \eqref{equation:cde} under the synthetic structural scenario obtained using the adaptive parallel tempering MCMC algorithm of \cite{Vousden2015} (left), and corresponding smoothed log-CDE estimated using Gaussian kernel bandwidth selected according to \cite{scott2015multivariate} (right).}
    \label{fig:ispt}
\end{figure}

Step (a) is achieved using the adaptive parallel tempering algorithm of \cite{Vousden2015} implemented in the \texttt{pyPESTO} module. We run $n_\text{Tm}=5$ parallel chains, supplying an initial temperature ladder $T_1, \ldots, T_5$ geometrically spaced between $T_1=1$ and $T_5 = 20$, with initial proposal variance $\sigma^2_{\text{MH}}=1$. The MCMC algorithm then adaptively tunes the temperature spacing and proposal variance, targetting equal acceptance probability of swaps between adjacent chains. Each of the five chains is run for $n_\text{PT}=400$ time steps, with periodic swaps between chains proposed according to \cite{Vousden2015}, requiring $n_\text{Tm}\times n_\text{PT} = 2000$ expensive function evaluations in total. An example trace plot from the $T_1$ chain is given in SM3.1. The chain at temperature $T_1=1$ is retained, and burn-in length $n_\text{Br}$ automatically chosen using Geweke's diagnostic \citep{geweke1991evaluating}. When $n_\text{Br}<n_\text{PT}$, this burn-in period is discarded, leaving a sample of length $n_\text{PT} - n_\text{Br}$. For step (b), the sample is then used to provide a Gaussian kernel smoothed estimate of the CDE, with kernel bandwidth chosen according to Scott's rule of thumb \citep{scott2015multivariate}, see SM3.1 for details. Step (c) consists of evaluating importance sampling probability estimate $\hat{p}_\text{IS}$ given by \eqref{equation:paper_p1IS}, using $n_\text{IS}=100$ draws from proposal density $g_\text{Pr}$ found in step (b), requiring a further 100 expensive function evaluations. Figure~\ref{fig:ispt} shows a typical sample obtained using this approach together with resulting CDE estimate $g_\text{Pr}$. The (root mean square error) RMSE \eqref{eq:rmse} is estimated to be $\text{RMSE}(\hat{p}_\text{IS}) = 2.20 \times 10^{-4}$, using $n_\text{Rp}=100$ replicates of the IS-PT analysis, with \textit{each} of the $n_\text{Rp}$ IS-PT estimates requiring $n_\text{Ev} = n_\text{Tm} \times n_\text{PT} + n_\text{IS} = 2100$ expensive function evaluations. The bias in the $\hat{p}_\text{IS}$ estimate over the 100 replicates is small, equal to $\text{Bias}(\hat{p}_\text{IS})=5.32\times 10^{-5}$.

\subsubsection{Adaptive Gaussian emulation (AGE)} \label{section:age_results}
\begin{figure}[t]
    \centering
    \includegraphics[width=.9\linewidth]{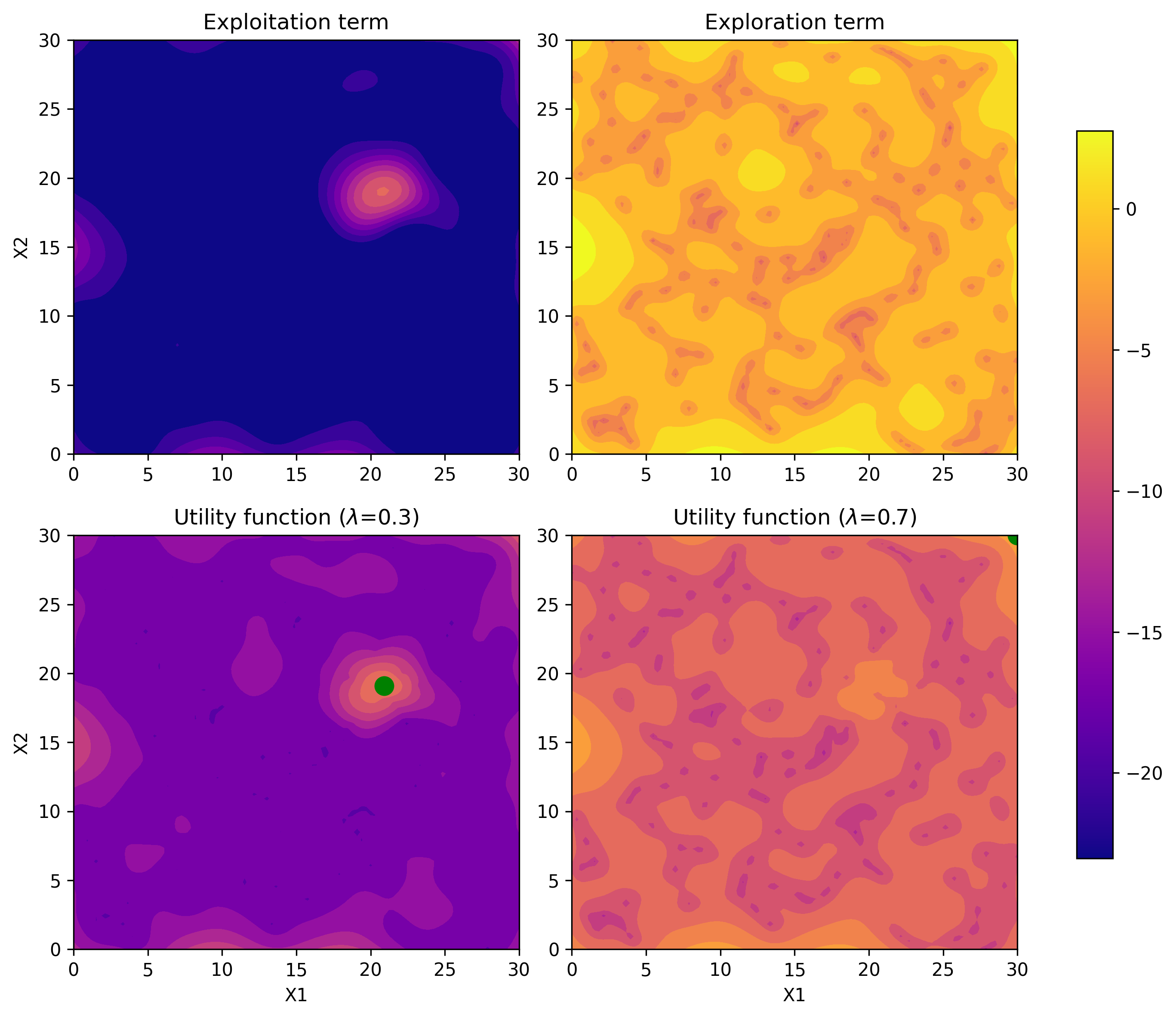}
    \caption{Behaviour of utility function $U^{(1)}(\longx; \lambda)$ over the environment space $\longE$, for synthetic scenario. Upper panels show exploitation and exploration terms obtained from GP emulator \eqref{equation:gp_log} trained on initial Latin hypercube set $\mathcal{D}_0$ of size $n_\text{Tr1}=144$. Lower panels show resulting utility functions for weights $\lambda=0.3$ and $\lambda=0.7$. In each lower panel, the optimal sampling point $\longx^* =\argmax_{\longx \in \longE} U^{(1)}(\longx; \lambda)$ is indicated in green. In the lower right hand panel, $\longx^*$ is located in the upper right corner of $\longE$.}
    \label{fig:utility_gramstad}
\end{figure}

The GP emulator \eqref{equation:gp_log} is used to model the log-CDE under this synthetic scenario, following the AGE procedure of Section~\ref{section:actv_learn}. It is iteratively trained as in \eqref{equation:post_update} on training sets $\mathcal{D}_1, \ldots, \mathcal{D}_{n_\text{It1}}$ for $n_\text{It1}$ iterations, with training set $\mathcal{D}_{n+1} = \{\mathcal{D}_n, \longx^*\}$, $n>1$, constructed with $\longx^*$ chosen according to either $U^{(1)}$ (\ref{equation:paper_acq_function}, Variance case) or $U^{(2)}$ (\ref{eq:u2}, ALC case). In each case, the initial training set $\mathcal{D}_0$ is a simple space-filling Latin hypercube design of $n_\text{Tr1}=144$ points, chosen as a low, but adequate, number of starting points found to provide stable kernel parameter convergence at iteration zero. At each subsequent iteration, we begin the kernel parameter optimisation at the previous iteration's estimates. 

Figure~\ref{fig:utility_gramstad} shows an example of how utility $U^{(1)}$ is constructed using the emulator \eqref{equation:gp_log} trained on initial set $\mathcal{D}_0$, for two example values of $\lambda$. The upper panels shows the exploration $\Sigma_0$ and exploitation $\text{M}_0$ terms as defined in \eqref{equation:paper_acq_function}, with lower panels showing utility functions obtained by prioritising exploration ($\lambda=0.7$) or exploitation ($\lambda=0.3$). Green points in the lower panels indicate the maximum $\longx^* =\argmax_{\longx \in \longE} U^{(1)}(\longx; \lambda)$, illustrating that the choice of this tuning parameter can alter the design of the training set $\mathcal{D}_1= \{\mathcal{D}_0, \longx^*\}$ (and thus subsequent training sets $\mathcal{D}_2, \mathcal{D}_3,\ldots$). In the lower right panel, the maximum is located on the edge of the environment space, due to the Latin hypercube sampling used to construct $\mathcal{D}_0$ placing no points on the boundary. (This can be prevented by adding initial training points along the boundary, however, this isn't necessary as subsequent iterations move away from the edge of the space once it has been explored.) See SM3.2 for an equivalent example for $U^{(2)}$.
\begin{figure}[t]
    \centering
    \includegraphics[width=.95\linewidth]{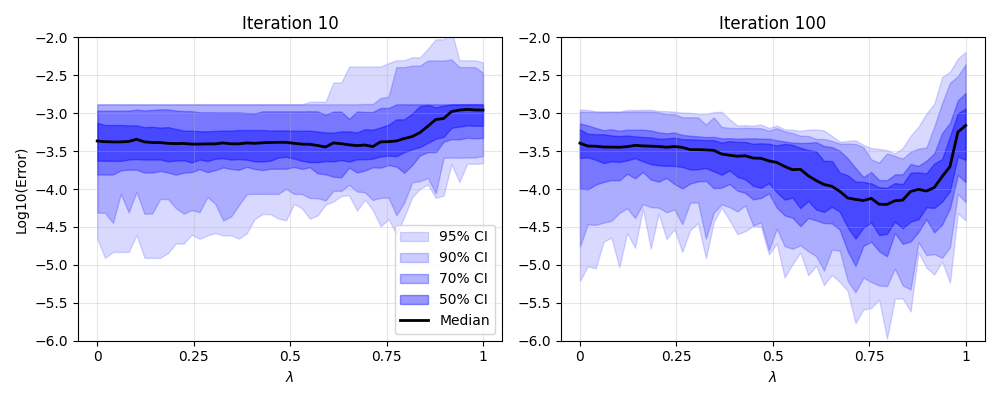}
    \caption{Log-scale absolute error $\Delta_\text{GP}$ of the GP probability estimate $\hat{p}_\text{GP}$ at specified iterations, for emulator \eqref{equation:gp_log} trained using $U^{(1)}$ over the range of weight $\lambda \in [0.01, 0.99]$ for the synthetic scenario. At iteration 100, the weight that minimises median error is $\lambda^*=0.80$.}
    \label{fig:nu_sens1}
\end{figure}

The GP emulator is trained using both utility functions $U^{(1)}$ and $U^{(2)}$, for a range of $n_\lambda=50$ values of weight parameter $\lambda$, equally spaced on the interval $[0.01, 0.99]$. For each value of $\lambda$, we perform $n_\text{It1}=100$ iterations of the GP update \eqref{equation:post_update} for each utility. This yields posterior estimates $\mu_n^*$ and $k^*_n$ for $n=1,\ldots, n_\text{It1}$. This analysis is replicated $n_\text{Rp}=100$ times, with randomised initial set $\mathcal{D}_0$ and conditional response scale $\eta$ \eqref{equation:noisy_scale} at each replicate. 

\begin{figure}[t]
    \centering
    \includegraphics[width=0.75\linewidth]{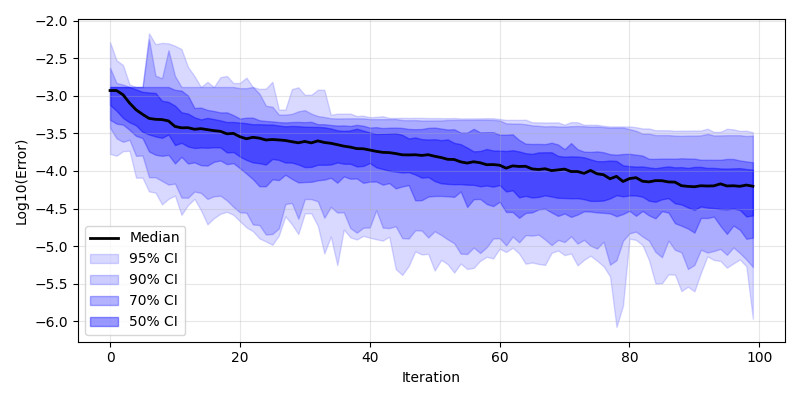}
    \caption{Distribution of log-scale absolute error $\Delta_\text{GP}$ in the GP probability estimate with respect to iteration, trained using $U^{(1)}$ with $\lambda=\lambda^*$. The trend in median error is indicated in black, with various confidence intervals shown in blue.}
    \label{fig:opt_lambda_error}
\end{figure}

Each of the $n_\text{Rp}$ replicate analyses produces $n_\lambda \times n_\text{It1}$ values of the failure probability estimate $\hat{p}_\text{GP}$ and error $\Delta_\text{GP}$ for each utility. Figure~\ref{fig:nu_sens1} shows the distribution of the resulting $\Delta_\text{GP}$ values under variance utility $U^{(1)}$ \eqref{equation:paper_acq_function} with respect to $\lambda$, at iterations 10 and 100. Errors are plotted on the log scale, with 50\%, 70\%, 90\% and 95\% confidence bands indicated in different shades of blue. The median log-error trend with respect to $\lambda$ is given as a black line. The GP emulator converges to the truth for weights in the interval $I^*$, which in this case corresponds approximately to $[0.5, 1]$. The location of $I^*$ on the unit interval is determined by the bimodal nature of the synthetic response. For some initial training sets $\mathcal{D}_0$, at iteration zero, the emulator detects one peak in response but fails to detect the other; this can be seen in the top left panel of Figure~\ref{fig:utility_gramstad}, where the mode at $\longx_\text{Pk2}=(20, 20)$ is found, but that at $\longx_\text{Pk1}=(5, 5)$ is not. For low values of $\lambda$, the utility function $U^{(1)}$ sometimes does not place enough weight on the exploration term for the algorithm to detect the second peak in subsequent iterations (e.g., the lower left panel of Figure~\ref{fig:utility_gramstad} shows a low value for utility at $\longx_\text{Pk1}$, whereas the lower right panel has a higher utility there). That is, for low values of $\lambda$, the iterative algorithm tends not to allow the GP to `discover' the second mode. The value of $\lambda$ minimising the median error at the final iteration is $\lambda^*=0.80$. Figure~\ref{fig:opt_lambda_error} shows the distribution of $\Delta_\text{GP}$ across all iterations when $\lambda=\lambda^*$. In general, there is a decrease in error with iteration, with `spike' at around iteration 10 for some replicates; these spikes shows where the algorithm tends to detect the second mode, causing a temporary increase in bias due to the uncertainty in \eqref{equation:noisy_scale}. Figures corresponding to Figures~\ref{fig:nu_sens1} and \ref{fig:opt_lambda_error} for ALC utility $U^{(2)}$ can be found in SM3.2. For $U^{(2)}$, a minimum of $\Delta_\text{GP}$ is found in $I^*=[0.2, 0.5]$, and comparison of errors $\Delta_\text{GP}$ at the final iteration indicates that $\Delta_\text{GP}$ for $U^{(2)}$ is somewhat larger than for $U^{(1)}$.

\begin{figure}[h]
    \centering
    \includegraphics[width=\linewidth]{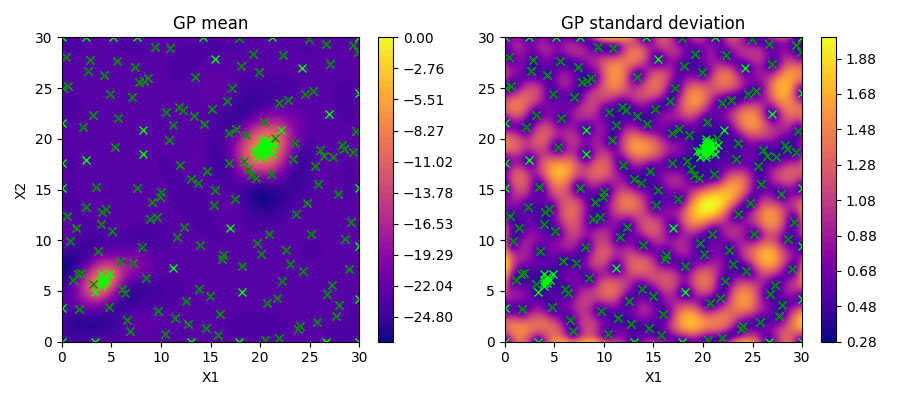}
   \caption{GP emulator at iteration 100 for variance utility $U^{(1)}$ \eqref{equation:paper_acq_function}, trained using the optimal value $\lambda^*=0.80$ minimising median of error $\Delta_\text{GP}$. The panels from left to right show: the posterior GP mean $\mu^*_{100}(\longx)$ over $\longx \in \longE$; the posterior GP standard deviation $k^*_{100}(\longx, \longx)^{1/2}$. The initial random Latin hypercube training set $\mathcal{D}_0$ is shown as dark green crosses, and the iteratively selected new training points $\mathcal{D}_{100} \setminus \mathcal{D}_0$ are shown as light green crosses.}
    \label{fig:opt_emulator}
\end{figure}

Figure~\ref{fig:opt_emulator} shows an example GP emulator at the final iteration, trained on set $\mathcal{D}_{100}$ selected using $U^{(1)}$ with $\lambda^*=0.80$. The left panel shows the posterior GP mean $\mu^*_{100}(\longx)$ and the right the posterior GP standard deviation $k^*_{100}(\longx, \longx)^{1/2}$, both over $\longx \in \longE$. The initial Latin hypercube training set $\mathcal{D}_0$ is shown as dark green crosses, and the iteratively selected new training points $\mathcal{D}_{100} \setminus \mathcal{D}_0$ are shown as light green crosses. The light green crosses, iteratively selected using the utility function, mostly cluster around the high-density regions of the synthetic CDE, whilst allowing some exploration into low-density regions of $\longE$. 

We evaluate the RMSE \eqref{eq:rmse} of the AGE approach using $\hat{p}_\text{GP}$ obtained from emulators trained under $U^{(1)}$ with $\lambda= \lambda^*$ over $n_\text{Rp}=100$ replicate analysis. The resulting estimates yield $\text{RMSE}(\hat{p}_\text{GP}) = 1.16\times 10^{-4}$, a similar value to $\text{RMSE}(\hat{p}_\text{IS})$ reported in Section~\ref{section:is_results}. For the AGE approach, each replicate analysis involves a total of $n_\text{Ev}=|\mathcal{D}_{0}| + n_\text{Itr1}=244 $ expensive function evaluations, assuming $\lambda^*$ is known. The bias in the $\hat{p}_\text{GP}$ estimate over the 100 replicates is $\text{Bias}(\hat{p}_\text{GP}) = 3.18\times 10^{-5}$, comparable in size to that of $\hat{p}_\text{IS}$. Corresponding results using $U^{(2)}$ are similar, and summarised in the next section.

\subsubsection{Comparison of IS-PT and AGE results} \label{section:synth_comp}

\begin{figure}
    \centering
    \includegraphics[width=\linewidth]{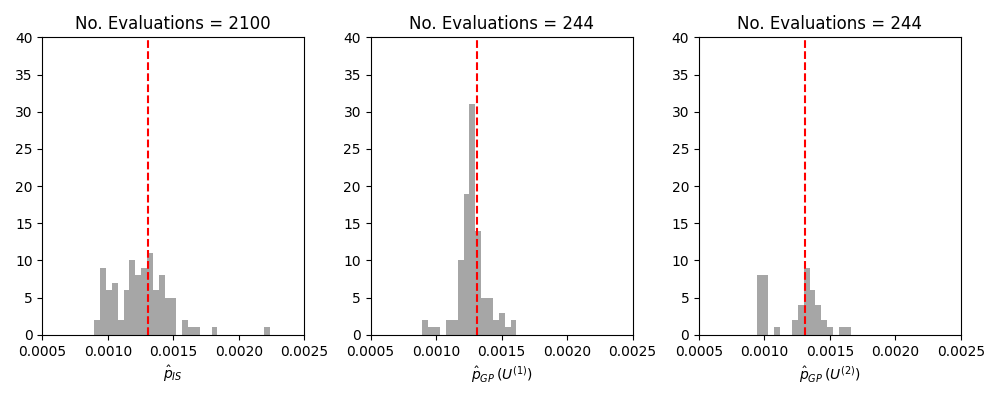}
    \caption{Distribution of $n_\text{Rp}=100$ estimates $\hat{p}_\text{IS}$ (left), $\hat{p}_\text{GP}$ for $U^{(1)}$ iteration 100 with $\lambda=\lambda^*$ (centre) and $\hat{p}_\text{GP}$ for $U^{(2)}$ iteration 100 with $\lambda=\lambda^*$ (right), for true failure probability $p_\text{Sn}$ (red). The number of function evaluations required for a single replicate analysis is indicated in the panel titles.}
    \label{fig:synth_est_comp}
\end{figure}

Figure~\ref{fig:synth_est_comp} shows the distribution of the IS-PT estimate $\hat{p}_\text{IS}$ and the AGE estimates $\hat{p}_\text{GP}$ (from $U^{(1)}$ and $U^{(2)}$ at iteration 100, $\lambda=\lambda^*)$ around the target failure probability $p_\text{Sn}$. A summary of the RMSEs and biases for these estimates can be seen in Table~\ref{tab:rmse_synth}, along with the number of expensive function evaluations $n_\text{Ev}$ required for each replicate analysis. Both variants of the $\hat{p}_\text{GP}$ estimate show an equivalent performance to $\hat{p}_\text{IS}$ for around 12\% of required expensive function evaluations, provided we have knowledge of the optimal weight parameter $\lambda^*$. The AGE approach with utility $U^{(2)}$ is computationally somewhat more demanding than that using $U^{(1)}$, due to the required calculation of ALC \eqref{equation:alc} at each iteration.

However, if $\lambda^*$ is unknown, and cannot be reliability estimated, we see that IS-PT provides a useful if computationally more demanding alternative. The current analysis shows that approximately 2000 expensive function iterations using IS-PT are sufficient to estimate a bimodal CDE well in two dimensions, avoiding the need to specify problematic hyperparameters such as $\lambda$. {To provide some context, we also considered estimating failure probability using a simple standard Monte Carlo sampling algorithm, with a total number of 2100 Monte Carlo samples to match that used for IS-PT and AGE variants in Table~1. We found corresponding RMSE and bias values for failure probability of $3.61 \times 10^{-4}$ and $7.29 \times 10^{-5}$ respectively. Clearly in terms of RMSE, standard Monte Carlo sampling provides considerably poorer estimates of failure probability than both PI-IS and the two AGE variants considered here.}
We explore the relative merits of IS-PT and AGE methodologies further for the monopile structure scenario of Section~\ref{section:application}.

% Bias of AKMC estimates: 0.00015000377827420342
% Bias of IS estimates: -1.8268833632317587e-06
% RMSE of AKMC estimates: 0.00024457310686122976
% RMSE of IS estimates: 5.095787403278413e-05
\begin{table}[h]
    \centering
    \resizebox{\textwidth}{!}{
    \begin{tabular}{|c||c|c|c|}
    \hline
         & IS-PT& $U^{(1)}$ AGE & $U^{(2)}$ AGE \\ \hline \hline
        RMSE& $2.20\times 10^{-4}$& $1.16\times 10^{-4}$ & $2.40 \times 10^{-4}$ \\ \hline 
 Bias& $5.32\times 10^{-5}$&$3.18\times 10^{-5}$ & $1.50 \times 10^{-4}$ \\ \hline
        Number of function evaluations, $n_\text{Ev}$&  $ n_\text{Tm} \times n_\text{PT} + n_\text{IS} = 2100$ & $|\mathcal{D}_{0}| + n_\text{Itr1}=244$&$|\mathcal{D}_{0}| + n_\text{Itr1}=244$ \\ \hline
    \end{tabular}}
    \caption{RMSEs and biases of $\hat{p}_\text{IS}$ and $\hat{p}_\text{GP}$ when targetting failure probability $p_\text{Sn}$, calculated for $n_\text{Rp}=100$ replicate analyses. The true value of probability of failure is $p_\text{Sn}= 1.3\times 10^{-3}$. Also shown is the number of expensive response function evaluations required for a single replicate analysis for each of IS-PT and AGE.}
    \label{tab:rmse_synth}
\end{table}

% n = 2100
% Monte Carlo Estimates
% RMSE: 3.61e-04
% Bias: 7.29e-05

\section{Application to monopile response models}\label{section:application}

\subsection{Overview of case study}
We now apply the IS-PT and AGE methodologies of Section~\ref{section:methodology} to a real-world case study, using hindcast data from a location around 1km offshore of Albany, Western Australia, produced by the Centre for Australian Weather and Climate Research, see Section~\ref{section:data} to estimate a model for the extreme ocean environment. We consider a model monopile structure situated in this environment, subject to wave-induced loading which in turn induces some resonant effect. To construct the test scenario, we first use the extreme value methods of \cite{davison1990models} and \cite{38} to model the joint behaviour of a bivariate ocean $\longX$ at this location, see Section~\ref{section:env_model}. This is followed in Section~\ref{section:TFNV_wave} with numeric simulation from the T-FNV model of \cite{taylor2024transformed} to approximate the inertial load placed on an offshore wind turbine at this location. Finally, this load is propagated through the linear response function for a damped harmonic oscillator, yielding realisations of the harmonic response on our model structure. The results of these simulations are given in Section~\ref{section:case_results} to provide a `baseline' estimate of the CDE. This baseline is then used to assess the performance of IS-PT and AGE methodologies in Sections~\ref{section:tfnv_ispt} and \ref{section:tfnv_age}. These methods are then compared in Section~\ref{section:tfnv_comparison}.

{The reader is referred to articles including \cite{Riise_Grue_Jensen_Johannessen_2018},  \cite{jmse7120430}, \cite{jmse11030628}, \cite{LIU2025122355} and \cite{YANG2025123120}, as well as \cite{Orszaghova2025}, for relevant discussion and illustration of wind turbine response characteristics.}

\subsection{Albany hindcast data} \label{section:data}

The data includes hourly hindcast observations over the period 1980-2017, consisting of sea state variables significant wave height $H_s$, peak wave period $T_p$, energy wave period $T_e$, and mean wave period $T_m$. There are a total of 333120 observations. We preprocess the data by isolating storm peak values of the sea state $H_s$. Given storm events that are sufficiently well spaced in time, this removes any temporal correlation in the storm peak data, simplifying the modelling process whilst retaining the observations most likely to induce structural failure.

To isolate the storms peaks, we follow the procedure of \cite{ewans2008effect}. Firstly, a wave height $h_\text{St}$ (in metres) is chosen as the storm threshold, such that an upcrossing above this height is considered the beginning of a storm event. The subsequent downcrossing of this height is considered the end of the storm event. We also merge any two storm events that occur within 48 hours of one another, retaining only the largest storm peak value. The value of $h_\text{St}$ is determined by assessing the number of storm peaks recovered from the dataset using a given threshold against the practically of observed storm lengths; this creates a trade-off between retaining enough points for statistical modelling, whilst avoiding identifying storms of unrealistically long duration. We choose to limit occurrences of storms lasting longer than three days, selecting $h_\text{St}=4$ yielding a total of 976 storm peak observations, with around $6\%$ of identified storm durations exceeding three days.

\begin{figure}[t]
    \centering
    \includegraphics[width=\linewidth]{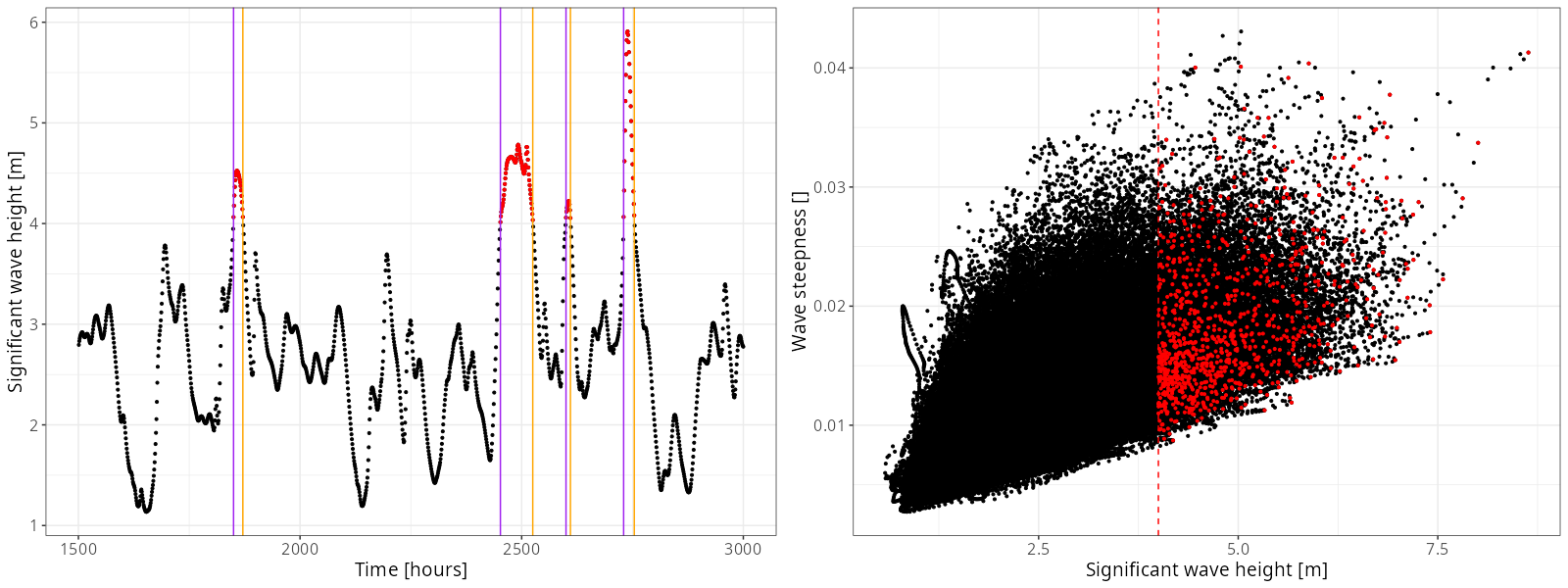}
    \caption{Illustration of storm peak isolation for storm threshold $h_\text{St}=4$. The left panel shows $H_s$ value against hourly index, with the beginning of each storm (defined as the first upcrossing of 4m) indicated in purple. The end of each storm (defined as the first downcrossing of 4m) is shown in orange. Sequences of within-storm $H_s$ values are highlighted in red. In the right panel, the entire hindcast data of $S_e$ against $H_s$ are shown in black, with chosen storm peak values indicated in red. The storm threshold $h_\text{St}=4$ is indicated as a dashed red vertical line.}
    \label{fig:storm_peak_comb}
\end{figure}

Figure~\ref{fig:storm_peak_comb} illustrates this process, with the left panel showing identified storms. Given selection of storm peak $H_s$ values, these can be matched with the corresponding $T_p$, $T_e$ or $T_m$ values to obtain a joint storm peak environment. We focus our attention on storm peak significant wave height $H_s$ and (significant) wave steepness
\begin{equation}
    S_e = \frac{2 \pi H_s}{gT_e^2},
\end{equation}
for gravitational acceleration $g=9.81\text{ms}^{-2}$, modelling the 2{-}dimensional environment $(H_s, S_e)$. We choose to model $S_e$ over $T_e$ (or other wave period variables) because the most extreme sea states tend to be the steepest. Using $S_e$, our interest therefore lies in characterising the pair of positively valued variables $(H_s, S_e)$, when at least one of the pair is very large, an appropriate setting for application of the conditional extremes method of \cite{38}. Going forward, we let $\longX = (H_s, S_e)$, referring to the joint storm peak values seen in the right panel of Figure~\ref{fig:storm_peak_comb}, rather than the original hourly data.

% \begin{figure}[H]
%     \centering
% \includegraphics[width=0.75\linewidth]{albany/albany_EDA.png}
%     \caption{Hindcast data from Albany, 1980-2017.}
%     \label{fig:albany_eda}
% \end{figure}

% \begin{figure}[H]
%     \centering
% \includegraphics[width=0.75\linewidth]{albany/storm_peaks.png}
%     \caption{Storm Peak Hindcast data from Albany, 1980-2017.}
%     \label{fig:peaks}
% \end{figure}

\subsection{Joint storm peak variable modelling} \label{section:env_model}

\subsubsection{Outline of long term environment model}

We describe the joint behaviour a long term environment, using the conditional extreme value model of \cite{38}. This approach facilitates modelling of the joint extremes of $\longX$, facilitating the extrapolation of joint behaviour beyond the range of the sample data. This asymptotically justified framework has been widely applied in capturing the tail dependence of environmental data due to its flexibility in capturing different extremal dependence types and its ease of use (e.g., \citealt{jonathan2014non, Towe2019model, shooter2021basin, tendijck2023modeling}). It is also simple to extend this model to account for seasonality or long term trends in an environment through the addition of covariates (e.g., \citealt{ewans2008effect}), however, we omit the inclusion of covariate effects as this is not the focus of this work. Furthermore, whilst our synthetic environment and example data are of dimension $d=2$, we present the conditional extremes method in the general case of $d>1$. 

The conditional extremes method consists of a two stage modelling process: first, the transformation of environment variable $\longX$ to a standard marginal scale (typically Laplace); and second, modelling of the joint structure of the standardised variables. The first step is achieved using univariate extreme value techniques (e.g., \citealt{davison1990models}), and the second via a series of $d$ pairwise non-linear regressions. Details of these steps are discussed below.

\subsubsection{Marginal modelling and transformation}

We use the peaks over threshold method of \cite{davison1990models} when modelling the marginal distributions of $X_1, \ldots, X_d$. That is, for $X_j$, $j=1, \ldots, d$, we fit a generalised Pareto distribution (GPD) to sample exceedances of $X_j$ above some high threshold $u_j$, modelling non-exceedances empirically. The full model for the marginal distribution $F_{X_j}$ of $X_j$ is 
\begin{equation} \label{eq:gpd}
F_{X_j}(x) = 
\begin{cases}
    \Tilde{F}_{X_j}(x) & x \leq u_j \\
    \Tilde{F}_{X_j}(u_j) + \{1-\Tilde{F}_{X_j}(u_j)\} F_{\text{GPD},j}(x; u_j, \sigma_j, \xi_j) & x > u_j,
\end{cases}
\end{equation}
for empirical distribution $\Tilde{F}_{X_j}$ of $X_j$, and GPD distribution function
\begin{equation}
F_{\text{GPD},j}(x;u_j, \sigma_j, \xi_j) = 1 - \left( 1+ \frac{\xi_j (x-u_j)}{\sigma_j}\right)_+^{-1/\xi_j}, \quad x>u_j,
\end{equation}
for scale and shape parameters $\sigma_j>0$ and $\xi_j \in \mathbb{R}$, with $y_+ = \max(y, 0)$ for $y \in \RR$. The conditioning thresholds $u_j$, $j=1,\ldots, d$, are chosen so the asymptotic behaviour justifying the use of the GPD tail distribution holds approximately. Appropriate values of these thresholds are typically selected by either manually examining the stability of $\sigma_j$ and $\xi_j$ when fitting to exceedances above candidate values for $u_j$, or using automated methods such as those of \cite{varty2021inferenceextremeearthquakemagnitudes} and \cite{Murphy03042025}. We find parameter stability tests to be satisfactory for our data, see SM4.1. Given a choice of $u_j$, parameters $\sigma_j$ and $\xi_j$ are found using maximum likelihood techniques.

The marginal model $F_{X_j}$ is used to map $X_j$ onto $X'_j$ with Laplace margins, via the probability integral transform
\begin{equation}\label{eq:marg_trs}
    X'_j = \begin{cases}\log \left\{2 F_{X'_j}\left(X'_j\right)\right\} &  X'_j\leq F_{X'_j}^{-1}(0.5) \\ -\log \left\{2\left[1-F_{X'_j}\left(X'_j\right)\right]\right\} &  X'_j > F_{X'_j}^{-1}(0.5),\end{cases}
\end{equation}
for $j=1, \ldots, d$, obtaining the multivariate Laplace-scale environment variable $\longX' = (X'_1, \ldots, X'_d)$.

\subsubsection{Joint dependence modelling}

We now apply the conditional extremes framework to the Laplace-scale environment variable $\longX' \in \RR^d$. This requires specifying a conditioning environmental variable $X_j' \in \RR$, followed by modelling the remaining variables $\longX'_{-j} \in \RR^{d-1}$, conditional on the event $X'_j>v_j$ for $v_j>0$, $j=1,\ldots,d$. Fitting this model allows simulation of new multivariate events with extremal dependence structure representative of the original process $\longX'$, facilitating estimation of joint extreme event set probabilities.

Broadly following \cite{Keef2013}, \cite{38} assume that, for $j=1,\ldots,d$, there exist unique values $\bm{\alpha}_{|j}\in [-1,1]^{d-1}$, $\bm{\beta}_{|j}\in (-\infty, 1]^{d-1}$ and $\mathbf{z}_{|j} \in \RR^{d-1}$, such that
\begin{equation} \label{eq:ht_limit}
    \lim_{v_j\rightarrow \infty }\Prbb{\frac{\longX'_{-j} -\bm{\alpha}_{|j} X'_j}{X_j'^{\bm{\beta}_{|j}}} < \mathbf{z}_{|j}, X'_j - v_j >x | X'_j >v_j } = e^{-x} G_{|j}(\mathbf{z}_{|j}),
\end{equation}
for $x>0$ and distribution function $G_{|j}: \RR^{d-1}\mapsto \RR$ with non-degenerate marginals, where componentwise operations are assumed. In practice, the limit \eqref{eq:ht_limit} is assumed to hold for some suitably large finite threshold $v_j$, yielding the regression 
\begin{equation} \label{eq:ht_regression}
    \longX'_{-j}|\{X'_j = x\} = \bm{\alpha}_{|j}x + x^{\bm{\beta_{|j}}}\mathbf{Z}_{|j},
\end{equation}
for $x > v_j$, and residual random variable $\mathbf{Z}_{|j}$ independent of $X'_j$ given $X'_j>v_j$, where element-wise operations are assumed. Regression \eqref{eq:ht_regression} is then used to model all data in the region $\{\longX'_j \in \RR^d: X'_j>v_j\}$, and parameters $\bm\alpha_{|j}$ and $\bm\beta_{|j}$ are estimated using standard maximum likelihood estimation (MLE) techniques. For this estimation we utilise the additional parameter constraints of \cite{Keef2013} ensuring consistency of conditional return level values between extremal dependence types. For model fitting only, it is assumed that $G_{|j}$ corresponds to independent Gaussian distributions with unknown means and variances. Once parameter estimates have been obtained, we follow \cite{winter2017kth} and model $G_{|j}$ using the Gaussian kernel smoothed density estimate of the observed values of residual
\begin{equation} \label{eq:ht_res}
    \mathbf{Z}_{|j} = \frac{\longX'_{-j} - \bm{\alpha}_{|j} X'_j}{X_j^{\bm\beta_{|j}}},
\end{equation}
for $X'_j>v_j$, smoothed using kernel bandwidth $\delta_\text{HT}>0$. The conditioning threshold $v_j$ is chosen by studying parameter stability above candidate values, see SM4.1. The selection of $\delta_\text{HT}$ is considered in SM4.2. This model is fitted for all choices of the conditioning variable $\longX'_j$, allowing simulation of $\longX'$ in each of the corresponding regions $\{\longX' \in \RR^d: X'_j > v_j\}$ as described by \cite{38}.

The environment density $f_\longX$ is then estimated as in \cite{speers2024estimating}, using prediction from fitted models in each of the the upper tail regions $\longE^{(j)}= \{\longX' \in \RR^d: X'_j > v_j\}$, $j=1,\ldots, d$, followed by empirical estimation in the remaining lower region $\longE^\text{Lw}=\{\longX' \in \RR^d: X_j \leq v_j \forall j\}$. In a given upper region $\longE^{(j)}$, we make Laplace-scale simulations from the joint dependence model \eqref{eq:ht_limit} (using the parameter estimates found via MLE), followed by marginal transformation back to the physical scale using the inverse of transformation \eqref{eq:marg_trs}. During simulation in $\longE^{(j)}$, we reject realisations for which $\max_{j': j'\neq j}X'_{j'} > X'_j$. This simulation yields a set of $n_\text{Sm}$ realisations of $\longX$ within $\longE^{(j)}$, from which we may empirically estimate the probability density $f_\longX$ over a gridded set of subregions of $\longE^{(j)}$. Specifically, for a set $D$ of feasible values of $\longX$ such that $\Prbb{\longX \in \longE \setminus D} \approx 0$, we partition $D$ using grid $(D_1, \ldots, D_{n_\text{Gr}})$. We then assume that each $|D_i|$, $i=1,\ldots, n_\text{Gr}$, is small enough for the approximation  
\begin{equation}\label{eq:prob_approx}
    \Prbb{\mathbf{X} \in D_i} =\int_{\mathbf{s}\in D_i} {f}_\mathbf{X}(\mathbf{s}) \wrt{\mathbf{s}}\approx |D_i|{f}_\mathbf{X}(\mathbf{x}),
\end{equation}
to be suitable, assuming that $f_\longX$ is reasonably constant for all $\longx \in D_i$. For any $D_i$ within an upper tail region $\longE^{(j)}$, $j=1,\ldots,d$, we estimate the joint density with 
\begin{equation}
    \hat{f}^{(j)}_\longX(\longx) = \frac{n_{\text{Sm}}^{(i)}}{n_\text{Sm}|D_i|}, \quad \longx \in D_i \subset \longE^{(j)},
\end{equation}
where $n^{(i)}_\text{Sm}$ is the number of simulated values of $\longx$ in $D_i$. Combining these estimates with the empirical density $\tilde{f}_\longX$ used in the lower region $\longE^\text{Lw}$, the full density estimate for $\longx \in \longE$ is
\begin{equation}\label{eq:dens_est}
    \hat{f}_\longX(\longx) = \begin{cases}
        \tilde{f}_\longX(\longx) & \longx \in \longE^\text{Lw}\cap D , \\
        \hat{f}^{(j)}_\longX(\longx) & \longx \in  \longE^{(j)} \cap D, \quad j =1,\ldots,d, \\
        0 & \longx \notin D.
    \end{cases}
\end{equation}

\subsubsection{Estimate of the environment density} \label{section:dens_est}

\begin{figure}[t]
    \centering
    \includegraphics[width=0.75\linewidth]{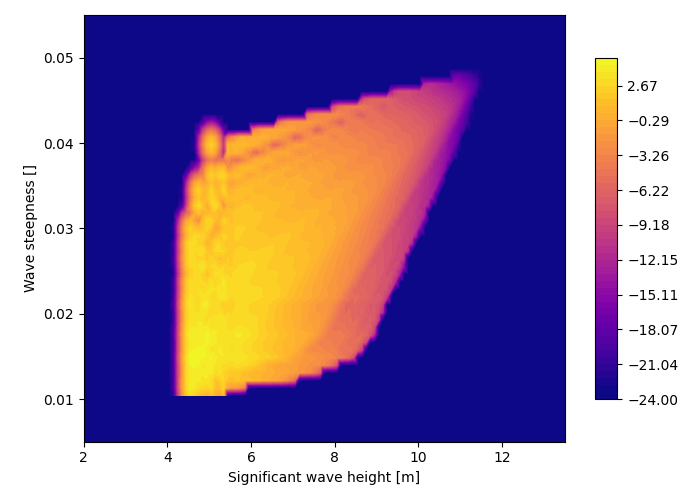}
    \caption{Estimate of the joint density $f_\longX$ of environment variable $\longX=(H_S, S_e)$, using the conditional extremes model of \cite{38} fitted to storm peak data from a location 1km offshore of Albany.}
    \label{fig:dens_0.4}
\end{figure}

Figure \ref{fig:dens_0.4} shows the resulting estimate \eqref{eq:dens_est} of the environment density $f_\longX$, found using the conditional extremes model. We take marginal thresholds $u_1=\tilde{F}^{-1}_{H_s}(0.7)$ and $u_2=\tilde{F}^{-1}_{S_e}(0.7)$, where $\tilde{F}_{H_s}$ and $\tilde{F}_{S_e}$ are the empirical distribution functions of $H_s$ and $S_e$. The conditioning threshold for $H_s$ is chosen as  $v=\tilde{F}^{-1}_{H_s}(0.6)$. When simulating joint values of $\longX$ conditional on $H_s>v$, we take $\delta_\text{HT}=0.4$, $n_\text{Sm}=10^{5}$, $D=[3, 12] \times [0.01, 0.05]$ and $n_\text{Gr} = 90 \times 45$; these values yield $|D_i|$ small enough for approximation \eqref{eq:prob_approx} to be reasonable, for negligible computational cost when estimating density $f_\longX$. We choose not to fit the conditional extremes model to the region where $S_e$ is large since in typical offshore applications, large values of $H_s$ rather than $S_e$ dominate structural failure. We model the density empirically for values of $H_s$ below the conditioning threshold $v$, and apply a Gaussian kernel smoother with bandwidth chosen according to \cite{scott2015multivariate}.

\subsection{Non-linear harmonic structural response simulation} \label{section:TFNV_wave}

\subsubsection{Overview of response simulation}

We now obtain empirical distributions of the structural response $R|\{\longX=\longx\}$, $\longX=(H_s, S_e)$, in our model monopile scenario, given a fixed environment $\longx \in D$. This is achieved using realisations $R_j|\{\longX=\longx\}$, $j = 1,\ldots, n_\text{Rl}$, of the structural response, obtained via repeated direct simulation using physical models of environmental loading on the monopile. For each of $n_\text{Rl}$ realisations of fluid loading, this requires: (a) simulation of hour-long time series realisations of the stochastic linear wave elevation $\{E_j(t;\longx): t \in [0,60^2]\}$ for an underlying environment $\longx$; (b) conversion of the surface elevation $E_j(t; \longx)$ to load $L_j(t; \longx)$ induced on a monopile; (c) transformation of this load via a linear response function to resonant response ${R}_j(t; \longx)$ time series observed on the model structure; and (d) isolation of maximum response $R_j|\{\longX=\longx\}$ from the time series ${R}_j(t; \longx)$. Note that all physical quantities are given in SI units throughout this section. Further, ${R}_j(t; \longx)$ refers to a time series of response whereas $R_j|\{\longX=\longx\}$ is the maximum response observed over the time series,
\begin{equation}\label{eq:max-resp}
    R_j|\{\longX= \longx\} = \max_{t\in [0, 60^2]} {R}_j(t; \longx).
\end{equation}
Step (a) is achieved by modelling the surface elevation according to linear wave theory (see e.g., \citealt{holthuijsen2010waves}). We obtain the load in (b) using linear surface elevation as input to the methods of \cite{taylor2024transformed} and \cite{Orszaghova2025}, outputting an approximation to the non-linear inertial load $L_j(t; \longx)$ that $E_j(t; \longx)$ induces on a monopile. For (c) we pass this load through the linear response function for a damped harmonic oscillator, obtaining a realisation of harmonic response time series ${R}_j(t; \longx)$. Steps (a)-(d) are repeated for all $n_\text{Rl}$ realisations, yielding the a numerical estimate of the environment conditioned response $R|\{\longX=\longx\}$. 

\subsubsection{Simulation details} \label{section:sim_details}

Under linear wave theory, the surface elevation $E_j(t; \longx)$ at time $t>0$, in a sea state with parameter $\longX =\longx$, is modelled as the finite sum of Fourier components at $n_\text{Fr}$ evenly spaced frequencies $f_1, \ldots, f_{n_\text{Fr}}>0$, $f_2 - f_1 = \Delta_\text{Fr}$, with contributions determined by underlying wave spectrum $S(f; \longx)$. We take $n_\text{Fr}=60^2$, $f_1=10^{-3}$ and $f_{n_\text{Fr}}=1$, and the JONSWAP \citep{hasselmann1973measurements} parametric form for $S(f; \longx)$ (see SM2) and then model the surface elevation at the location of the structure as
\begin{equation}\label{eq:eta}
    E_j(t; \longx) = \sum_{i=1}^{n_\text{Fr}} \Big\{A_i|\{\longX=\longx\} \cdot \cos (2 \pi f_it) + B_i|\{\longX=\longx\} \cdot \sin (2 \pi f_it)\Big\}, \quad t>0,
\end{equation}
where $A_i|\{\longX=\longx\}$, $B_i|\{\longX=\longx\} \sim N(0, \Delta_{\text{Fr}} S(f_i;\longx))$, $i = 1,\ldots,n_\text{Fr}$, are random Gaussian coefficients with variance equal to the wave energy in frequency band $(f_i - \Delta_{\text{Fr}}/2, f_i + \Delta_{\text{Fr}}/2)$ of the discretised wave spectrum. Model \eqref{eq:eta} assumes the monopile is placed at the spatial origin and is concentrated at this point with no spatial dimensions. The wave surface elevation \eqref{eq:eta} is stochastic due to the random Gaussian coefficients, requiring multiple realisations of hourly time series $\{E_j(t;\longx): t \in [0,60^2]\}$ to capture the full behaviour of the wave surface when $\longX=\longx$.

The method of \cite{Orszaghova2025} takes these linear surface elevation time series $E_j(t; \longx)$, recovers the non-linear higher-order harmonics of the wave signal (see SM4.3) and outputs a time series of non-linear horizontal monopile loading $L_j(t; \longx)$ using the T-FNV model of \cite{taylor2024transformed}. We omit the full details of this methodology here as it is beyond the scope of our case study; in short, the method allows evaluation of structural load without the need to calculate full wave-kinematic profiles (see e.g., \citealt{speers2024estimating}), greatly increasing computational efficiency. For this reason, the T-FNV approach is well-suited for our example scenario, as it provides physically-accurate model output (thus testing our methodology in a realistic setting) at a low computational cost (allowing us to generate the full true CDE \eqref{equation:cde} as the target). This method requires specifying a water depth $d$, for which we take $d=30$.

To approximate the effect of wave-induced oscillation on the model monopile, we pass the load $L_j(t; \longx)$ through the linear response, or transfer, function of a damped harmonic oscillator. For input signal $L_j(t; \longx)$, the output signal ${R}_j(t; \longx)$ is then defined as 
\begin{equation}
    \chi_{{R}_j}(f) =  \chi_T(f;\gamma) \chi_{L_j}(f), \quad f>0,
\end{equation}
where transfer function $\chi_T(f;\gamma)$, the ratio of Fourier transform of the output to the input, takes the form
\begin{equation} \label{eq:trans}
    \chi_T(f;\gamma) = \frac{1}{f_0 - f^2 + i \gamma f},
\end{equation}
Alternatively, 
\begin{equation}\label{eq:harmonic_response}
    {R_j}(t; \longx) = \mathcal{F}^{-1}\{\chi_T(f) \cdot\mathcal{F}(L_j(t;\longx)\},
\end{equation}
where $\mathcal{F}: \RR \mapsto \RR^+$ is the Fourier transform mapping functions in the time domain to the frequency domain. See SM4.3 for further discussion of the transfer function \eqref{eq:trans}. From time series \eqref{eq:harmonic_response}, we obtain a realisation of maximum response \eqref{eq:max-resp}. Steps (a)-(d) are carried out over a grid of environment values $\longx_1, \ldots, \longx_{n_\text{Gr}}$, chosen as the centre points of the cells $D_i$, $i=1,\ldots,n_\text{Gr}$, of $\longE$ used to estimate the environment density via \eqref{eq:dens_est}. At each grid point $\longx_i$, we obtain $n_\text{Rl}=1000$ realisations of the response ${R|\{\longX=\longx_i\}}$, for centre value $\longx_i$ of $D_i$. This procedure provides an empirical estimate $\tilde{F}_{R|\longX}(\cdot|\longx)$ of the distribution of response $R|\{\longX=\longx\}$, for $\longx = \longx_1, \ldots, \longx_{n_\text{Gr}}$, from $n_\text{Rl}=1000$ realisations $R_j|\{\longX=\longx\}$, $j=1,\ldots, n_\text{Rl}$, for each $\longx$.

\subsection{Benchmarking: obtaining a good estimate of CDE and probability of failure} \label{section:case_results}

\begin{figure}[t]
    \centering
    \includegraphics[width=.75\linewidth]{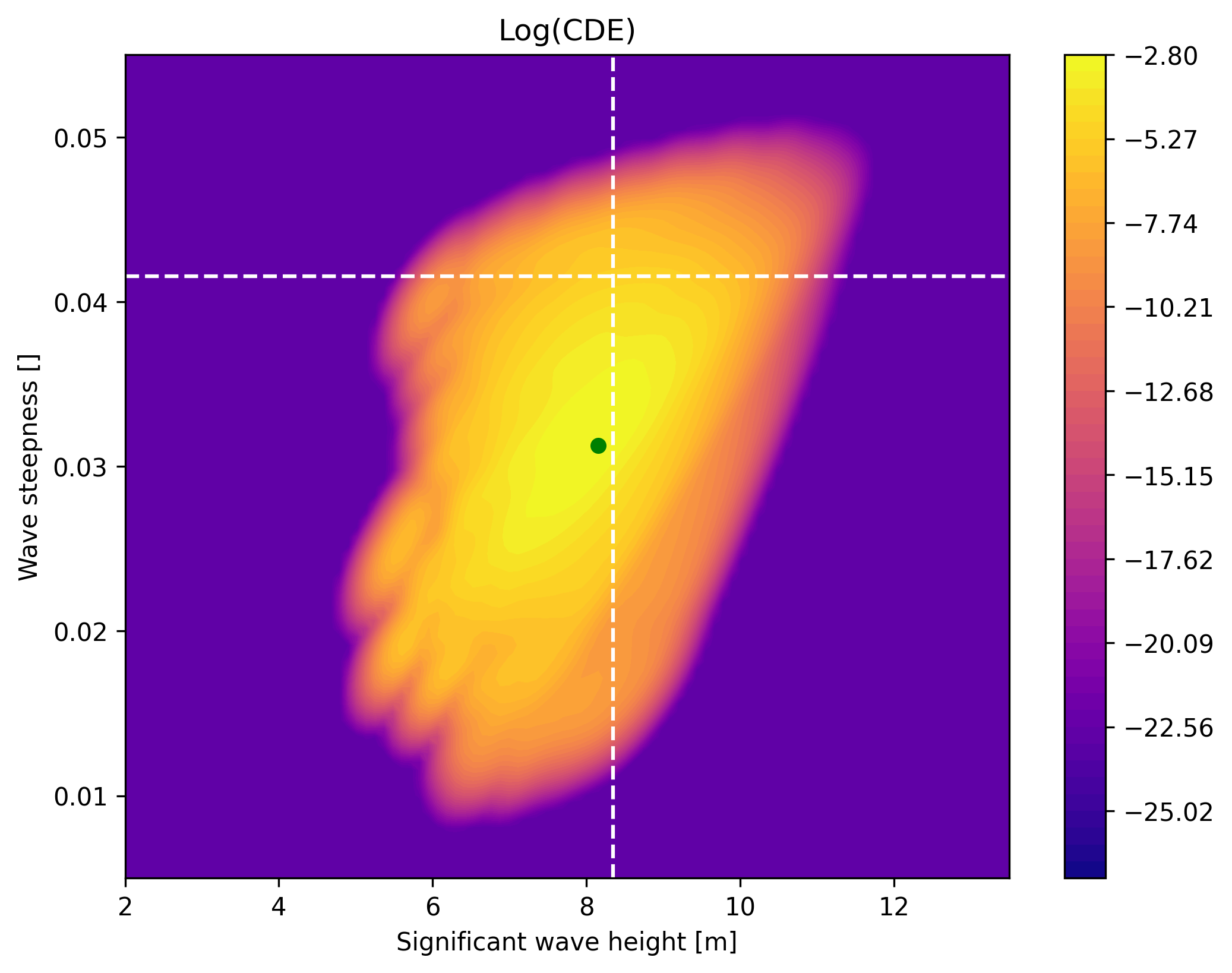}
    \caption{Conditional density of the environment (CDE) for the oscillating monopile scenario, conditioned on exceedance of the 50-year response event. White lines show the marginal 50-year events. The mode of the CDE is indicated in green.}
    \label{fig:tfnv_cde}
\end{figure}

We choose to exploit knowledge of the distribution of $R|\{\longX=\longx\}$ over the full grid of $\longx_{1}, \ldots, \longx_{n_\text{Gr}} \in \longE$ in order to obtain a good estimate of CDE. In practical application, we would not attempt this estimation, since it requires a prohibitively expensive total of $n_\text{Rl} \times n_\text{Gr}$ function evaluations of $R|\{\longX=\longx\}$. However, with this good estimate of CDE, we are able to evaluate the performance of the IS-PT and AGE procedures, the main objective of this section, reported in Section~\ref{section:tfnv_age}. The CDE is estimated using the simulation of Section~\ref{section:TFNV_wave} as 
\begin{equation}\label{eq:tfnv_cde}
    \tilde{f}_\longX(\longx; r_\text{Cr}) = \left\{ 1-\tilde{F}_{R|\longX}(r_\text{Cr}|\longx) \right\} \times \hat{f}_\longX(\longx),
\end{equation}
for empirical distribution $\tilde{F}_{R|\longX}(\cdot|\longx)$. Further, $\hat{f}_\longX$ is the estimated environment density \eqref{eq:dens_est} and $r_\text{Cr}$ is the critical response. The resulting CDE estimate, smoothed using a Gaussian kernel smoother with \cite{scott2015multivariate} bandwidth, is shown in Figure \ref{fig:tfnv_cde}. The white dashed lines show the marginal 50-year events for both $H_s$ and $S_e$, found using the marginal extreme value models \eqref{eq:gpd}. The modal point of the estimated CDE \eqref{eq:tfnv_cde} is indicated in green. 

In practice, the critical response $r_\text{Cr}$ is specified by domain experts from detailed knowledge of the structure, and the corresponding failure probability then estimated as discussed in Section~\ref{section:introduction}. Here, we set the value of $r_\text{Cr}$ in order to yield a known failure probability for testing purposes. The critical response $r_\text{Cr}$ is set to $r_\text{Cr} = \tilde{F}_{R_A}^{-1}(1-1/50)$, the 50-year response event, where 
\begin{equation}
    \tilde{F}_{R_A} = \sum_{m=0}^\infty [\tilde{F}_{R}(r)]^m \frac{\rho_\text{St} e^{-\rho_\text{St}}}{m!} = \exp \big[ -\rho_\text{St}(1-\tilde{F}_{R}(r) \big],
\end{equation}
is the empirical distribution of the annual maximum response $R_A$. Further, $\rho_\text{St}=26$ is the expected number of storms per annum estimated empirically from the data, and
\begin{equation}
    \tilde{F}_R(r) = \int_\longE \tilde{F}_{R|\longX}(r|\longx) \hat{f}_\longX(\longx) \wrt{\longx},
\end{equation}
the empirical distribution of the marginal response $R$ for a single storm event. See Section 3.1.2 of \citealt{speers2024estimating} for further discussion of annual response distribution estimation. The resulting `single storm' failure probability in this case becomes $p_\text{TFNV} = 1.1\times10^{-3}$.

\subsection{IS-PT results} \label{section:tfnv_ispt}
\begin{figure}[t]
    \centering
    \includegraphics[width=\linewidth]{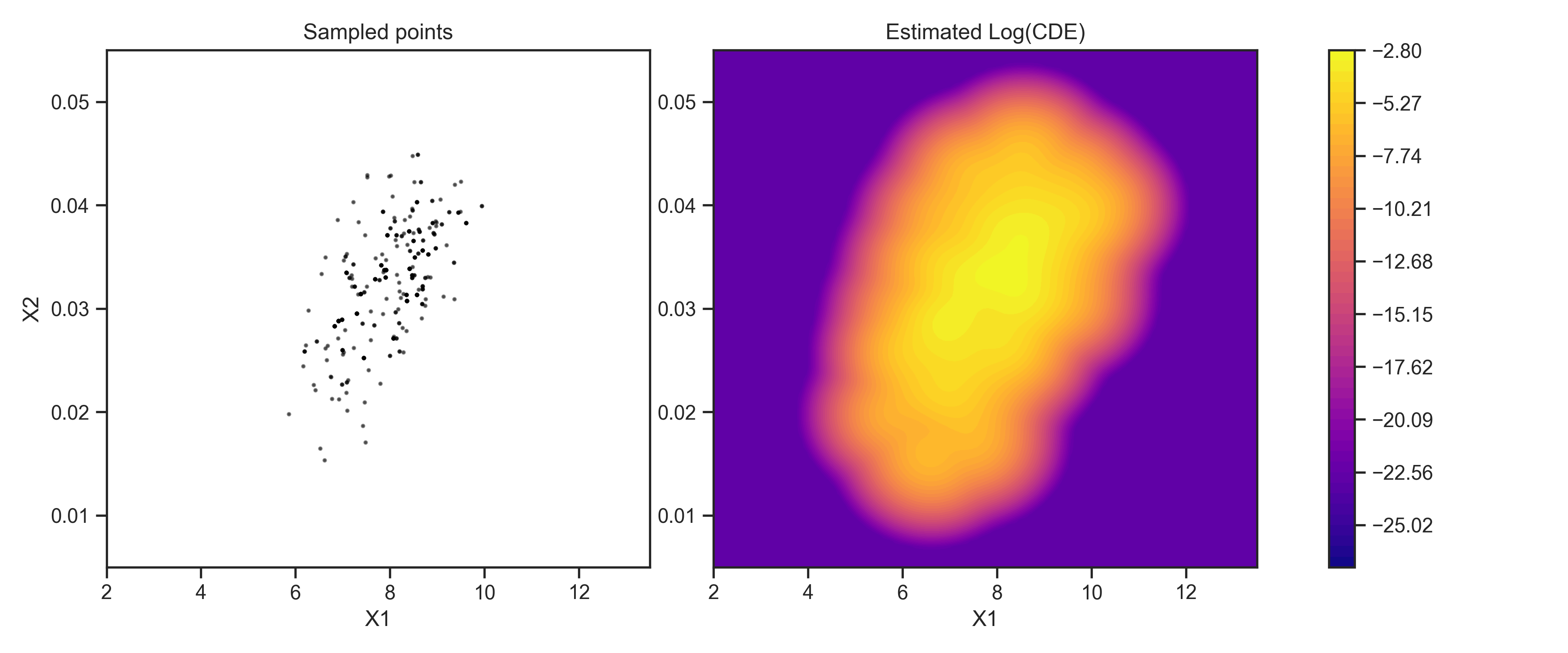}
    \caption{Example sample from CDE \eqref{equation:cde} under the synthetic structural scenario obtained using the adaptive parallel tempering MCMC algorithm of \cite{Vousden2015} (left), and corresponding smoothed log-CDE estimated using Gaussian kernel bandwidth selected according to \cite{scott2015multivariate} (right).} 
    \label{fig:ispt_tfnv}
\end{figure}

We use the IS-PT approach of Section~\ref{section:methodology} to emulate the CDE \eqref{eq:tfnv_cde}, taking sampling parameter values $n_\text{Tm} = 5$, $n_\text{PT}=400$, $n_\text{IS}=100$, initial proposal variance $\sigma^2_\text{MH}=1$, and bounding temperatures $T_1=1$, $T_5 = 20$ seen to perform well in Section~\ref{section:is_results}. For $n_\text{Rp}=100$ replicates, the adaptive algorithm of \cite{Vousden2015} is used to obtain a sample of size $n_\text{PT}=400$ (minus burn-in length $n_\text{Br}$) from the CDE. A Gaussian kernel smoothed estimate with Scott's bandwidth is then used as proposal density $p_\text{Pr}$ in importance sampling estimate \eqref{equation:paper_p1IS}. Figure~\ref{fig:ispt_tfnv} shows an example MCMC sample and resulting proposal estimate of the CDE at a single replicate. Over all $n_\text{IS}=100$ replicates, we obtain $\text{RMSE}(\hat{p}_\text{IS}) = \left({\sum_{r=1}^{n_\text{Rp}}{(\hat{p}_{\text{IS}}^{(r)} - p_\text{TFNV})^2}/{n_\text{Rl}}}\right)^{1/2} = 5.10\times10^{-5}$ where $\hat{p}_{\text{IS}}^{(r)}$ is the probability estimate \eqref{equation:paper_p1IS} obtained at replicate $r$, for true $p_\text{TFNV} = 1.1\times10^{-3}$. The corresponding bias $\text{Bias}(\hat{p}_\text{IS})=1.83\times 10^{-6}$ is also small.

\begin{figure}[H]
    \centering
    \includegraphics[width=0.95\linewidth]{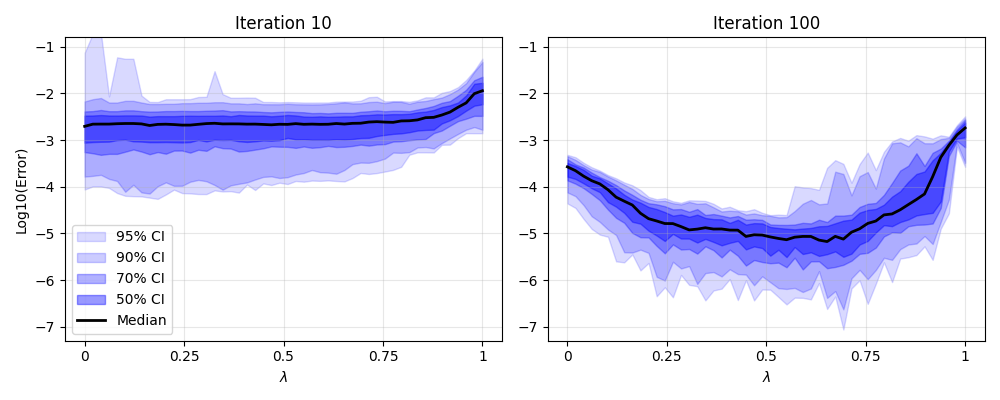}
   \caption{Log-scale absolute error $\Delta_\text{GP}$ of the GP probability estimate $\hat{p}_\text{GP}$ at specified iterations, for emulator \eqref{equation:gp_log} trained using $U^{(1)}$ over the range of weight $\lambda \in [0.01, 0.99]$ for the TFNV scenario. At iteration 100, the weight that minimises median error is $\lambda^*=0.67$.}
    \label{fig:nu_sens_tfnv}
\end{figure}

\begin{figure}[H]
    \centering
    \includegraphics[width=0.75\linewidth]{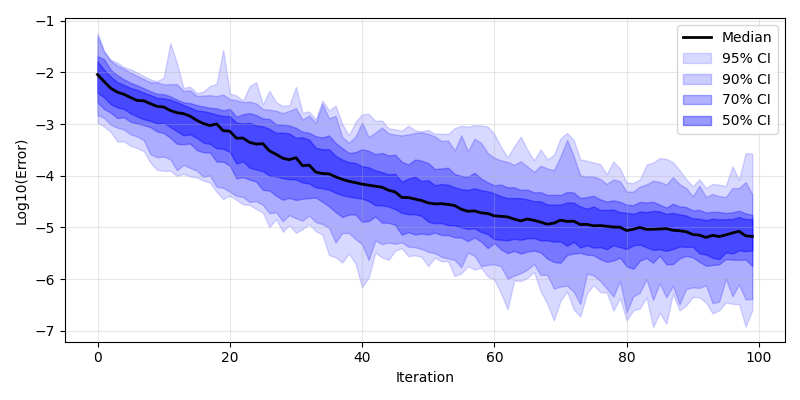}
    \caption{Distribution of log-scale absolute error $\Delta_\text{GP}$ in the GP probability estimate with respect to iteration, trained using $U^{(1)}$ with $\lambda=\lambda^*$. The trend in median error is indicated in black, with various confidence intervals shown in blue.}
    \label{fig:opt_error_tfnv}
\end{figure}

\subsection{AGE results} \label{section:tfnv_age}

\begin{figure}[t]
    \centering
    \includegraphics[width=\linewidth]{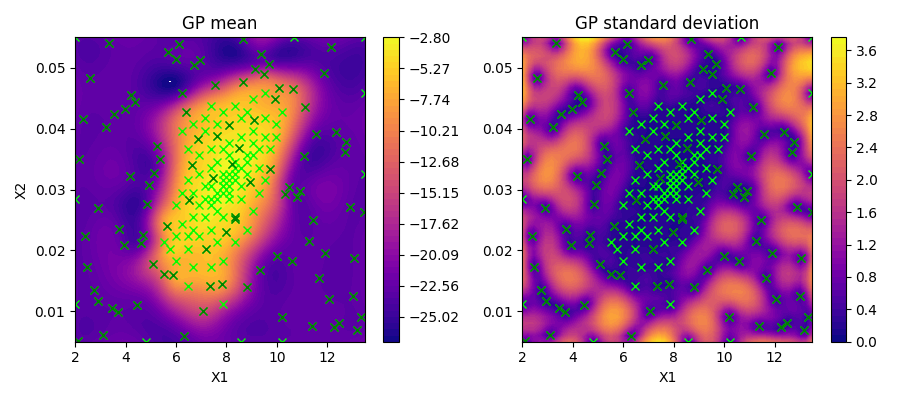}
   \caption{GP emulator at iteration 100 for variance utility $U^{(1)}$ \eqref{equation:paper_acq_function}, trained using the optimal value $\lambda^*=0.67$ minimising median of error $\Delta_\text{GP}$. The panels from left to right show the posterior GP mean $\mu^*_{100}(\longx)$ over $\longx \in \longE$, and the posterior GP standard deviation $k^*_{100}(\longx, \longx)^{1/2}$. The initial random Latin hypercube training set $\mathcal{D}_0$ is shown as dark green crosses, and the iteratively selected new training points $\mathcal{D}_{100} \setminus \mathcal{D}_0$ are shown as light green crosses.}
    \label{fig:opt_emulator_tfnv}
\end{figure}

We now use the AGE methods of Section~\ref{section:methodology} to emulate the log-CDE, seeking a reasonable estimate of probability of failure with considerably fewer than the $n_\text{Gr} \times n_\text{Rl}$ function evaluations of $R|\{\longX=\longx\}$ used for the IS-PT estimate in Section~\ref{section:case_results}. The emulator for CDE \eqref{fig:tfnv_cde} is defined as in \eqref{equation:gp_log}. For $n_\text{Rp}=100$ replicates, it is initialised using Latin hypercube sample $\mathcal{D}_0$, $|\mathcal{D}_0|=100$, then trained inductively over $n_\text{Rl}$ realisations of $n_\text{It2}=100$ iterations, using $U^{(1)}$ and $U^{(2)}$ for a range of weights $\lambda \in [0.01, 0.99]$. For utility $U^{(1)}$, Figure~\ref{fig:nu_sens_tfnv} shows the relationship between error $\Delta_\text{GP}$ and the value of $\lambda$, by comparison with the estimate of $p_\text{TFNV}$ from Section~\ref{section:case_results}. The weight $\lambda^*=0.67$ is found to minimise the median value of error $\Delta_\text{GP}$ at iteration 100. Figure~\ref{fig:opt_error_tfnv} shows $\Delta_\text{GP}$ with iteration for $\lambda^*$, and Figure~\ref{fig:opt_emulator_tfnv} shows the emulator trained using $\lambda^*$ at iteration 100. For $|\mathcal{D}_{100}|=200$ total function evaluations at $\longx \in \mathcal{D}_{100}\setminus \mathcal{D}_0$ chosen by $U^{(1)}$ with $\lambda=\lambda^*$, we obtain $\text{RMSE}(\hat{p}_\text{GP}) = \left({\sum_{r=1}^{n_\text{Rp}}{(\hat{p}_{\text{GP}}^{(r)} - p_\text{TFNV})^2}/{n_\text{Rl}}}\right)^{1/2} = 6.99\times10^{-5}$ where $\hat{p}_{\text{GP}}^{(r)}$ is the probability estimate \eqref{equation:gp_target} obtained at iteration 100 and replicate $r$, for true $p_\text{TFNV} = 1.1\times10^{-3}$. The corresponding bias $\text{Bias}(\hat{p}_\text{GP}) = 1.57 \times 10^{-5}$ is also small. Results using utility $U^{(2)}$ are reported in \ref{section:tfnv_alc}, and summarised in the next section.

\subsection{Comparison of IS-PT and AGE performance} \label{section:tfnv_comparison}

\begin{figure}[ht]
    \centering
    \includegraphics[width=\linewidth]{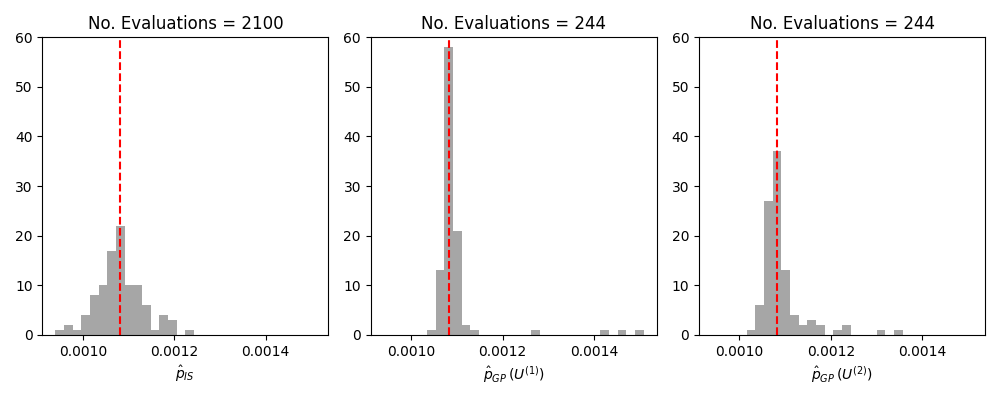}
    \caption{Distribution of $n_\text{Rp}=100$ estimates $\hat{p}_\text{IS}$ (left), $\hat{p}_\text{GP}$ for $U^{(1)}$ iteration 100 with $\lambda=\lambda^*$ (centre), and $\hat{p}_\text{GP}$ for $U^{(2)}$ iteration 100 with $\lambda=\lambda^*$ (right) for true failure probability $p_\text{TFNV}$ (red). The number of function evaluations required for a single replicate analysis is indicated in the panel titles.}
    \label{fig:tfnv_comp}
\end{figure}

Figure~\ref{fig:tfnv_comp} shows the distribution of the IS-PT estimate $\hat{p}_\text{IS}$ and the AGE estimates $\hat{p}_\text{GP}$ (based on variance and ALC utilities $U^{(1)}$ and $U^{(2)}$ at iteration 100, $\lambda=\lambda^*)$ around the target failure probability $p_\text{TFNV}$. These results are summarised in Table~\ref{tab:rmse_synth}. For the given budgets of expensive function evaluation set, as in Section~\ref{section:synth_comp}, methods demonstrate essentially equivalent performance. Again, the key issue is specification of $\lambda$ for AGE procedures. With $\lambda$ known, AGE procedures are computationally more efficient. However, specification of $\lambda$ is in general problematic, suggesting that IS-PT is a more reliably applicable approach.

% n=2100
% Monte Carlo Estimates
% RMSE: 3.74e-04
% Bias: 2.48e-04

\begin{table}[h]
    \centering
    \resizebox{\textwidth}{!}{\begin{tabular}{|c||c|c|c|}
    \hline
         & IS-PT& AGE $U^{(1)}$ & AGE $U^{(2)}$  \\ \hline \hline
        RMSE& $5.10\times 10^{-5}$& $6.99\times 10^{-5}$ & $4.99 \times 10^{-5}$ \\ \hline 
 Bias& $1.83\times 10^{-6}$&$1.57\times 10^{-5}$ & $ 9.40\times10^{-6}$\\ \hline
        Number of function evaluations, $n_\text{Ev}$&  $ n_\text{Tm} \times n_\text{PT} + n_\text{IS} = 2100$ & $|\mathcal{D}_{0}| + n_\text{Itr1}=244$ & $|\mathcal{D}_{0}| + n_\text{Itr1}=244$\\ \hline
    \end{tabular}}
    \caption{RMSEs and biases of $\hat{p}_\text{IS}$ and $\hat{p}_\text{GP}$ when targetting failure probability $p_\text{TFNV}$, calculated for $n_\text{Rp}=100$ replicate analyses. The true value of probability of failure is $p_\text{TFNV}= 1.3\times 10^{-3}$. Also shown is the number of expensive response function evaluations required for a single replicate analysis for each of IS-PT and AGE.}
    \label{tab:rmse_comp}
\end{table}

\section{Discussion}  \label{section:discussion}

Estimation of failure probability for marine structures can be a computationally demanding task. In earlier work \citep{speers2024estimating} we showed that the conditional density of the environment (CDE) for a structure is a useful design diagnostic, preferable to design contours. Moreover, CDE also provides a natural starting point for estimation of failure probability: the integral of the CDE over the environment space is the probability of structural failure. 

The mode of the CDE represents the combination of long term environmental conditions most likely to induce structural failure at the 50-year level. In practice, the location of the mode depends both on the extremal dependence characteristics of the environment variables, and the nature of the fluid-structure interaction. Interestingly, for the oscillating monopile application discussed in Section~\ref{section:application}, the location of the mode calculated by brute force in Section~\ref{section:case_results}, corresponds approximately to the combination of the 50-year storm peak significant wave height and the 1-year storm peak significant wave steepness. 

In this work, we develop, demonstrate and compare two methods to estimate the CDE and hence failure probability for simple monopile structures. The first methodology (IS-PT) incorporates parallel tempering MCMC to estimate the CDE, together with importance sampling to estimate failure probability. The second methodology uses adaptive Gaussian emulation (AGE) to estimate the CDE and thence Bayesian quadrature to estimate failure probability. 

Whereas use of either methodology requires the specification of hyperparameters, the AGE approach is particularly problematic, necessitating the specification of a key weight ($\lambda$ in e.g., \eqref{eq:acq_general}) to control the extent to which adaptive emulation is encouraged to explore the environmental space as opposed to exploiting already-identified informative structure in that space. Specification of $\lambda$ is in general case dependent.

The computational complexity of each methodology is typically dictated by the number of expensive evaluations of the structural response $R$ given long term environmental conditions $\longX$. If the value of explore-exploit $\lambda$ is known, then procedures adopting AGE provide a good estimate of failure probability requiring an order of magnitude fewer expensive function evaluations than IS-PT. We demonstrate the good performance of IS-PT and two AGE procedures on a simple synthetic structure with complex fluid loading behaviour (and bimodal CDE), and on a more realistic monopile structure (with more straightforward unimodal CDE). Good performance for AGE procedures requires knowledge of the optimal choice of explore-exploit $\lambda$, which was evaluated by us in this work using assumed knowledge of the true structural response; obviously, this information will not generally be available to the structural designer. Nevertheless, if it is anticipated that the CDE is likely to be unimodal, we speculate that the choice of $\lambda$ is likely to be less critical than for more complex CDEs. This can be seen, e.g., by comparison of the intervals $I^*$ of minimum median error in the right hand panels of Figure~\ref{fig:nu_sens1} (bimodal CDE, AGE with variance utility $U^{(1)}$, $I^* \approx [0.6,0.9]$), Figure~\ref{fig:nu_sens_tfnv} (unimodal, $U^{(1)}$, $I^* \approx [0.25,0.75]$), Figure~\ref{fig:alc_nu_sens} (bimodal, AGE with ALC utility $U^{(2)}$, $I^* \approx [0.2, 0.35]$) and Figure~\ref{fig:alc_nu_sens2} (unimodal CDE, $U^{(2)}$, $I^* \approx [0.1, 0.5]$). Intervals $I^*$ of acceptable values for $\lambda$ are wider for unimodal CDEs, and moreover the intervals corresponding to utilities $U^{(1)}$ and $U^{(2)}$ overlap. However, for bimodal CDEs, the intervals $I^*$ corresponding to $U^{(1)}$ and $U^{(2)}$ are disjoint. Given the importance of selecting $\lambda$ well, investigation into the performance acquisition functions not reliant on the specification of weight parameter $\lambda$ (e.g., \citealt{NIPS2012_6364d3f0}, \citealt{NIPS2014_a0d08267}) is warranted. These methods, however, incur additional theoretical assumptions and computational complexity, which may limit their usefulness in general offshore applications. 

{More generally, sampling from high-dimensional, multimodal distributions is receiving increasing attention in the academic literature, since it is a fundamental issue in modern statistical learning. Tempering methods are often used when multimodality is suspected, yet these can in principle be difficult to tune as noted by e.g., \cite{Miasojedow01072013}, especially in high dimensional settings. Solutions have been proposed, including \cite{graham2017}, \cite{Luo2018} and \cite{park2025} in the context of Hamiltonian Monte Carlo sampling. Other authors (e.g., \citealt{Qiu2024}) consider sampling using tempered distribution flows. In fact, the parallel tempering package exploited in the current work has already been applied successfully in five dimensions by \cite{Vousden2015}; this may be sufficient for less complex metocean design studies. Nevertheless, exploration of the performance of IS-PT in higher dimensions for more complex structural types is an interesting areas for future work; studies along these lines are currently being conducted by the authors.}

If the optimal value of $\lambda$ {for AGE} is unknown (as will generally be the case), IS-PT provides a reliable general-purpose approach to estimation of CDE and failure probability useful even for challenging multimodal CDEs. In the current work, we consider univariate responses in a two-dimensional environment. We anticipate that, for higher-dimensional responses and environments, the structure of the CDE will be more complex (and multimodal) in general. Given this, it appears reasonable to assume that IS-PT will prove a more reliable route to estimation of CDE and probability of structural failure. {In contrast, we view AGE as a method requiring further development before it can be considered a mature engineering solution.}

{
The current work should be considered in the context of the growing body of academic literature on the application of machine learning and related techniques in ocean engineering, in particular to the design and reanalysis of offshore structures. \cite{elhamahmy2025integrating} provides a recent review, comparing different methodologies with examples of specific use cases, and a discussion of current and future challenges. \cite{ALVESRIBEIRO2025126294} provides a review specific to the design of offshore wind facilities. One popular area of application is the use of machine learning to estimate surrogate models (e.g., using deep neural networks, Gaussian processes) to emulate the output of computationally-complex physical model simulators, e.g., for fluid loading and soil mechanics (\citealt{yu2024construction}, \citealt{ALVESRIBEIRO2025126294}, \citealt{jmse12112001}, \citealt{https://doi.org/10.1002/we.2851}, \citealt{MERT2025122952}). The review of \cite{Zhou2025} lists a number of outstanding issues which need to be addressed before modern machine learning tools can be adopted reliably in practical application. These include: (a) the need for adequate databases to train machine learning and deep learning models in offshore engineering contexts; (b) know-how to guide the appropriate choice of machine learning model architectures and to estimate the often large number of hyperparameters involved; (c) the need to develop interpretable machine learning models which can be understood and diagnosed by the ocean engineering practitioner; and (d) improved collaboration between specialists in machine learning and structural engineering, bridging the gap between machine learning theory and academic application, and the real world. 
}

\section*{Acknowledgments}
The work was completed while Matthew Speers was part of the EPSRC funded STOR-i
centre for doctoral training (grant no. EP/S022252/1), with part-funding from the ARC
TIDE Industrial Transformational Research Hub at the University of Western Australia.
{The authors wish to acknowledge the support of Jana Orszaghova and Paul Taylor at the University of Western Australia, and David Randell at Shell.}

%%~~~~~~~~~~~~~~~~~bibliography~~~~~~~~~~~~~~~~~%%
\bibliographystyle{elsarticle-harv}
\bibliography{bibli.bib}

@article{speers2024estimating,
title = {Estimating metocean environments associated with extreme structural response to demonstrate the dangers of environmental contour methods},
journal = {Ocean Engineering},
volume = {311},
pages = {118754},
year = {2024},
issn = {0029-8018},
author = {Matthew Speers and David Randell and Jonathan Angus Tawn and Philip Jonathan}
}

@article{gramstad2020sequential,
  title={Sequential sampling method using {G}aussian process regression for estimating extreme structural response},
  author={Gramstad, Odin and Agrell, Christian and Bitner-Gregersen, Elzbieta and Guo, Bingjie and Ruth, Eivind and Vanem, Erik},
  journal={Marine Structures},
  volume={72},
  pages={102780},
  year={2020},
  publisher={Elsevier}
}

@article{genton2001classes,
  title={Classes of kernels for machine learning: a statistics perspective},
  author={Genton, Marc G},
  journal={Journal of machine learning research},
  volume={2},
  number={Dec},
  pages={299--312},
  year={2001}
}

@article{lystad2023full,
  title={Full long-term extreme buffeting response calculations using sequential {G}aussian process surrogate modeling},
  author={Lystad, Tor M and Fenerci, Aksel and {\O}iseth, Ole},
  journal={Engineering Structures},
  volume={292},
  pages={116495},
  year={2023},
  publisher={Elsevier}
}

@article{castellon2023full,
  title={Full long-term buffeting analysis of suspension bridges using {G}aussian process surrogate modelling and importance sampling {M}onte {C}arlo simulations},
  author={Castellon, Dario Fernandez and Fenerci, Aksel and Petersen, {\O}yvind Wiig and {\O}iseth, Ole},
  journal={Reliability Engineering \& System Safety},
  volume={235},
  pages={109211},
  year={2023},
  publisher={Elsevier}
}

@article{wang2024comparison,
  title={Comparison of probabilistic structural reliability methods for ultimate limit state assessment of wind turbines},
  author={Wang, Hong and Gramstad, Odin and Sch{\"a}r, Styfen and Marelli, Stefano and Vanem, Erik},
  journal={Structural Safety},
  volume={111},
  pages={102502},
  year={2024},
  publisher={Elsevier}
}

@article{castellon2022investigations,
  title={Investigations of the long-term extreme buffeting response of long-span bridges using importance sampling {M}onte {C}arlo simulations},
  author={Castellon, Dario Fernandez and Fenerci, Aksel and {\O}iseth, Ole and Petersen, {\O}yvind Wiig},
  journal={Engineering Structures},
  volume={273},
  pages={114986},
  year={2022},
  publisher={Elsevier}
}

@article{Wang2021,
   author = {Jian Wang and Zhili Sun and Runan Cao},
   issn = {0951-8320},
   journal = {Reliability Engineering \& System Safety},
   month = {12},
   pages = {107953},
   publisher = {Elsevier},
   title = {An efficient and robust Kriging-based method for system reliability analysis},
   volume = {216},
   year = {2021},
}

@article{Yang2018,
   author = {Xufeng Yang and Yongshou Liu and Caiying Mi and Chenghu Tang},
   issn = {0951-8320},
   journal = {Reliability Engineering \& System Safety},
   month = {1},
   pages = {235-241},
   publisher = {Elsevier},
   title = {System reliability analysis through active learning Kriging model with truncated candidate region},
   volume = {169},
   year = {2018},
}

@article{Xiao2020,
   author = {Ning Cong Xiao and Hongyou Zhan and Kai Yuan},
   issn = {0045-7825},
   journal = {Computer Methods in Applied Mechanics and Engineering},
   month = {12},
   pages = {113336},
   publisher = {North-Holland},
   title = {A new reliability method for small failure probability problems by combining the adaptive importance sampling and surrogate models},
   volume = {372},
   year = {2020},
}

@book{holthuijsen2010waves,
  title={Waves in {O}ceanic and {C}oastal {W}aters},
  author={Holthuijsen, Leo H},
  year={2010},
  publisher={Cambridge university press}
}

@article{hasselmann1973measurements,
  title={Measurements of wind-wave growth and swell decay during the {J}oint {N}orth {S}ea {W}ave {P}roject ({JONSWAP}).},
  author={Hasselmann, Klaus and Barnett, Tim P and Bouws, E and Carlson, H and Cartwright, David E and Enke, K and Ewing, JA and Gienapp, A and Hasselmann, DE and Kruseman, P and others},
  journal={{E}rgaenzungsheft zur {D}eutschen {H}ydrographischen {Z}eitschrift, {R}eihe A},
  year={1973}
}

@article{0f232807-5558-3c25-9446-537f6b18f086,
 ISSN = {10170405, 19968507},
 author = {Xiao-Li Meng and Wing Hung Wong},
 journal = {Statistica Sinica},
 number = {4},
 pages = {831--860},
 publisher = {Institute of Statistical Science, Academia Sinica},
 title = {SIMULATING RATIOS OF NORMALIZING CONSTANTS VIA A SIMPLE IDENTITY: A THEORETICAL EXPLORATION},
 urldate = {2024-10-07},
 volume = {6},
 year = {1996}
}

@article{peherstorfer2016multifidelity,
  title={Multifidelity importance sampling},
  author={Peherstorfer, Benjamin and Cui, Tiangang and Marzouk, Youssef and Willcox, Karen},
  journal={Computer Methods in Applied Mechanics and Engineering},
  volume={300},
  pages={490--509},
  year={2016},
  publisher={Elsevier}
}

@article{Marrel2024,
   author = {Amandine Marrel and Bertrand Iooss},
   issn = {0951-8320},
   journal = {Reliability Engineering \& System Safety},
   month = {7},
   pages = {110094},
   publisher = {Elsevier},
   title = {Probabilistic surrogate modeling by {G}aussian process: A review on recent insights in estimation and validation},
   volume = {247},
   year = {2024},
}

@article{Moustapha2022,
   author = {Maliki Moustapha and Stefano Marelli and Bruno Sudret},
   issn = {0167-4730},
   journal = {Structural Safety},
   month = {5},
   pages = {102174},
   publisher = {Elsevier},
   title = {Active learning for structural reliability: survey, general framework and benchmark},
   volume = {96},
   year = {2022},
}

@article{wang2022recent,
  title={Recent advances in surrogate modeling methods for uncertainty quantification and propagation},
  author={Wang, Chong and Qiang, Xin and Xu, Menghui and Wu, Tao},
  journal={Symmetry},
  volume={14},
  number={6},
  pages={1219},
  year={2022},
  publisher={MDPI}
}

@article{tabandeh2022review,
  title={A review and assessment of importance sampling methods for reliability analysis},
  author={Tabandeh, Armin and Jia, Gaofeng and Gardoni, Paolo},
  journal={Structural Safety},
  volume={97},
  pages={102216},
  year={2022},
  publisher={Elsevier}
}

@article{chib1995understanding,
  title={Understanding the {M}etropolis-{H}astings algorithm},
  author={Chib, Siddhartha and Greenberg, Edward},
  journal={The American Statistician},
  volume={49},
  number={4},
  pages={327--335},
  year={1995},
  publisher={Taylor \& Francis}
}

@article{earl2005parallel,
  title={Parallel tempering: theory, applications, and new perspectives},
  author={Earl, David J and Deem, Michael W},
  journal={Physical Chemistry Chemical Physics},
  volume={7},
  number={23},
  pages={3910--3916},
  year={2005},
  publisher={Royal Society of Chemistry}
}

@article{taylor2024transformed,
  title={Transformed-{FNV}: wave forces on a vertical cylinder—A free-surface formulation},
  author={Taylor, Paul H and Tang, T and Adcock, Thomas AA and Zang, Jun},
  journal={Coastal Engineering},
  volume={189},
  pages={104454},
  year={2024},
  publisher={Elsevier}
}

@inproceedings{Orszaghova2025,
  author    = {Jana Orszaghova and Paul H. Taylor and Hugh Wolgamot and Guy McCauley and Adi Kurniawan and Qinming Wu and Bryan Tan and Ajay Ebey George},
  title     = {Wave Loads on Monopile Foundations Revisited – New High-Quality Experiments for Validation of a Novel Engineering Model},
  booktitle = {Proc. ASME OMAE 2025},
  year      = {2025},
  address   = {Vancouver, British Columbia, Canada},
  month     = jun,
  organization = {ASME},
  note      = {OMAE2025-156732}
}

@inproceedings{cohn1993neural,
 author = {Cohn, David},
 booktitle = {Advances in Neural Information Processing Systems},
 editor = {J. Cowan and G. Tesauro and J. Alspector},
 pages = {},
 publisher = {Morgan-Kaufmann},
 title = {Neural Network Exploration Using Optimal Experiment Design},
 volume = {6},
 year = {1993}
}

@book{rubinstein2016simulation,
  title={Simulation and the Monte Carlo method},
  author={Rubinstein, Reuven Y and Kroese, Dirk P},
  year={2016},
  publisher={John Wiley \& Sons}
}

@book{abramowitz1965handbook,
  title={Handbook of {M}athematical {F}unctions: with {F}ormulas, {G}raphs, and {M}athematical {T}ables},
  author={Abramowitz, Milton and Stegun, Irene A},
  volume={55},
  year={1965},
  publisher={Courier Corporation}
}

@INPROCEEDINGS{861310,
  author={Sambu Seo and Wallat, M. and Graepel, T. and Obermayer, K.},
  booktitle={Proceedings of the IEEE-INNS-ENNS International Joint Conference on Neural Networks. IJCNN 2000. Neural Computing: New Challenges and Perspectives for the New Millennium}, 
  title={Gaussian process regression: active data selection and test point rejection}, 
  year={2000},
  volume={3},
  number={},
  pages={241-246 vol.3}}

@article{10.1093/bioinformatics/btad711,
    author = {Schälte, Yannik and Fröhlich, Fabian and Jost, Paul J and Vanhoefer, Jakob and Pathirana, Dilan and Stapor, Paul and Lakrisenko, Polina and Wang, Dantong and Raimúndez, Elba and Merkt, Simon and Schmiester, Leonard and Städter, Philipp and Grein, Stephan and Dudkin, Erika and Doresic, Domagoj and Weindl, Daniel and Hasenauer, Jan},
    title = {py{PESTO}: a modular and scalable tool for parameter estimation for dynamic models},
    journal = {Bioinformatics},
    volume = {39},
    number = {11},
    pages = {711},
    year = {2023},
    month = {11},
    issn = {1367-4811},
}

@article{10.1093/gji/ggt342,
    author = {Sambridge, Malcolm},
    title = {A Parallel Tempering algorithm for probabilistic sampling and multimodal optimization},
    journal = {Geophysical Journal International},
    volume = {196},
    number = {1},
    pages = {357-374},
    year = {2013},
    month = {10},
    issn = {0956-540X},
}

@book{bishop2006pattern,
  title={Pattern {R}ecognition and {M}achine {L}earning},
  author={Bishop, Christopher M and Nasrabadi, Nasser M},
  volume={4},
  year={2006},
  publisher={Springer}
}

@article{JMLR:v12:pedregosa11a,
  author  = {Fabian Pedregosa and Ga{{\"e}}l Varoquaux and Alexandre Gramfort and Vincent Michel and Bertrand Thirion and Olivier Grisel and Mathieu Blondel and Peter Prettenhofer and Ron Weiss and Vincent Dubourg and Jake Vanderplas and Alexandre Passos and David Cournapeau and Matthieu Brucher and Matthieu Perrot and {{\'E}}douard Duchesnay},
  title   = {Scikit-learn: Machine Learning in {P}ython},
  journal = {Journal of Machine Learning Research},
  year    = {2011},
  volume  = {12},
  number  = {85},
  pages   = {2825-2830},
}

@article{MATHISEN199093,
title = {Joint distributions for significant wave height and wave zero-up-crossing period},
journal = {Applied Ocean Research},
volume = {12},
number = {2},
pages = {93-103},
year = {1990},
issn = {0141-1187},
author = {Jan Mathisen and Elzbieta Bitner-Gregersen}
}

@article{38,
  title={A conditional approach for multivariate extreme values (with discussion)},
  author={Heffernan, Janet E and Tawn, Jonathan A},
  journal={Journal of the Royal Statistical Society: Series B (Statistical Methodology)},
  volume={66},
  number={3},
  pages={497--546},
  year={2004},
  publisher={Wiley Online Library}
}

@article{davison1990models,
  title={Models for exceedances over high thresholds (with discussion)},
  author={Davison, Anthony C and Smith, Richard L},
  journal={Journal of the Royal Statistical Society Series B},
  volume={52},
  number={3},
  pages={393--425},
  year={1990},
  publisher={Oxford University Press}
}

@article{Murphy03042025,
author = {Conor Murphy and Jonathan A. Tawn and Zak Varty},
title = {Automated Threshold Selection and Associated Inference Uncertainty for Univariate Extremes},
journal = {Technometrics},
volume = {67},
number = {2},
pages = {215--224},
year = {2025},
publisher = {ASA Website}}

@article{shooter2021basin,
  title={Basin-wide spatial conditional extremes for severe ocean storms},
  author={Shooter, Rob and Tawn, Jonathan A. and Ross, Emma and Jonathan, Philip},
  journal={Extremes},
  volume={24},
  pages={241--265},
  year={2021},
  publisher={Springer}
}

@article{towe2019model,
  title={Model-based inference of conditional extreme value distributions with hydrological applications},
  author={Towe, Ross Paul and Tawn, Jonathan Angus and Lamb, Rob and Sherlock, Christopher Gerrard},
  journal={Environmetrics},
  volume={30},
  number={8},
  pages={e2575},
  year={2019},
  publisher={Wiley Online Library}
}

@article{jonathan2014non,
  title={Non-stationary conditional extremes of northern {N}orth {S}ea storm characteristics},
  author={Jonathan, P and Ewans, K and Randell, D},
  journal={Environmetrics},
  volume={25},
  number={3},
  pages={172--188},
  year={2014},
  publisher={Wiley Online Library}
}

@article{tendijck2023modeling,
  title={Extremal characteristics of conditional models},
  author={Tendijck, Stan and Tawn, Jonathan and Jonathan, Philip},
  journal={Extremes},
  volume={26},
  number={1},
  pages={139--156},
  year={2023},
  publisher={Springer}
}

@book{hennig2022probabilistic,
  title={Probabilistic Numerics: Computation as Machine Learning},
  author={Hennig, Philipp and Osborne, Michael A and Kersting, Hans P},
  year={2022},
  publisher={Cambridge {U}niversity {P}ress}
}

@article{pollatsek1970theory,
  title={A theory of risk},
  author={Pollatsek, Alexander and Tversky, Amos},
  journal={Journal of Mathematical Psychology},
  volume={7},
  number={3},
  pages={540--553},
  year={1970},
  publisher={Elsevier}
}

@techreport{geweke1991evaluating,
  title={Evaluating the accuracy of sampling-based approaches to the calculation of posterior moments},
  author={Geweke, John},
  year={1991},
  institution={Federal Reserve Bank of Minneapolis}
}

@book{scott2015multivariate,
  title={Multivariate {D}ensity {E}stimation: {T}heory, {P}ractice, and {V}isualization},
  author={Scott, David W},
  year={2015},
  publisher={John Wiley \& Sons}
}

@misc{varty2021inferenceextremeearthquakemagnitudes,
      title={Inference for extreme earthquake magnitudes accounting for a time-varying measurement process}, 
      author={Zak Varty and Jonathan A. Tawn and Peter M. Atkinson and Stijn Bierman},
      year={2021},
      eprint={2102.00884},
      archivePrefix={arXiv},
      primaryClass={stat.ME},
}

@article{Keef2013,
   author = {Caroline Keef and Ioannis Papastathopoulos and Jonathan A. Tawn},
   issn = {0047259X},
   journal = {Journal of Multivariate Analysis},
   month = {3},
   pages = {396-404},
   title = {Estimation of the conditional distribution of a multivariate variable given that one of its components is large: Additional constraints for the {H}effernan and {T}awn model},
   volume = {115},
   year = {2013},
}

@article{ewans2008effect,
  title={The effect of directionality on Northern {N}orth {S}ea extreme wave design criteria},
  author={Ewans, Kevin and Jonathan, Philip},
  journal={Journal of Offshore Mechanics and Arctic Engineering},
  year={2008},
volume={130},
pages={041604 }
}

@article{winter2017kth,
  title={kth-order {M}arkov extremal models for assessing heatwave risks},
  author={Winter, Hugo C and Tawn, Jonathan A},
  journal={Extremes},
  volume={20},
  number={2},
  pages={393--415},
  year={2017},
  publisher={Springer}
}

@inproceedings{NIPS2012_6364d3f0,
 author = {Osborne, Michael and Garnett, Roman and Ghahramani, Zoubin and Duvenaud, David K and Roberts, Stephen J and Rasmussen, Carl},
 booktitle = {Advances in Neural Information Processing Systems},
 editor = {F. Pereira and C.J. Burges and L. Bottou and K.Q. Weinberger},
 pages = {},
 publisher = {Curran Associates, Inc.},
 title = {Active Learning of Model Evidence Using {B}ayesian Quadrature},
 volume = {25},
 year = {2012}
}

@inproceedings{NIPS2014_a0d08267,
 author = {Gunter, Tom and Osborne, Michael A. and Garnett, Roman and Hennig, Philipp and Roberts, Stephen J.},
 booktitle = {Advances in Neural Information Processing Systems},
 editor = {Z. Ghahramani and M. Welling and C. Cortes and N. Lawrence and K.Q. Weinberger},
 pages = {},
 publisher = {Curran Associates, Inc.},
 title = {Sampling for Inference in Probabilistic Models with Fast {B}ayesian Quadrature},
 volume = {27},
 year = {2014}
}

@Article{jmse11030628,
AUTHOR = {Wan, Jian-Hong and Bai, Rui and Li, Xue-You and Liu, Si-Wei},
TITLE = {Natural Frequency Analysis of Monopile Supported Offshore Wind Turbines Using Unified Beam-Column Element Model},
JOURNAL = {Journal of Marine Science and Engineering},
VOLUME = {11},
YEAR = {2023},
NUMBER = {3},
pages = {628},
}

@article{YANG2025123120,
title = {Higher-harmonic resonance response of a monopile-supported offshore wind turbine in waves and wind},
journal = {Ocean Engineering},
volume = {342},
pages = {123120},
year = {2025},
author = {Hui Yang and Yi Zhang and Jinbo Lin and Hongfei Mao},
}

@article{LIU2025122355,
title = {High-frequency resonance of bottom-fixed monopile-supported offshore wind turbines under nonlinear waves},
journal = {Ocean Engineering},
volume = {340},
pages = {122355},
year = {2025},
author = {Jiawang Liu and Shutong Xu and Hongfei Mao and Chengwei Han and Jianbo Han},
}

@Article{jmse7120430,
AUTHOR = {He, Rui and Zhu, Tao},
TITLE = {Model Tests on the Frequency Responses of Offshore Monopiles},
JOURNAL = {Journal of Marine Science and Engineering},
VOLUME = {7},
YEAR = {2019},
NUMBER = {12},
pages = {430},
}

@misc{graham2017,
      title={{Continuously tempered Hamiltonian Monte Carlo}}, 
      author={Matthew M. Graham and Amos J. Storkey},
      year={2017},
      eprint={1704.03338},
      archivePrefix={arXiv},
      primaryClass={stat.CO},
}

@inproceedings{Luo2018,
 author = {Luo, Rui and Wang, Jianhong and Yang, Yaodong and Wang, Jun and Zhu, Zhanxing},
 booktitle = {Advances in Neural Information Processing Systems},
 editor = {S. Bengio and H. Wallach and H. Larochelle and K. Grauman and N. Cesa-Bianchi and R. Garnett},
 publisher = {Curran Associates, Inc.},
 title = {{Thermostat-assisted continuously-tempered Hamiltonian Monte Carlo for Bayesian learning}},
 year = {2018}
}

@article{Riise_Grue_Jensen_Johannessen_2018, 
title={High frequency resonant response of a monopile in irregular deep water waves}, 
volume={853}, 
journal={Journal of Fluid Mechanics}, 
author={Riise, Bjorn Hervold and Grue, John and Jensen, Atle and Johannessen, Thomas B.}, 
year={2018}, 
pages={564–586},
}

@misc{park2025,
      title={{Sampling from high-dimensional, multimodal distributions using automatically tuned, tempered Hamiltonian Monte Carlo}}, 
      author={Joonha Park},
      year={2025},
      eprint={2111.06871},
      archivePrefix={arXiv},
      primaryClass={stat.CO},
}

@article{Qiu2024,
author = {Yixuan Qiu and Xiao Wang},
title = {Efficient Multimodal Sampling via Tempered Distribution Flow},
journal = {Journal of the American Statistical Association},
volume = {119},
number = {546},
pages = {1446--1460},
year = {2024},
}

@article{Vousden2015,
    author = {Vousden, W. D. and Farr, W. M. and Mandel, I.},
    title = {Dynamic temperature selection for parallel tempering in Markov chain Monte Carlo simulations},
    journal = {Monthly Notices of the Royal Astronomical Society},
    volume = {455},
    number = {2},
    pages = {1919-1937},
    year = {2015},
}

@article{Miasojedow01072013,
author = {Blazej Miasojedow and Eric Moulines and Matti Vihola},
title = {An Adaptive Parallel Tempering Algorithm},
journal = {Journal of Computational and Graphical Statistics},
volume = {22},
number = {3},
pages = {649--664},
year = {2013},
}

@article{Faltinsen_Newman_Vinje_1995, title={Nonlinear wave loads on a slender vertical cylinder}, volume={289}, journal={Journal of Fluid Mechanics}, author={Faltinsen, O. M. and Newman, J. N. and Vinje, T.}, year={1995}, pages={179–198}}

@article{elhamahmy2025integrating,
  title={Integrating Artificial Intelligence in Offshore Platform Design: Enhancing Efficiency, Safety, and Structural Integrity},
  author={ElHamahmy, Dr Ahmed},
  journal={Safety, and Structural Integrity (June 01, 2025)},
  year={2025}
}

@article{yu2024construction,
  title={Construction high precision neural network proxy model for ship hull structure design based on hybrid datasets of hydrodynamic loads},
  author={Yu, Ao and Li, Yunbo and Li, Shaofan and Gong, Jiaye},
  journal={Journal of Marine Science and Application},
  volume={23},
  number={1},
  pages={49--63},
  year={2024},
  publisher={Springer}
}

@article{ALVESRIBEIRO2025126294,
title = {{Offshore wind turbine tower design and optimization: a review and AI-driven future directions}},
journal = {Applied Energy},
volume = {397},
pages = {126294},
year = {2025},
issn = {0306-2619},
author = {João {Alves Ribeiro} and Bruno {Alves Ribeiro} and Francisco Pimenta and Sérgio {M.O. Tavares} and Jie Zhang and Faez Ahmed}
}

@Article{jmse12112001,
AUTHOR = {Martzikos, Nikolas and Ruzzo, Carlo and Malara, Giovanni and Fiamma, Vincenzo and Arena, Felice},
TITLE = {Applying Neural Networks to Predict Offshore Platform Dynamics},
JOURNAL = {Journal of Marine Science and Engineering},
VOLUME = {12},
YEAR = {2024},
NUMBER = {11},
ARTICLE-NUMBER = {2001},
ISSN = {2077-1312},
}

@article{Zhou2025, 
author = {Xiaoguang Zhou and Chao Hou and Yantao Yu and Yifan Zhou},
title = {Machine learning-based techniques for marine structures: A state-of-the-art review},
year = {2025},
journal = {Ocean},
volume = {1},
number = {1},
pages = {9470005}
}

@article{https://doi.org/10.1002/we.2851,
author = {Kirby, Andrew and Briol, François-Xavier and Dunstan, Thomas D. and Nishino, Takafumi},
title = {Data-driven modelling of turbine wake interactions and flow resistance in large wind farms},
journal = {Wind Energy},
volume = {26},
number = {9},
pages = {968-984},
year = {2023}
}

@article{MERT2025122952,
title = {A reliability-based framework for offshore monopile design using CPT data and deep learning enhanced adaptive metamodeling},
journal = {Ocean Engineering},
volume = {342},
pages = {122952},
year = {2025},
issn = {0029-8018},
author = {Ahmet Can Mert and Xiangfeng Guo and Zhichao Shen and Huayun Pan and Daniel Dias}
}

\newpage

\setcounter{section}{0}
\setcounter{equation}{0}

\begin{center}
    \Large
    \textbf{Supplementary Material to `Sequential Design for the Efficient Estimation of Offshore Structure Failure Probability'}
\end{center}
\begin{center}
    \large 
    Matthew Speers, Philip Jonathan, Jonathan Tawn 
    \vspace{.3cm}
    
    \small \textit{School of Mathematical Sciences, Lancaster University LA1 4YF, United Kingdom}	
\end{center}

% for SM labelling %
\renewcommand{\theequation}{SM.\arabic{equation}}
\renewcommand{\thesection}{SM\arabic{section}} 
\renewcommand{\thefigure}{SM\arabic{figure}}
\setcounter{figure}{0}
\renewcommand{\theHfigure}{SM\arabic{figure}}
\renewcommand{\thetable}{SM\arabic{table}}
\setcounter{table}{0}

\section{Alternative methodology to that presented in Section~\ref{section:methodology} of the main text}
\subsection{Gaussian process-informed importance sampling} \label{section:akmc}

Referring to Section~\ref{section:monte_carlo} of the main text,  authors \cite{Xiao2020} and \cite{lystad2023full}, use Gaussian emulation to inform their choice of importance sampling density. We briefly propose a similar approach using the GP emulators of the main article. This avoids the need to run the MCMC sampler described in the main article, allowing either (a) more budget allocation to the evaluation of importance sampling estimate $\hat{p}_\text{IS}$, or (b) a reduction in total computational cost. 

Our approach mirrors that of \cite{lystad2023full}, who create a uniform proposal density with support informed by a GP estimate of the CDE. Given an estimate $\hat{f}_{\longX|R>r_\text{Cr}}^{(n)}$ of the CDE, found via \eqref{equation:cde_est} using the $n$th-iterate GP emulator defined in the main article, we define a proposal density
\begin{equation} \label{eq:isgp_prpsl}
    g^{(n)}_\text{Pr}(\longx) = \frac{1}{A_n} \quad \text{for} \quad A_n = \int_{\mathcal{E}_\longX} \mathbb{I} \left\{\hat{f}_{\longX|R>r_\text{Cr}}^{(n)}(\longx)>\delta\right\} \wrt{\longx},
\end{equation}
for some $\delta\in[0,1]$. That is, we draw a proposal sample $\longx_1^*, \ldots, \longx_N^*$ uniformly on the region where the estimate of the CDE at the current iteration $n$ is greater than some $\delta$.

\subsection{Gaussian process emulation of failure probability} \label{sm:logistic}

We consider an alternate emulator construction to that shown in Section~\ref{section:emulation} of the main article. Here, we emulate only the conditional failure probability $\Prb{R>r_\text{Cr}|\{\longX=\longx\}}$, rather than the entire integrand $\Prb{R>r_\text{Cr}|\{\longX=\longx\}} f_\longX(\longx)$. Since probabilities must always be observed on the unit interval, we map the Gaussian emulator output $w(\longx)\in \RR$ onto the range $[0,1]$ via the logistic function $g_\text{Lg}:\RR \mapsto [0,1]$, $g_\text{Lg}(w) =  e^w/(1+e^{w})$, modelling
\begin{equation}
\label{equation:paper_gp_posterior}
    w(\longx) = g^{-1}_\text{Lg} ( \Pr (G_R(\longx) >r_\text{Cr}))\sim  \mathcal{GP}(\mu_\text{GP}(\longx), k(\longx, \longx')), \quad w:\mathcal{E}_\longX \mapsto \RR,
\end{equation}
for mean and covariance functions
\begin{align} 
\mu_\text{GP}(\longx)&=\mathbb{E}[w(\longx)], \quad \mu_\text{GP}: \mathcal{E}_\longX \rightarrow \RR, \\
     k\left(\longx, \longx'\right)&=\mathbb{E}\left[\left(w(\longx)-\mu(\longx)\right)\left(w(\longx')-\mu\left(\longx'\right)\right)\right], \quad k(\longx, \longx'): \mathcal{E}_\longX \times \mathcal{E}_\longX \rightarrow \mathbb{R}.
\end{align}
This emulator may then be trained according to the posterior update steps defined in the main article. The target marginal failure probability estimate $\hat{p}_\text{GP}$ can then be summarised using
\begin{align}
        \hat{p}_\text{GP} &= \mathbb{E}_{W,\longX}(\{g_\text{Lg}(w(\longx))\})  \\
       &= \int_{\mathcal{E}_\longX} \left\{ \int_{\RR}  g_\text{Lg}(w) \phi(w ; \mu^*_\text{GP}(\mathbf{\longx}), k^*(\longx, \longx)) \wrt{w} \right\}  f_{\longX} (\longx)\wrt\longx,\label{eq:gp_target}
\end{align}
for $W|\{\longX=\longx\} \sim N(\mu^*_\text{GP}(\mathbf{\longx}), k^*(\longx, \longx))$ with parameters obtained from \eqref{equation:paper_gp_posterior}. The estimate \eqref{eq:gp_target} can be written
\begin{equation}\label{equation:gp_target_approx}
    \hat{p}_\text{GP} \approx \int_\longE g_\text{Lg} \left( \frac{\mu^*_\text{GP}(\longx)}{\sqrt{1 + \pi k(\longx, \longx')/8}} \right) f_\longX(\longx) \wrt{\longx},
\end{equation}
by the approximation for the convolution of a logistic sigmoid function with a Gaussian density given in Section 4.5.2 of \cite{bishop2006pattern}. Values for \eqref{equation:gp_target_approx} may then be obtained via numerical integration. In this setting, the CDE estimate of the GP emulator at iteration $n$ becomes
\begin{equation}\label{eq:cde_gp_est_full}
    \hat{f}^{(n)}_{\longX|R>r_\text{Cr}}(\longx) = \frac{g_\text{Lg} \left( {\mu_n^*(\longx)(1 + \pi k_n^*(\longx, \longx')/8)^{-\frac{1}{2}}} \right) f_\longX(\longx) }{\int_{\mathcal{E}_\longX} g_\text{Lg} \left( {\mu_n^*(\longx)(1 + \pi k_n^*(\longx, \longx')/8)^{-\frac{1}{2}}} \right) f_\longX(\longx) \wrt\longx}.
\end{equation}

% That is, similar to as in \eqref{equation:paper_acq_function}, we use the delta method to account for the logistic mapping in \eqref{equation:paper_gp_posterior}. The ALC criteria \eqref{equation:alc} then becomes
% %
% \begin{equation}\label{eq:alc_trns}
%     \text{ALC}'(\mathbf{x}) = \frac{1}{N_\text{ref}} \sum^{n_\text{ref}}_{i=1} [g'_\text{Lg}(\mu_n(\longx))]^2k_n(\longx_i, \longx_i) - [g'_\text{Lg}(\mu_{n+1}(\longx))]^2\tilde{k}_{n+1}(\longx_i, \longx_i; \longx).
% \end{equation}
% %
% Since we have no knowledge of $\mu_{n+1}(\longx)$ at iteration $n$, we make the approximation
% \begin{align}
%     &[g'_\text{Lg}(\mu_n(\longx))]^2k_n(\longx_i, \longx_i) - [g'_\text{Lg}(\mu_{n+1}(\longx))]^2\tilde{k}_{n+1}(\longx_i, \longx_i; \longx) = \\
%     &= [g'_\text{Lg}(\mu_n(\longx))]^2[k_n(\longx_i, \longx_i)- \tilde{k}_{n+1}(\longx_i, \longx_i; \longx) ] \\
%     &+ \tilde{k}_{n+1}(\longx_i, \longx_i; \longx)\left\{[g'_\text{Lg}(\mu_n(\longx))]^2 - [g'_\text{Lg}(\mu_{n+1}(\longx))]^2\right\}  
%     \\&\approx [g'_\text{Lg}(\mu_n(\longx))]^2[k_n(\longx_i, \longx_i)- \tilde{k}_{n+1}(\longx_i, \longx_i; \longx) ] ,
% \end{align}
% by assuming the change in variance term dominates the change in squared mean term. Numerical experiments found this assumption to be acceptable, see SM. This allows us to write 
% \begin{equation}\label{eq:alc_approx}
%      \text{ALC}'(\mathbf{x}) \approx \frac{1}{N_\text{ref}} \sum^{N_\text{ref}}_{i=1}[g'_\text{Lg}(\mu_n(\longx))]^2[k_n(\longx_i, \longx_i)- \tilde{k}_{n+1}(\longx_i, \longx_i; \longx) ].
% \end{equation}

\section{JONSWAP wave spectrum discussed in Section~\ref{section:TFNV_wave} of the main text} \label{SM:jonswap}
The JONSWAP spectral density \citep{hasselmann1973measurements} is used to simulate linear random waves in the monopile application of the main article. It is defined, in terms of angular frequency $\omega = 2 \pi f$, as
\begin{equation}
S(\omega ; \mathbf{x})=\alpha \omega^{-r} \exp \left\{-\frac{r}{4}\left(\frac{|\omega|}{\omega_p(\mathbf{x})}\right)^{-4}\right\} \gamma^{\delta(\omega ; \mathbf{x})},
\end{equation}
for $\omega>0$, where $\mathbf{X}=(H_s, S_e)$ and $\omega_p(\mathbf{x}) = 2\pi/t(\mathbf{x})$, where $t(\mathbf{x})$ is the observed value of the second spectral moment wave period $T_2 = \sqrt{({2\pi H_S})/({g S_e})}$ in sea state $\mathbf{X}=\mathbf{x}$, with 
\begin{equation}
\delta(\omega ; \mathbf{x})=\exp \left\{-\frac{1}{2\left(0.07+0.02 \cdot I\{{\omega_p(\mathbf{x})}>|\omega|\}\right)^2}\left(\frac{|\omega|}{\omega_p(\mathbf{x})}-1\right)^2\right\},
\end{equation}
and constants $r, \alpha, \gamma > 0$. The Phillips constant $\alpha$ is chosen so that 
\begin{equation}
4\cdot\left\{\int_{-\infty}^\infty S(\omega; \mathbf{x})\wrt{\omega}\right\}^{\frac{1}{2}} = h(\mathbf{x}),
\end{equation}
where $h(\mathbf{x})$ is the observed value of significant wave height $H_S$ in sea state $\mathbf{X}=\mathbf{x}$.

\section{Supplementary results to case studies of Section~\ref{section:synth_study} of the main text}

\subsection{Importance sampling coupled with parallel tempering MCMC (IS-PT)} \label{SM:ispt}

\begin{figure}[H]
    \centering
    \includegraphics[width=.95\linewidth]{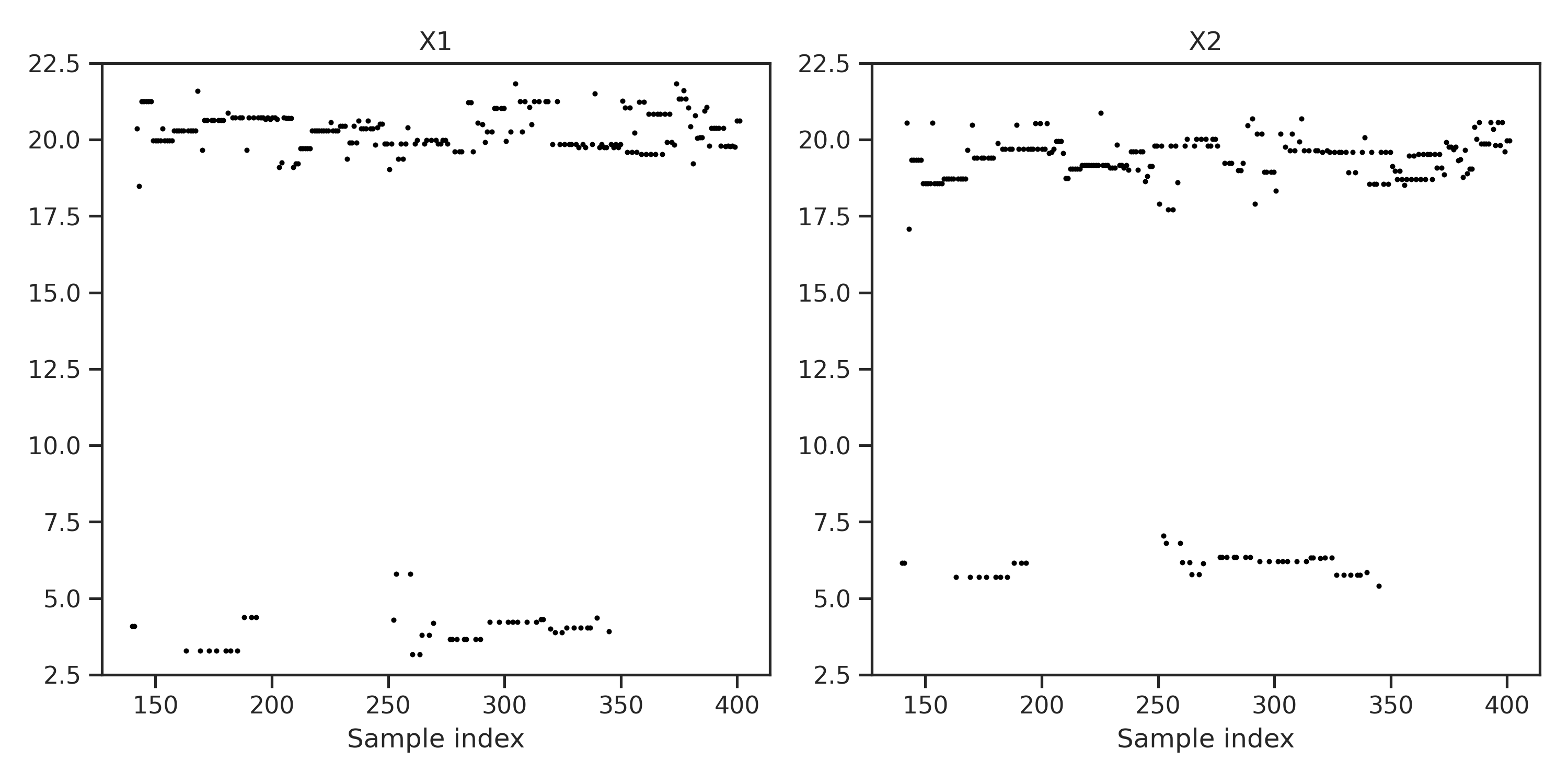}
    \caption{Example trace plot of sample from synthetic CDE  discussed in Section~\ref{section:synthetic_study}, obtained from adaptive parallel tempering algorithm of \cite{Vousden2015} at temperature $T_1=1$, used in IS-PT approach for estimation of proposal density. Sample is initially of length $n_\text{PT}=400$, with $n_\text{Br}=144$ discarded (not plotted).}
    \label{fig:trace_synth}
\end{figure}

\begin{figure}[H]
    \centering
    \includegraphics[width=.95\linewidth]{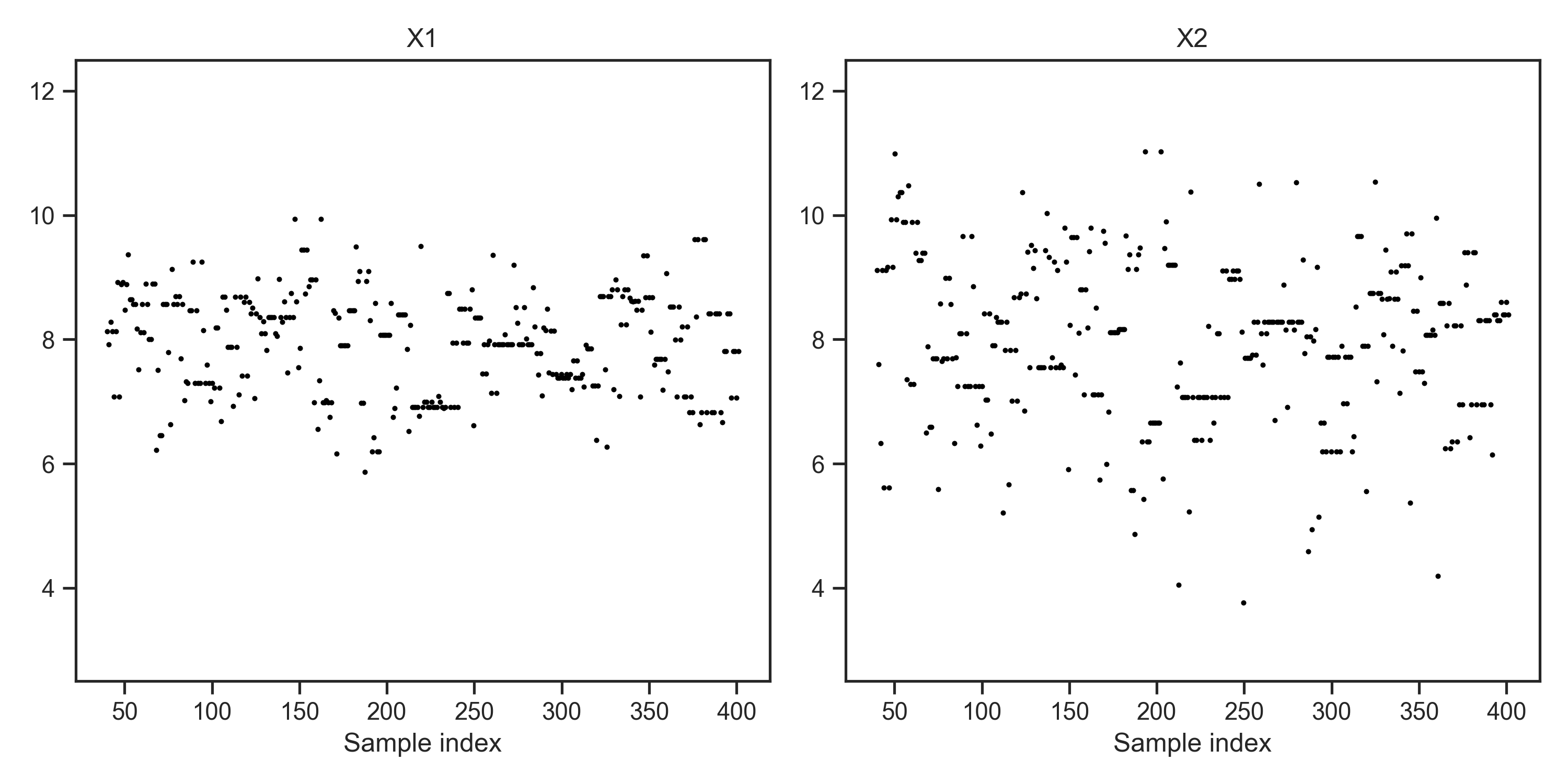}
  \caption{Example trace plot of sample from monopile CDE discussed in Section~\ref{section:application}, obtained from the adaptive parallel tempering algorithm of \cite{Vousden2015} at temperature $T_1=1$, used in IS-PT approach for estimation of proposal density. Sample is initially of length $n_\text{PT}=400$, with $n_\text{Br}=40$ discarded (not plotted).}
    \label{fig:trace_tfnv}
\end{figure}

\newpage

\subsection{Adaptive Gaussian emulation (AGE)} \label{SM:alc}

\begin{figure}[H]
    \centering
    \includegraphics[width=1\linewidth]{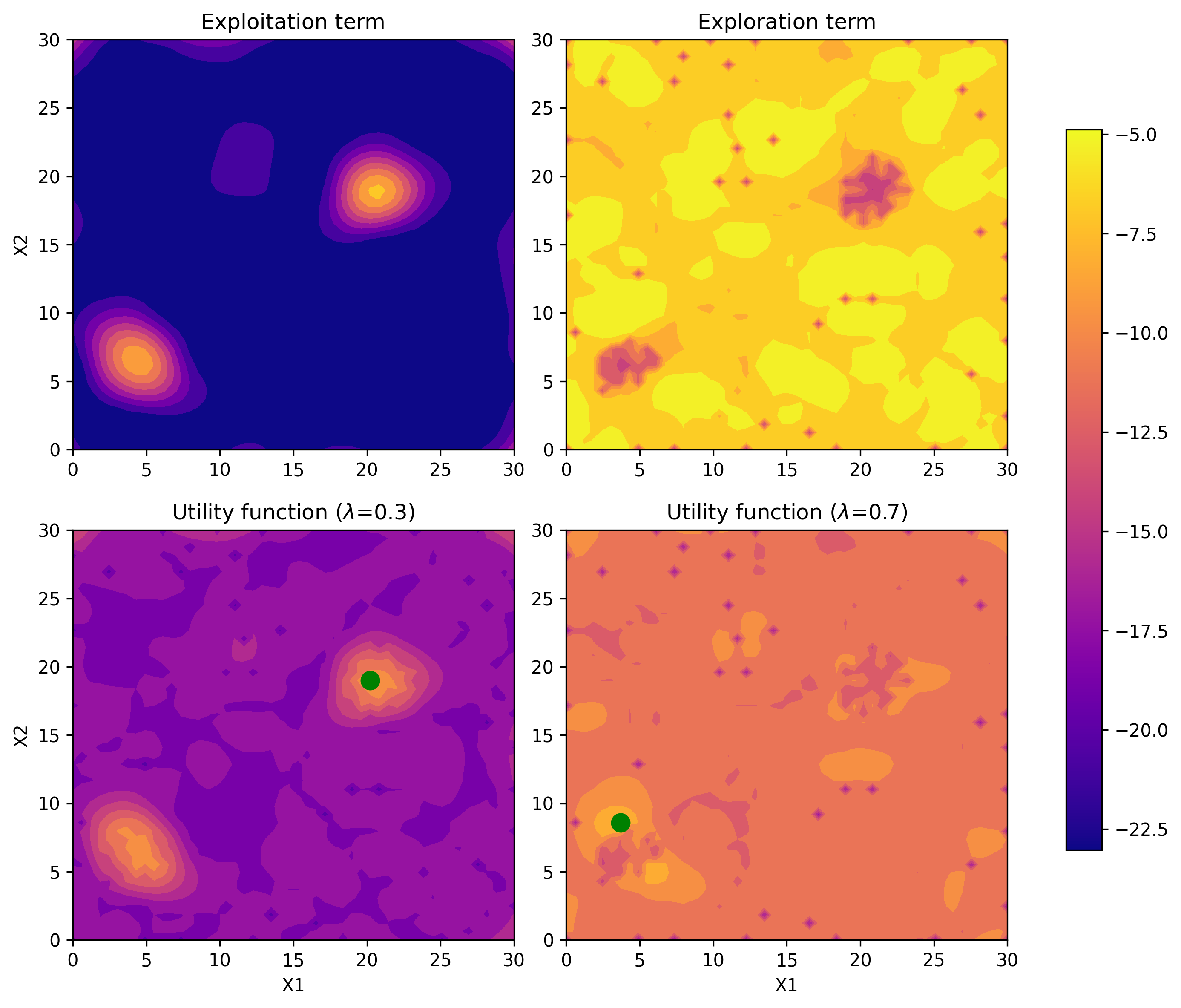}
     \caption{Behaviour of utility function $U^{(2)}(\longx; \lambda)$ over the environment space $\longE$, for synthetic scenario. Upper panels show exploitation and exploration terms obtained from GP emulator \eqref{equation:gp_log} trained on initial Latin hypercube set $\mathcal{D}_0$ of size $n_\text{Tr1}=144$. Lower panels show resulting utility functions for weights $\lambda=0.3$ and $\lambda=0.7$. In each lower panel, the optimal sampling point $\longx^* =\argmax_{\longx \in \longE} U^{(2)}(\longx; \lambda)$ is indicated in green. To be compared with Figure~\ref{fig:utility_gramstad} of the main text.}
    \label{fig:alc_utility}
\end{figure}

\begin{figure}[H]
    \centering
    \includegraphics[width=1\linewidth]{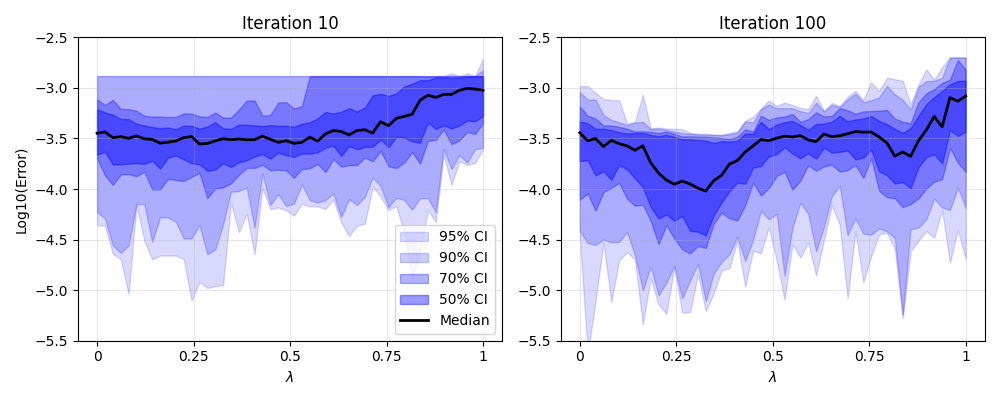}
  \caption{Log-scale absolute error $\Delta_\text{GP}$ of the GP probability estimate $\hat{p}_\text{GP}$ at specified iterations, for emulator \eqref{equation:gp_log} trained using $U^{(2)}$ over the range of weight $\lambda \in [0.01, 0.99]$ for the synthetic scenario. At iteration 100, the weight that minimises median error is $\lambda^*=0.33$. To be compared with Figure~\ref{fig:nu_sens1} of the main text.}
    \label{fig:alc_nu_sens}
\end{figure}

\begin{figure}[H]
    \centering
    \includegraphics[width=.75\linewidth]{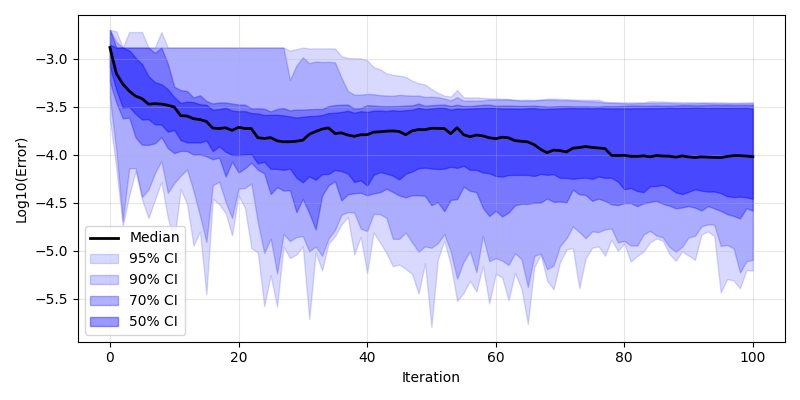}
  \caption{Distribution of log-scale absolute error $\Delta_\text{GP}$ in the GP probability estimate with respect to iteration, trained using $U^{(2)}$ with $\lambda=\lambda^*$. The trend in median error is indicated in black, with various confidence intervals shown in blue. To be compared with Figure~\ref{fig:opt_lambda_error} of the main text.}
    \label{fig:enter-label}
\end{figure}

\newpage

\section{Supplementary results to monopile case study of Section~\ref{section:application} of the main text}
\subsection{Extreme value model threshold selection supporting the discussion of Section~\ref{section:dens_est} of the main text}\label{SM:threshold}

\begin{figure}[H]
    \centering
    \includegraphics[width=1\linewidth]{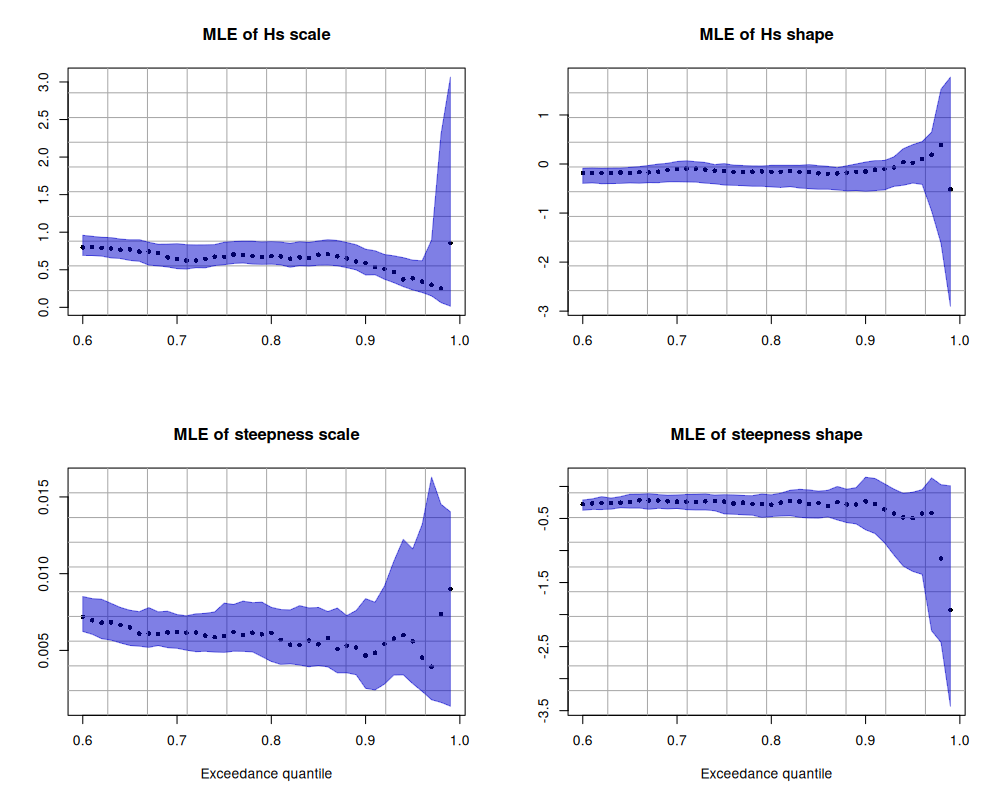}
    \caption{Threshold stability plots for estimates of the generalised Pareto scale and shape parameters $\sigma$ and $\xi$ when fitted to $H_s$ and $S_e$ above a range of values of the conditioning threshold $u>0$. The estimates of $\sigma$ and $\xi$ are given on the y-axes, with the respective threshold quantile  $q_u= \tilde{F}_{H_s}^{-1}(u)$ and $q_u= \tilde{F}^{-1}_{S_e}(u)$ indicated on the x-axes, for empirical distributions $\tilde{F}_{H_s}$ of $H_s$ and $\tilde{F}_{S_e}$ of $S_e$. Point estimates from original Albany data are given in black, and bootstrapped 95\% confidence intervals are shown as a blue region. Stability of estimates for $\xi$ and linearity of estimates of $\sigma$ above a threshold quantile $q_u$ indicates that $u$ is a suitable choice for GPD model threshold. Following visual analysis of the four panels we select threshold $u_1 = \tilde{F}_{H_s}^{-1}(0.7)$ and $u_2 = \tilde{F}_{S_e}^{-1}(0.7)$ for marginal modelling of $H_s$ and $S_e$ respectively.}
    \label{fig:marg_thresh}
\end{figure}

\begin{figure}[H]
    \centering
    \includegraphics[width=1\linewidth]{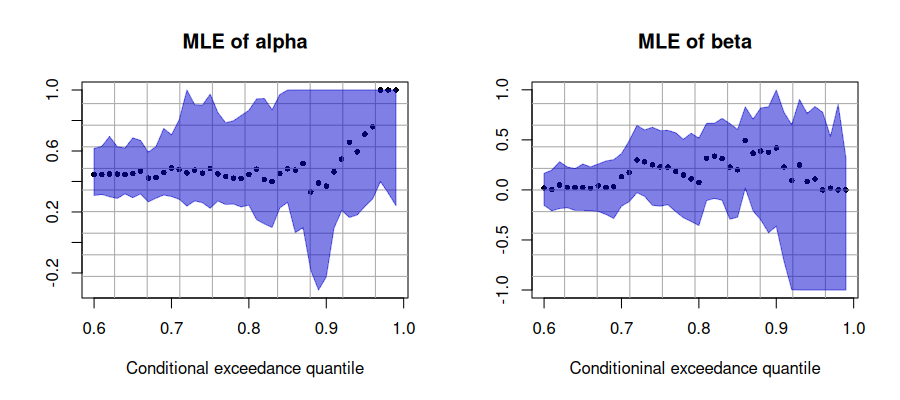}
    \caption{Threshold stability plots for estimates of the conditional extremes parameters $\alpha$ and $\beta$ when fitted to $(H_s, S_e)|\{H_s>v\}$, for a range of values of the conditioning threshold $v>0$. The estimates of $\alpha$ and $\beta$ are given on the y-axes and the respective quantile $q_v =  \tilde{F}_{H_s}^{-1}(u)$ on the x-axes. Point estimates from original Albany data are given in black, and bootstrapped 95\% confidence intervals are shown as a blue region. Stability of parameter estimates above a threshold quantile $q_v$ indicates that $v$ is a suitable choice for conditional model threshold. Following visual analysis of the two panels, we select threshold $v = \tilde{F}_{H_s}^{-1}(0.6)$ for joint modelling of $H_s$ and $S_e$.}
    \label{fig:joint_thresh}
\end{figure}

{The Heffernan-Tawn model assumes Gaussian residual form to facilitate model fitting only, but adopts the empirically-evaluated residuals from the model fit to the sample for inferences under the model. For this reason, assessment of model adequacy using residual diagnostics using e.g., Q-Q plots is of no relevance for the Heffernan-Tawn model.}  

\subsection{Extreme value density estimation supporting the discussion of Section~\ref{section:dens_est} of the main text.}\label{SM:dens}

\begin{figure}[h]
    \centering
    \includegraphics[width=.8\linewidth]{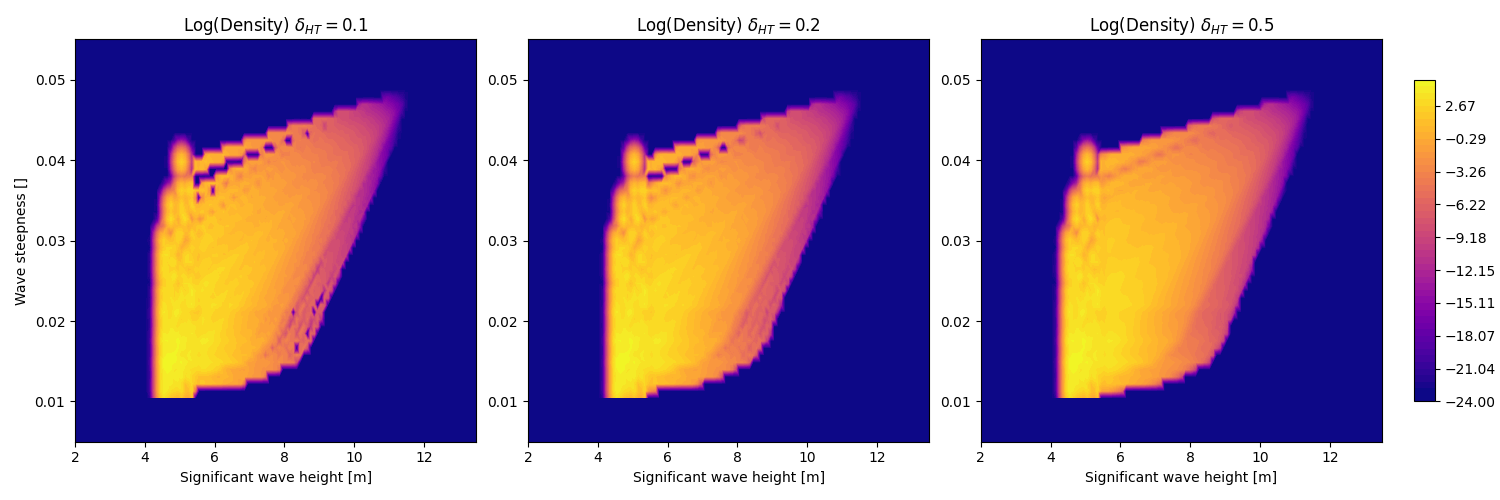}
    \caption{Sensitivity analysis of estimated joint density $\hat{f}_\longX$ of $\longX= (H_s, S_e)$ with conditional extremes smoothing parameter $\delta_\text{HT}$. We aim to obtain the smallest value of $\delta_\text{HT}$ which eliminates `gaps' in the extrapolated region. Following visual inspection of the three panels and Figure \ref{fig:dens_0.4} of the main article, we take $\delta_\text{HT}=0.4$.} 
\end{figure}

\newpage

\subsection{Non-linear harmonic response simulation} \label{SM:tfnv}

\begin{figure}[H]
    \centering
    \includegraphics[width=1\linewidth]{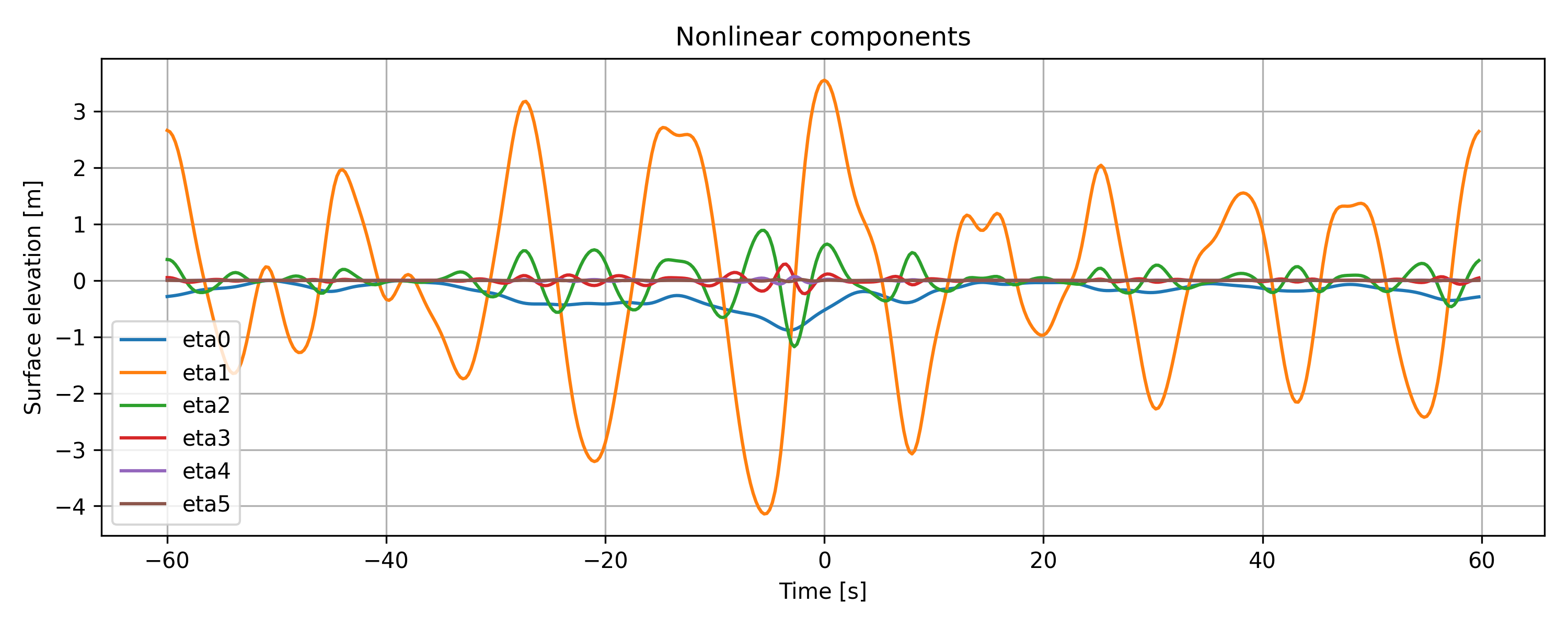}
    \caption{Harmonic signals constructed using the method of \cite{Orszaghova2025}, from a linear surface elevation input. The 0-5th order harmonics are shown. To support the discussion in Section~\ref{section:sim_details}.}
    \label{fig:higher_order}
\end{figure}

\begin{figure}[H]
    \centering
    \includegraphics[width=.75\linewidth]{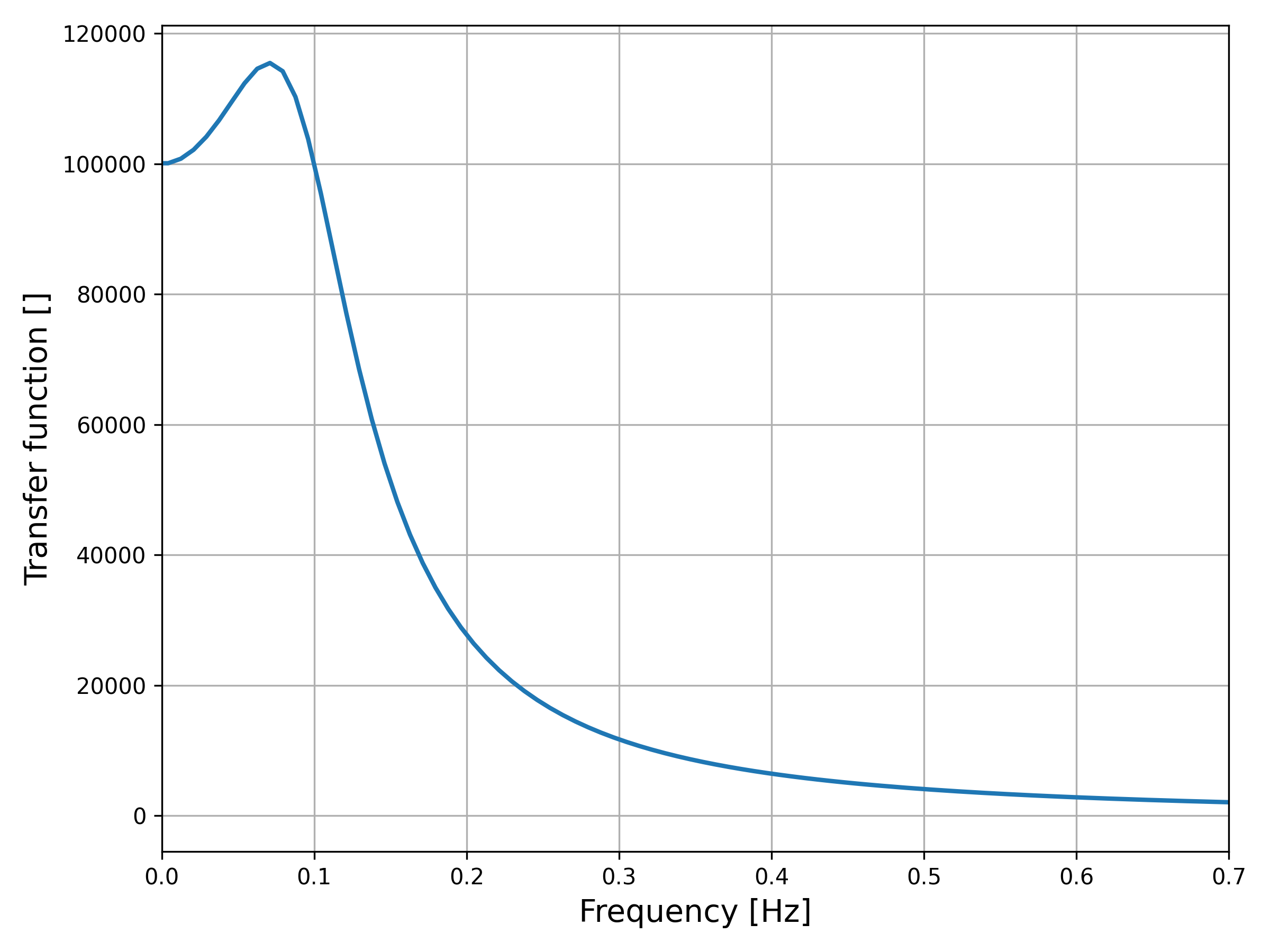}
    \caption{Transfer function of a damped harmonic oscillator  \eqref{eq:trans}, used in the case study of Section~\ref{section:TFNV_wave} of the main text. The transfer function is plotted against input frequency [Hz], with resonant frequency $f_0=1/10$. The output of the transfer function is assumed unit-less for our case study.}
    \label{fig:trans}
\end{figure}

\section{Supplementary results to the AGE results of Section~\ref{section:tfnv_age} of the main text} \label{section:tfnv_alc}

\begin{figure}[H]
    \centering
    \includegraphics[width=1\linewidth]{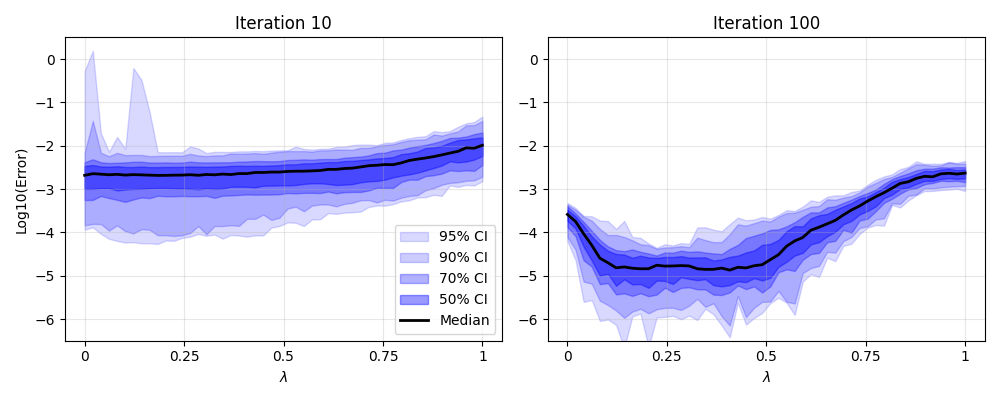}
  \caption{Log-scale absolute error $\Delta_\text{GP}$ of the GP probability estimate $\hat{p}_\text{GP}$ at specified iterations, for emulator \eqref{equation:gp_log} trained using $U^{(2)}$ over the range of weight $\lambda \in [0.01, 0.99]$ for the TFNV scenario. At iteration 100, the weight that minimises median error is $\lambda^*=0.41$. To be compared with Figure~\ref{fig:nu_sens_tfnv} of the main text.}
    \label{fig:alc_nu_sens2}
\end{figure}

\begin{figure}[H]
    \centering
    \includegraphics[width=.75\linewidth]{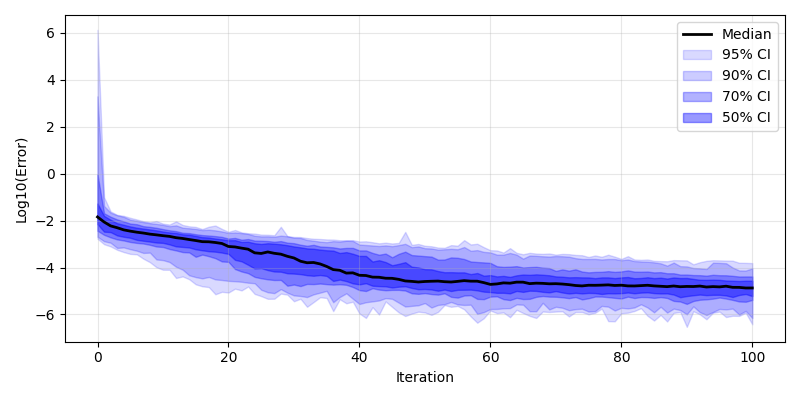}
  \caption{Distribution of log-scale absolute error $\Delta_\text{GP}$ in the GP probability estimate with respect to iteration, trained using $U^{(2)}$ with $\lambda=\lambda^*$. The trend in median error is indicated in black, with various confidence intervals shown in blue. To be compared with Figure~\ref{fig:opt_error_tfnv} of the main text.}
    \label{fig:alc_best_tfnv}
\end{figure}

\end{document}